\documentclass[11pt]{article}
\pdfoutput=1

\usepackage{jheppub}
\usepackage{amssymb,amsfonts,amsmath,amsthm,lineno}
\usepackage{enumerate}
\usepackage{mathrsfs}
\usepackage{bbold}
\usepackage{verbatim}
\usepackage[dvipsnames]{xcolor}
\usepackage{tikz, adjustbox}
\usepackage{tikz-cd}
\usetikzlibrary{arrows.meta, bending}
\usetikzlibrary{shapes.geometric, arrows}
\usepackage{hyperref}
\usepackage{mathtools}

\usepackage{mathabx,graphicx}
\def\Circlearrowleft{\ensuremath{%
  \rotatebox[origin=c]{90}{$\circlearrowright$}}}

\makeatletter
\newcommand{\Vast}{\bBigg@{4.75}}
\makeatother

\newcommand{\be}{\begin{equation}}
\newcommand{\ee}{\end{equation}}
\newcommand{\bea}{\begin{eqnarray}}
\newcommand{\eea}{\end{eqnarray}}

\newcommand{\CA}{\mathcal{A}}
\newcommand{\CB}{\mathcal{B}}
\newcommand{\CC}{\mathcal{C}}

\newcommand{\CF}{\mathcal{F}}
\newcommand{\CG}{\mathcal{G}}
\newcommand{\CH}{\mathcal{H}}

\newcommand{\CN}{\mathcal{N}}
\newcommand{\CM}{\mathcal{M}}

\newcommand{\CR}{\mathcal{R}}

\newcommand{\lr}{\left (}
\newcommand{\rr}{\right )}

\newcommand\qt\tau

\newcommand{\wt}{\widetilde}

\newcommand{\p}{\partial}

\renewcommand{\tilde}[1]{\widetilde{#1}}
\newcommand{\tr}{\text{tr}}
\renewcommand{\emph}[1]{\textit{#1}}

\makeatletter
\renewcommand{\@seccntformat}[1]{\csname the#1\endcsname.\,\,}
\makeatother
\allowdisplaybreaks

\DeclareMathOperator{\sgn}{sgn}

\let \savenumberline \numberline
\def \numberline#1{\savenumberline{#1.}}

\makeatletter
\def\@fpheader{\relax}
\makeatother

\usetikzlibrary{shapes,arrows,chains}
\usetikzlibrary{decorations.markings}
\usetikzlibrary{decorations.pathmorphing}
\usetikzlibrary{shapes.multipart}
\usetikzlibrary{positioning}
\tikzset{snake it/.style={decorate, decoration=snake}}

\newcommand{\dd}{\mathrm{d}}

\newcommand{\gM}{\mathcal{M}}
\newcommand{\cH}{\mathcal{H}}

\newcommand{\om}{\boldsymbol{\alpha}}
\newcommand{\on}{\boldsymbol{\beta}}
\newcommand{\op}{\boldsymbol{\gamma}}
\newcommand{\oq}{\boldsymbol{\delta}}

\usepackage{mdframed}
\newmdenv[
topline=true, bottomline=true, rightline=true, leftline=true,
linewidth=1pt, linecolor=red!75, skipabove=1em, skipbelow=0em, backgroundcolor=red!10
]{rmrkred}
\newmdenv[
topline=true, bottomline=true, rightline=true, leftline=true,
linewidth=1pt, linecolor=vub!75, skipabove=1em, skipbelow=0em, backgroundcolor=vub!10
]{rmrkblue}
\newmdenv[
topline=true, bottomline=true, rightline=true, leftline=true,
linewidth=1pt, linecolor=PineGreen!75, skipabove=1em, skipbelow=0em, backgroundcolor=PineGreen!10
]{rmrkgreen}
\newmdenv[
topline=true, bottomline=true, rightline=true, leftline=true,
linewidth=1pt, linecolor=black!75, skipabove=1em, skipbelow=0em, backgroundcolor=black!10
]{rmrkblack}

\definecolor{vub}{RGB}{0,52,154}
\definecolor{vubo}{RGB}{255,102,0}
\definecolor{redd}{RGB}{255,40,40}
\definecolor{r}{RGB}{228,32,20}
\definecolor{o}{RGB}{238,69,4}
\definecolor{y}{RGB}{253,228,1}
\definecolor{g}{RGB}{108,160,0}
\definecolor{b}{RGB}{0,162,203}
\definecolor{i}{RGB}{120,42,117}
\definecolor{vred}{rgb}{0.78, 0.03, 0.08}

\newcommand{\CCpoly}{\mathcal{C}_{\text{Poly}}}

\title{Dual Non-Lorentzian Backgrounds for Matrix Theories
}
\author[a]{Chris D. A. Blair,}

\emailAdd{c.blair@csic.es}

\author[b]{Johannes Lahnsteiner,}
\emailAdd{j.m.lahnsteiner@outlook.com}
\author[b,c]{Niels A. Obers,}

\emailAdd{obers@nbi.ku.dk}
\author[b]{and Ziqi Yan\medskip}
\emailAdd{ziqi.yan@su.se}
\affiliation[a]{Instituto de F\'{i}sica Te\'{o}rica UAM/CSIC, Universidad Aut\'{o}noma de Madrid, Cantoblanco, Madrid 28049, Spain \smallskip}
\affiliation[b]{Nordita, KTH Royal Institute of Technology and Stockholm University,
Hannes Alfv\'{e}ns v\"{a}g 12, SE-106 91 Stockholm, Sweden \smallskip}
\affiliation[c]{Center of Gravity, The Niels Bohr Institute, Copenhagen University, Blegdamsvej 17,
DK-2100 Copenhagen \O, Denmark}

\preprint{IFT-UAM/CSIC-25-18, NORDITA 2025-014}

\abstract{
We study properties of non-Lorentzian geometries arising from BPS decoupling limits of string theory that are central to matrix theory and the AdS/CFT correspondence.
We focus on duality transformations between ten-dimensional non-Lorentzian geometries coupled to matrix theory on D-branes.
We demonstrate that T- and S-duality transformations exhibit novel asymmetric properties: 
depending not only on the choice of transformation but also on the value of the background fields,
the codimension of the foliation structure of the dual non-Lorentzian background may be different or the same.  
This \emph{duality asymmetry} underlies features observed in the study of non-commutativity and Morita equivalence in matrix and gauge theory.
Finally, 
we show how the holographic correspondence involving non-commutative Yang-Mills fits into our framework, from which we further obtain novel holographic examples with non-Lorentzian bulk geometries.}

\begin{document}

\maketitle

\section{Introduction}

In this paper, we uncover novel features of string theory dualities, which appear after taking BPS decoupling limits that are central to matrix theory and holography.
These limits zoom in on background BPS objects, and lead to the appearance of \emph{non-Lorentzian geometry} describing the resulting background.
One can often think of these limits as defining self-contained corners of string theory (and M-theory).

In previous work~\cite{Blair:2023noj, Gomis:2023eav} we have set out a roadmap to explore and unify the space of such limits. Building on this perspective, we showed in~\cite{Blair:2024aqz} that holography in string theory is remarkably just another example of 
an asymptotic 
BPS decoupling limit associated with an asymptotic ten-dimensional non-Lorentzian geometry, coupled to the branes on which the dual field theory lives.
We emphasise that this statement applies to standard AdS/CFT with a relativistic bulk. The perspective of~\cite{Blair:2024aqz} not only highlights the common origin of holographic~\cite{Maldacena:1997re, Itzhaki:1998dd}, matrix theoretic \cite{Banks:1996vh, Susskind:1997cw, Seiberg:1997ad, Sen:1997we} and `non-relativistic' limits \cite{Klebanov:2000pp, Gomis:2000bd, Danielsson:2000gi, Danielsson:2000mu} in string and M-theory, but also allows one to 
classify new non-Lorentzian versions of AdS/CFT.\,\footnote{Including those proposed in \cite{Lambert:2024uue,Lambert:2024yjk}, while see \cite{Gomis:2005pg, Fontanella:2024rvn, Fontanella:2024kyl} for another proposal with a different treatment of the coordinates.} See also~\cite{Danielsson:2000mu, Avila:2023aey, Guijosa:2023qym} for insights from the dual non-relativistic string side, and~\cite{Lambert:2024ncn, Harmark:2025ikv, Guijosa:2025mwh} for further recent developments.

String duality plays an essential role in generating and unifying the different decoupling limits in this program. 
Upon compactification, seemingly distinct non-Lorentzian geometries may turn out to be equivalent due to a duality relation inherited from string theory prior to the associated BPS decoupling limit(s).
We have used this fact extensively in~\cite{Blair:2023noj, Gomis:2023eav, Blair:2024aqz} to build the duality web of these decoupling limits and their resulting corners of string and M-theory.\,\footnote{A complementary perspective is to use the BPS mass spectrum to classify the U-dual orbits of these decoupling limits~\cite{bpslimits}.}

In this paper, we will take a further step and  
spell out a number of interesting intricacies of string duality applied to non-Lorentzian geometry for matrix theory.
We will see that the nature of the dual non-Lorentzian geometry, which is characterised by the codimension of its foliation structure, is not determined solely by the choice of duality transformation but also by the value of the original background fields.
We will explain how this behaviour is inherited by non-commutative gauge theory on D-branes, extending the understanding of T-duality as Morita equivalence \cite{Connes:1997cr, Seiberg:1999vs}.
Combining this connection with the framework of \cite{Blair:2024aqz}, we will further uncover the underlying non-Lorentzian prescriptions governing holographic duals of non-commutative Yang-Mills.
The reader may be surprised that a geometrically very different but arguably more complete viewpoint 
can still be uncovered in this well-studied field.

\vspace{3mm}

\noindent $\bullet$~\emph{Relation to matrix theory.} Our main focus will be the corners of type II superstring theory resulting from applying a single BPS decoupling limit where one zooms in on a background D$p$-brane. In~\cite{Blair:2023noj, Gomis:2023eav, Blair:2024aqz}, we introduced the nomenclature \emph{Matrix $p$-brane Theory} (M$p$T) to refer to these corners of string theory. The low-energy excitations in M$p$T are the D$p$-branes (instead of the fundamental strings), whose dynamics is described by matrix (gauge) theory. 

In the classic case of M0T, the low-energy excitations are captured by the bound state of $N$ D0-branes and we are led to the Banks-Fischler-Shenker-Susskind (BFSS) matrix theory~\cite{deWit:1988wri, Banks:1996vh}. At large $N$, BFSS matrix theory is conjectured to describe eleven-dimensional M-theory~\cite{Banks:1996vh} (and at finite $N$ to describe the null compactification of M-theory~\cite{Susskind:1997cw, Sen:1997we, Seiberg:1997ad}). The D0-branes in M0T are non-relativistic particles interacting with each other via an instantaneous gravitational force that is Newton-like. The covariantisation of such a Newton-like force leads to a curved Newton-Cartan geometry, where the ten-dimensional target space is foliated into an absolute time direction and nine-dimensional spatial hypersurfaces. These `longitudinal' worldline and `transverse' directions are mixed via a Galilean boost transformation. Such a target space geometry (that we loosely refer to as an M0T geometry) lacks any ten-dimensional metric description, and is therefore \emph{non-Lorentzian}. 

Compactifying the BFSS matrix theory over a $p$-torus and T-dualising along all the cycles on the torus, we are led to super Yang-Mills theory (SYM) on a $p$-torus, describing the dynamics of wrapped D$p$-branes in M$p$T. These wrapped D$p$-branes interact with each other via instantaneous Newton-like forces. In the decompactification limit where M$p$T typically becomes ten-dimensional, the target space is non-Lorentzian, equipped with a ($p$+1)-dimensional longitudinal and (9$-$$p$)-dimensional transverse sector, related via a D-brane generalisation of the Galilean boost transformation. Here, the longitudinal sector is aligned with the worldvolume of a background D$p$-brane on which the BPS decoupling limit zooms in. For example, in the case where $p=3$\,, we are led to $\CN = 4$ SYM on D3-branes, which is coupled to ten-dimensional non-Lorentzian M3T geometry.  

\vspace{3mm}

\noindent $\bullet$~\emph{Duality asymmetry.} In our earlier papers~\cite{Blair:2023noj, Gomis:2023eav, Blair:2024aqz}, we have shown that T-dualising M$p$T in a longitudinal (transverse) isometry direction leads to M($p-$1)T (M($p$+1)T), accompanied by a set of duality transformation rules relating the original and dual M$p$T background fields. 
These T-duality rules appear to formalise the intuition that the ten-dimensional non-Lorentzian geometries seen by different matrix theories are related to each other in the same way as D-branes on tori
(see also~\cite{Gomis:2000bd, Gopakumar:2000ep}). 

However, this expectation changes when one dualises on multiple isometry directions simultaneously, in the presence of arbitrary background gauge fields.
In particular,  
T-dualising M$p$T with $d$ transverse isometries leads to a non-Lorentzian geometry whose nature depends on the rank $r$ of the $B$-field in the isometry directions: M$p$T is mapped to M($p$+$d$$-$$r$)T under T-duality along all the transverse isometry directions. Here, $0 \leq r \leq d$\,. 
Conversely, T-dualising M($p$+$d$)T along $d$ longitudinal isometries, with or without a $B$-field, we are always led to M$p$T with vanishing $B$-field in the dual transverse isometry directions. We refer to this mismatched behaviour as \emph{duality asymmetry}.

This asymmetric behaviour is not restricted to T-duality, but a generic phenomenon associated with duality between corners arising from different decoupling limits of string theory.
An S-duality example has been previously studied in~\cite{Bergshoeff:2022iss,Bergshoeff:2023ogz, Ebert:2023hba}, which considered the asymmetric SL($2,\mathbb{Z}$) transformations involving M1T and \emph{non-relativistic string theory}, where the latter arises from a BPS decoupling limit zooming in on a background fundamental string~\cite{Klebanov:2000pp, Gomis:2000bd, Danielsson:2000gi} (see also recent reviews in~\cite{Oling:2022fft, Demulder:2023bux}). In this case, whether a given SL($2,\mathbb{Z}$) transformation maps between M1T and non-relativistic string theory, or is a self-duality map, depends on both the value of the Ramond-Ramond (RR) zero-form (\emph{i.e.}~the axion) and the  transformation parameters.
Algebraically, the full transformations are captured by an intriguing polynomial realisation of SL($2,\mathbb{Z}$). In this paper, we extend the previous analysis~\cite{Bergshoeff:2022iss, Bergshoeff:2023ogz, Ebert:2023hba} of SL($2,\mathbb{Z}$) duals of non-relativistic string theory by presenting the full SL($2,\mathbb{Z}$) transformation of M1T. Moreover, we will show that an identical structure arises when restricting to SL($2,\mathbb{Z}$) subgroups of two-dimensional T-duality between M$p$T on a transverse 2-torus and M($p$+2)T on a dual longitudinal 2-torus, with the $B$-field component on the torus playing the role of the RR zero-form in S-duality.

The same behaviour also extends to the M$p$Ts with $p < 0$\,, which are associated with the Ishibashi-Kawai-Kitazawa-Tsuchiya (IKKT) matrix theory~\cite{Ishibashi:1996xs} and target space geometries that are either Lorentzian ($p=-1$) or Carroll-like ($p < -1$)~\cite{Blair:2023noj, Gomis:2023eav}. A version of the T-duality asymmetry that maps M$(-1)$T to itself has been studied in the context of tensionless string theory with a $B$-field~\cite{Banerjee:2024fbi}.

\vspace{3mm}

\noindent $\bullet$~\emph{Non-commutative gauge theory and holography.}
The T-duality asymmetry between M$p$T geometries hinges on the presence of the $B$-field. 
Connes, Douglas and Schwarz famously showed how toroidal compactification of matrix theory leads to non-commutative geometry, where the non-commutativity is characterised by a constant $B$-field on the torus~\cite{Connes:1997cr}.
T-dualising matrix theory on such a `non-commutative torus' leads to non-commutative gauge theories~\cite{Douglas:1997fm, Seiberg:1999vs}.
In this setting, T-duality transformations between string theory backgrounds are viewed as Morita equivalence between algebras labeled by different non-commutativity parameters~\cite{Connes:1997cr, Schwarz:1998qj, Brace:1998ku, Pioline:1999xg, Seiberg:1999vs}.
For rational values of non-commutativity parameter, T-duality can be used to map to a commutative theory. 
We will explain how the duality asymmetry of the underlying ten-dimensional non-Lorentzian geometries accounts for these features.\,\footnote{Historically, the non-relativistic limits introduced in \cite{Gomis:2000bd, Danielsson:2000gi, Danielsson:2000mu} were developed as `closed string' versions of `open string' decoupling limits leading to non-commutative theories \cite{Seiberg:2000ms, Gopakumar:2000na, Gopakumar:2000ep, Klebanov:2000pp, Bergshoeff:2000ai, Harmark:2000ff} living on branes.
The former can be seen as inducing the latter when the appropriate brane configurations are explicitly included. 
The decoupling limit used by Seiberg and Witten \cite{Seiberg:1999vs} when studying non-commutative Yang-Mills similarly can be seen to arise in this context~\cite{Gopakumar:2000na,Danielsson:2000gi}. It is therefore natural that there is a non-Lorentzian geometric underpinning of the appearance of non-commutativity.
Furthermore, it is interesting to note in this context that S-duality asymmetry between M1T and non-relativistic string theory underlies the relationships between non-commutative open strings and non-commutative Yang-Mills on D3-branes discussed in~\cite{Lu:2000ys,Russo:2000zb,Cai:2000yk,Lu:2000vv,Gran:2001tk}. 
}

Inspired by the above observations, in the last part of this paper, we will examine explicit examples of backgrounds which are related holographically to non-commutative Yang-Mills~\cite{Hashimoto:1999ut, Maldacena:1999mh, 
Alishahiha:1999ci}.
In particular, we will revisit the solution analysed by Maldacena and Russo in~\cite{Maldacena:1999mh} which corresponds to D1-branes delocalised in D3-branes, with a non-vanishing $B$-field at infinity.
We will show that the decoupling limit leading to non-commutative Yang-Mills can be realised as an asymptotic M1T limit. This provides another example of how holography is governed by our overarching perspective of BPS decoupling limits and asymptotic non-Lorentzian geometry~\cite{Blair:2024aqz}. By taking a further BPS decoupling limit, we will generate new non-Lorentzian bulk geometries that underlie a dual field theory that should arise from a non-relativistic limit of non-commutative Yang-Mills. 

\vspace{3mm}

The paper is organised as follows. In Section~\ref{sec:nlbmt}, we collect the essential ingredients of the non-Lorentzian geometries that underlie matrix theories. In Section~\ref{sec:da}, we study both the T- and S-duality asymmetry.
In Section~\ref{sec:damt}, we discuss how this can be interpreted in terms of matrix theory and non-commutative geometry.
In Section~\ref{sec:mrh}, we apply the new insights from our study of non-Lorentzian geometry and duality asymmetry to the holographic correspondence associated with non-commutative gauge theories. We conclude the paper in Section~\ref{sec:concl}. We have also included two appendices: Appendix~\ref{sec:ReviewOdd} reviews the usual T-duality transformations, and Appendix~\ref{sec:appTMpT} provides further technical details of T-duality asymmetry.

\section{Non-Lorentzian Backgrounds for Matrix Theories} \label{sec:nlbmt}

We begin by summarising, following~\cite{Blair:2023noj, Gomis:2023eav, Blair:2024aqz}, the defining prescriptions for the BPS decoupling limit of type II superstring theory that leads to Matrix $p$-brane Theory (M$p$T). 
For simplicity, we first consider flat spacetime with coordinates $x^\text{M}$\,, $\text{M} = 0\,, \, \cdots, \, 9$.
Then, we reparametrise the NSNS sector such that the metric $G_\text{MN}$, $B$-field $B_\text{MN}$, and dilaton $\Phi$ become
 \begin{subequations} \label{eq:mptp}
 \begin{align}
 	\dd s^2 &\equiv G_\text{MN} \, \dd x^\text{M} \, \dd x^\text{N} = \omega \, \dd x^A \, \dd x^B \, \eta^{}_{AB} + \frac{1}{\omega} \, \dd x^{A'} \dd x^{A'}, \\[4pt]
	B_\text{MN} &= b_\text{MN}\,,
		\qquad\,\,\,\,%
	e^\Phi = \omega^{\frac{p-3}{2}} \, e^\varphi\,,
 \end{align}
where $A = 0\,, \, \cdots, \, p$ and $A' = p+1 \,, \, \cdots, \, 9$\,. Here, $\omega$ is a dimensionless parameter that we will send to infinity to define the BPS decoupling limit. Moreover, $b_\text{MN}$ and $\varphi$ will be the new $B$-field and dilaton in the resulting M$p$T. In the $\omega \rightarrow \infty$ limit, the ten-dimensional metric $G_\text{MN}$ becomes singular, which implies that the resulting geometry is non-Lorentzian. It is important that the divergences in the metric must be cancelled exactly for any physical observables, which is achieved by reparametrising the RR ($p$+1)-form potential as
\begin{align}
 	C^{(p+1)} &= \omega^2 \, e^{-\varphi} \, \dd x^0 \wedge \cdots \wedge \dd x^p + c^{(p+1)}\,, \label{eq:crrorp}
\end{align}
\end{subequations}
where $c^{(p+1)}$ will be the RR ($p$+1)-form in the resulting M$p$T after the $\omega \rightarrow \infty$ limit. For now we assume that \emph{no} other RR potentials are present. Physically, this $\omega \rightarrow \infty$ limit zooms in on a background D$p$-brane in type II superstring theory, in a way that the brane charge is fine tuned according to \eqref{eq:crrorp} such that it cancels the infinite brane mass (see below). As a result, we are led to the self-contained M$p$T corner, where the light excitations are D$p$-branes. This also implies that M$p$T is non-perturbative from the perspective of the fundamental string. 

For $p=0$, the above prescription can be directly related to the infinite boost limit of M-theory compactified on a spatial circle, with $\omega$ corresponding to the large Lorentz factor associated with the boost. 
From the M-theory perspective, the infinite boost leads to the discrete light cone quantisation (DLCQ), \emph{i.e.}~M-theory compactified over a lightlike circle. The reduction on this circle leads to M0T in type IIA. The idea of DLCQ forms the foundation for the BFSS matrix theory, with the D0-branes being identified with the Kaluza-Klein excitations in the lightlike compactification. 
The M$p$T limits with $p \neq 0$ can be reached from this DLCQ perspective by applying T-duality, as explained in~\cite{Blair:2023noj, Gomis:2023eav, Blair:2024aqz}. 
 
The above $\omega \rightarrow \infty$ limit manifests itself as a BPS decoupling limit when applied to the D$p$-branes. For simplicity, we only review the single brane case here, while the more general case of multiple coinciding branes has been discussed in~\cite{Blair:2024aqz} and leads to matrix (gauge) theory. Consider the bosonic sector of the D$p$-brane action, 
\be \label{eq:dpa}
	S_{\text{D}p} = - T_p \int \dd^{p+1} \sigma \, e^{-\Phi} \sqrt{-\det \Bigl[ \p_\alpha X^\text{M} \, \p_\beta X^\text{N} \, \bigl( G_\text{MN} + B_\text{MN} \bigr) + F_{\alpha\beta} \Bigr]} + T_p \int C^{(p+1)},
\ee
where $\sigma^\alpha$, $\alpha = 0\,, \, \cdots, \, p$ are the worldvolume coordinates and $X^\text{M} (\sigma)$ describe how the D$p$-brane is embedded within the ten-dimensional target space. Moreover, $F_{\alpha\beta} = \p_\alpha A_\beta - \p_\beta A_\alpha$ is field strength associated with the $U(1)$ gauge potential $A_\alpha$ on the brane and
\be
	C^{(p+1)} \equiv \frac{1}{(p+1)!} \, C^{}_{\text{M}_0\cdots\text{M}_p} \, \dd X^{\text{M}_0} \!\! \wedge \! \cdots \! \wedge \! \dd X^{\text{M}_p}
\ee
is the pullback of the target space RR ($p$+1)-form to the worldvolume of the D$p$-brane. 
Plugging the M$p$T prescription~\eqref{eq:mptp} into the D$p$-brane action~\eqref{eq:dpa} and sending $\omega$ to infinity, we find the following D$p$-brane action in M$p$T:\,\footnote{The same D$p$-brane actions in M$p$T can also be derived from dualising the D$p$-branes in non-relativistic string theory~\cite{Ebert:2021mfu}, where the brane worldvolume actions can be derived from first principles using the non-relativistic string worldsheet theory~\cite{Gomis:2020fui}.}
\be \label{eq:mptdp}
	S^\text{M$p$T}_{\text{D}p} = T_p \int \dd^{p+1} \sigma \, e^{-\phi} \, \sqrt{-\tau} \, \Bigl( \tfrac{1}{2} \, \tau^{\alpha\beta} \, \p_\alpha X^{A'} \p_\beta X^{A'} + \tfrac{1}{4} \, \tau^{\alpha\beta} \, \tau^{\gamma\delta} \, \CF_{\alpha\gamma} \, \CF_{\beta\delta} \Bigr)+ T_p \int c^{(p+1)}\,,
\ee
where $\tau_{\alpha\beta} \equiv \p_\alpha X^A \, \p_\beta X^B \, \eta_{AB}$\,, $\tau = \det (\tau_{\alpha\beta})$\,, $\tau^{\alpha\beta}$ is the inverse of $\tau_{\alpha\beta}$\,, and 
\be
	\CF_{\alpha\beta} \equiv \p_\alpha X^\text{M} \, \p_\beta X^\text{N} \, B_\text{MN} + F_{\alpha\beta}\,.
\ee 
The crucial point here is that the infinite $\omega$ divergences from the brane mass and charge contributions in \eqref{eq:dpa} precisely cancel each other, leaving us with the finite action~\eqref{eq:mptdp}. It is in this sense that the $\omega \rightarrow \infty$ limit is a BPS decoupling limit.  
In the case of coinciding D$p$-branes, the action~\eqref{eq:mptdp} generalises to that of (super) Yang-Mills theory which is identified as the associated matrix theory. We then observe that the target space isometry group is non-Lorentzian: it admits longitudinal SO($1,p$) and transverse SO($9-p$) transformations, as well as $p$-brane Galilean boosts,
\begin{subequations}
\begin{align}
	\delta^{}_\text{G} x^A &= 0\,,
		\qquad%
	\delta^{}_\text{G} x^{A'} = \Lambda^{A'}{}^{}_{\!A} \, x^A, \\[4pt]
	\delta^{}_\text{G} c^{(p+1)} &= \frac{1}{p!} \, e^{-\varphi} \, \Lambda^{A'A} \, \dd x^{A'} \wedge \dd x^{A_1} \cdots \wedge \dd x^{A_p} \, \epsilon^{}_{AA_1 \cdots A_p}\,.
\end{align}
\end{subequations}
Here the role of the absolute `time' direction in the standard Galilean transformation is replaced by the longitudinal directions $x^A$ aligned with the background D$p$-brane. 

The above discussion illustrated the essential features of M$p$T assuming a flat spacetime.
Now we discuss the generalisation to curved non-Lorentzian geometries.
This can be inferred by `covariantising' $\dd x^A$ and $\dd x^{A'}$ in M$p$T as
\be \label{eq:cov}
	\dd x^A \rightarrow \tau^A \equiv \tau^{}_\text{M}{}^A \, \dd x^\text{M}\,,
		\qquad%
	\dd x^{A'} \rightarrow E^{A'} \equiv E^{}_\text{M}{}^{A'} \dd x^\text{M}\,,
\ee
where $\tau^{}_\text{M}{}^A$ and $E^{}_\text{M}{}^{A'}$ are, respectively, the longitudinal and transverse vielbein fields that encode the M$p$T geometry. 
The $p$-brane Galilean boost now acts as:
\begin{subequations}
\begin{align}
	\delta^{}_\text{G} \tau^A &= 0\,,
		\qquad%
	\delta^{}_\text{G} E^{A'} = \Lambda^{A'}{}^{}_{\!A} \, \tau^A, \\[4pt]
	\delta^{}_\text{G} c^{(p+1)} &= \frac{1}{p!} \, e^{-\varphi} \, \Lambda^{A'A} \, E^{A'} \wedge \tau^{A_1} \wedge \cdots \wedge \tau^{A_p} \, \epsilon^{}_{AA_1 \cdots A_p}\,.
\end{align}
\end{subequations}
This is a $p$-brane generalisation of the Newton-Cartan geometry that covariantises the Newtonian force between particles. Combining \eqref{eq:cov} with the flat space M$p$T prescription~\eqref{eq:mptp}, we are led to the curved spacetime M$p$T prescription,
\begin{subequations} \label{eq:cmpta}
\begin{align}
	G_\text{MN} & = \omega \, \tau^{}_\text{M}{}^A \, \tau^{}_\text{N}{}^B \, \eta^{}_{AB} + \frac{1}{\omega} \, E^{}_\text{M}{}^{A'} E^{}_\text{N}{}^{A'}, 
		&
	C^{(p+1)} &= \omega^2 \, e^{-\varphi} \, \tau^0 \wedge \cdots \wedge \tau^p + c^{(p+1)}, \\[4pt]
	B^{}_\text{MN} &= b^{}_\text{MN}\,,
		\qquad%
	e^\Phi = \omega^{\frac{p-3}{2}} \, e^\varphi\,,
		&
	C^{(q)} &= c^{(q)}, 
		\qquad%
	q \neq p\,.
\end{align}
\end{subequations} 
Here, we have introduced the finite part $c^{(q)}$ of the other RR forms, which lead to the associated RR potentials
in M$p$T after the $\omega \rightarrow \infty$ limit. Moreover, applying the covariantisation~\eqref{eq:cov} to the M$p$T D$p$-brane action leads to the following curved spacetime generalisation:
\begin{align} \label{eq:mptdpc}
\begin{split}
	S^\text{M$p$T}_{\text{D}p} = & T_p \int \dd^{p+1} \sigma \, e^{-\phi} \, \sqrt{-\tau} \, \Bigl( \tfrac{1}{2} \, \tau^{\alpha\beta} \, \p_\alpha X^\text{M} \, \p_\beta X^\text{N} \, E_\text{MN} + \tfrac{1}{4} \, \tau^{\alpha\beta} \, \tau^{\gamma\delta} \, \CF_{\alpha\gamma} \, \CF_{\beta\delta} \Bigr) \\[4pt]
	& + T_p \int \sum_q c^{(q)} \wedge e^{\CF^{(2)}} \Big|_{p+1}\,,
		\qquad\quad%
	\tau^{}_{\alpha\beta} \equiv \p_\alpha X^\text{M} \, \p_\beta X^\text{N} \, \tau^{}_\text{M}{}^A \, \tau^{}_\text{N}{}^B \, \eta^{}_{AB}\,.
\end{split}
\end{align}
It is understood that only the $(p+1)$-forms are kept in the Chern-Simons term. 

In general, if there is \emph{no} further constraint on the longitudinal vielbein $\tau^{}_\text{M}{}^A$, the foliation in the target space is not preserved and could result in pathologies when it comes to causality. In fact, the BPS decoupling limit of the superalgebra does require a more constraining symmetry algebra. This M$p$T algebra contains the generators of the na\"{i}ve Galilei-like algebra from amalgamating the longitudinal SO($1,p$) and transverse SO(9-$p$) via the $p$-brane Galilean boost generator $G_{AA'}$\,. In addition, there is an extension $Z_A$ that renders nonzero the commutator between $G_{AA'}$ and the transverse momentum generator $P_{A'}$, with
\be
	\bigl[ G_{AA'}, P_{B'} \bigr] = \delta_{A'B'} \, Z_A\,.
\ee  
This $Z_A$ is a natural generalisation~\cite{Brugues:2004an, Brugues:2006yd} of the central extension from the Galilei to Bargmann algebra in the particle case, where it would be associated with the particle number conservation. Instead, $Z_A$ is now associated with the winding of the wrapped brane configuration. The $Z_A$ symmetry acts nontrivially on the RR form $c^{(p+1)}$ as follows:
\be
	\delta^{}_Z c^{(p+1)} = (p+1) \, e^{-\varphi} \, D\Sigma^A \wedge \tau^{A_1} \wedge \cdots \wedge \tau^{A_p} \, \epsilon^{}_{AA_1 \cdots A_p}\,, 
\ee
where $\Sigma^A$ is the Lie group parameter associated with the $Z_A$ generator and $D_\alpha \equiv \p_\alpha X^\text{M} \, D_\text{M}$\,, with $D_\text{M}$ the derivative covariantised with respect to the frame index $A$\,. Up to a boundary term, the D$p$-brane action~\eqref{eq:mptdpc} transforms under the $Z_A$ symmetry as
\be
	\delta_Z S^\text{M$p$T}_{\text{D}p} = - \frac{T_p}{(p-1)!} \int \dd^{p+1} \sigma \, e^{-\varphi} \, \Sigma^{A_0} \, \epsilon^{\alpha_0\cdots\alpha_p} \, D_{\alpha_0} \tau_{\alpha_1}{}^{A_1} \, \tau_{\alpha_2}{}^{A_2} \cdots \tau_{\alpha_p}{}^{A_p} \, \epsilon^{}_{A_0 \cdots A_p}\,. 
\ee
Therefore, the $Z_A$ invariance requires that we impose a constraint on the torsion. For instance, we could require the torsional constraint $D^{}_{[\text{M}} \tau^{}_{\text{N}]}{}^A = 0$\,, which would be sufficient to preserve the ($p$+1)-dimensional foliation structure in the target space. This is analogous (and dual) to the $Z_A$ symmetry in non-relativistic string theory as discussed in \emph{e.g.}~\cite{Andringa:2012uz, Bergshoeff:2018yvt, Bergshoeff:2018vfn}, where such a torsional constraint is needed for quantum consistency of the worldsheet theory~\cite{Yan:2021lbe, hete}. 
Note that working from the supergravity perspective, incorporating supersymmetry appears to require an involved set of constraints (both bosonic and fermionic)~\cite{Bergshoeff:2021tfn, Bergshoeff:2024nin}, which remain to be worked out in the case of M$p$T.
Resolving the important question of determining which constraints are needed for an allowed background is beyond the scope of this paper.

\section{Duality Asymmetry} \label{sec:da}

We now illustrate the phenomenon of T- and S-duality asymmetry in the web of Matrix $p$-brane Theories (M$p$Ts). 
We will first discuss T-duality between different M$p$Ts, and show how the presence of a background $B$-field in a given M$p$T background determines which M$q$T, possibly including $q=p$, one lands in after T-duality.
This is the focus of Section~\ref{sec:tda}, where we  first explain how this happens using a simple revealing example, and then provide details for more general backgrounds (with additional technical details to be found in the appendices). 
Then in Section~\ref{sec:sda} we turn to $\mathrm{SL}(2,\mathbb{Z})$ S-duality asymmetry between M$1$T and non-relativistic string theory, before highlighting how similar structures appear when one considers the $\mathrm{SL}(2,\mathbb{Z})$ subgroup of two-dimensional T-duality between M$p$T and M($p$+2)T, in Section~\ref{sec:sltztda}.
We briefly discuss the extension to M$(-1)$T that is associated with the IKKT matrix theory and furthermore Carrollian versions of M$p$T (with $p<-1$) in Section~\ref{sec:ikktm}.

\subsection{T-Duality Asymmetry} \label{sec:tda}

In this subsection, we illustrate how T-duality asymmetry appears in M$p$T.
We start our discussion of T-duality asymmetry in M$p$T with the simplest possible example, namely the M0T limit in flat spacetime, compactified on a transverse 2-torus with a $B$-field component on the internal torus. Take the toroidal directions to be $x^1$ and $x^2$\,, which come with the scaling factor $\omega^{-1}$ in \eqref{eq:mptp}. Under the standard Buscher rule, na\"{i}vely, the scaling factor $\omega^{-1}$ in front of $x^1$ and $x^2$ is inverted to be $\omega$ after T-dualising along these two internal directions. One may hence expect that the T-dual frame is described by M2T with $x^1$ and $x^2$ now being longitudinal. However, this will \emph{not} be the case if the internal $B$-field is nonzero.  
Note that the very same example we use was pinpointed in~\cite{Douglas:1997fm} and used to argue for the appearance of non-commutative gauge theories on D-branes. This relationship should be borne in mind and will become prominent in Sections~\ref{sec:damt} and~\ref{sec:mrh}.

We first discuss, in Section \ref{sec:mztttt}, the T-duality of the NSNS sector of our M0T example, before looking at the would-be inverse transformation of M2T on a longitudinal torus in Section \ref{subsecM2T}.
In Section \ref{subsecRR} we include the RR sector and discuss some subtleties.
Then in Section \ref{sec:gtc} we discuss generalisations to higher-dimensional tori and curved backgrounds.

\subsubsection{Matrix 0-Brane Theory on a Transverse 2-Torus} \label{sec:mztttt}

First, we focus on the NSNS sector in type IIA superstring theory. Imposing the M0T prescription, we write \eqref{eq:mptp} with $p = 0$ as
\begin{align} \label{eq:omzt}
	\dd s^2 = - \omega \, \dd t^2 + \frac{1}{\omega} \, \Bigl( \dd x_1^2 + \dd x_2^2 + \dd x^m \, \dd x^m \Bigr)\,, 
		\quad\,\,%
	B^{(2)} = \CB \, \dd x^1 \!\! \wedge \dd x^2\,,
		\quad\,\,%
	e^\Phi = \omega^{-\frac{3}{2}} \, g^{}_\text{s}\,,
\end{align}
where $m = 3\,, \, \cdots, \, 9$\,. We always take $\omega > 0$ and for simplicity assume that $\CB>0$ if non-zero. In the $\omega \rightarrow \infty$ limit, we are led to the M0T geometry encoded by the vielbein fields $\tau^0$ and $E^{A'}$, with $A' = 1\,, \, \cdots, \, 9$ and
\begin{equation} \label{eq:omztg}
	\textbf{M0T geometry\qquad}
	\begin{array}{ll}
		\,\,\,\tau^0 = \dd t\,, 
			&%
		\,\,E^{A'} = \dd x^{A'}\,, \\[4pt]
		b^{(2)} = \CB \, \dd x^1 \wedge \dd x^2\,,
			&%
		\quad e^\varphi = g^{}_\text{s}\,.
	\end{array}
\end{equation}
T-dualising $x^1$ and $x^2$ by applying the standard Buscher transformations leads to the following dual geometric data:
\begin{subequations} \label{eq:ocb}
\begin{align}
	\dd \tilde{s}^2 &= - \omega \, \dd t^2 + \frac{\omega}{1 + \omega^2 \, \CB^2} \, \Bigl( \dd \tilde{x}_1^{\,2} + \dd \tilde{x}_2^{\,2} \Bigr) + \frac{1}{\omega} \, \dd x^m \, \dd x^m\,, \\[4pt]
	\tilde{B}^{(2)} &= - \frac{\omega^2 \, \CB}{1 + \omega^2 \, \CB^2} \, \dd \tilde{x}^1 \! \wedge \dd \tilde{x}^2\,,
		\qquad%
	e^{\tilde{\Phi}} = \frac{\omega^{-\frac{1}{2}} \, g^{}_\text{s}}{\sqrt{1 + \omega^2 \, \CB^2}}\,,
\end{align}
\end{subequations}
where $\tilde{x}^1$ and $\tilde{x}^2$ are the T-dual coordinates. In the case where $\CB = 0$\,, we obtain the M2T prescription as expected, with 
\begin{subequations} 
\begin{align}
	\dd \tilde{s}^2 &= \omega \, \Bigl( - \dd t^2 + \dd \tilde{x}_1^{\,2} + \dd \tilde{x}_2^{\,2} \Bigr) + \frac{1}{\omega} \, \dd x^m \, \dd x^m\,, \\[4pt]
	\tilde{B}^{(2)} &= 0\,,
		\qquad%
	e^{\tilde{\Phi}} = \omega^{-\frac{1}{2}} \, g^{}_\text{s}\,,
\end{align}
\end{subequations}
In the $\omega \rightarrow \infty$ limit, we are led to the following M2T geometry in the T-dual frame:
\begin{equation} \label{eq:dmtt}
	\begin{array}{l}
		\,\,\textbf{M2T geometry} \\[4pt]
		\emph{T-dual with $\CB=0$} 
	\end{array}
		\qquad%
	\begin{array}{ll}
		\,\,\,\tilde{\tau}^A = \bigl( \dd t\,,\, \dd \tilde{x}^{\,1},\, \dd \tilde{x}^{\,2} \bigr)\,, 
			&%
		\,\,\,\,\tilde{E}^{m} = \dd x^{m}\,, \\[4pt]
		\,\tilde{b}^{(2)} = 0\,,
			&%
		\quad\, e^{\tilde{\varphi}} = g^{}_\text{s}\,.
	\end{array}
\end{equation}
This is what one would intuitively expect given the fact that D0-branes are T-dual to D2-branes. However, if $\CB \neq 0$\,, the nature of the $\omega \rightarrow \infty$ limit of the dual prescription~\eqref{eq:ocb} is changed fundamentally, and the resulting theory is now M0T instead of M2T. At finite $\omega$\,, we rewrite the prescription~\eqref{eq:ocb} as
\begin{align} \label{eq:ocbexp}
	\dd \tilde{s}^2 = - \omega \, \tilde{\tau}^{\,0} \, \tilde{\tau}^{\,0} + \omega \, \tilde{E}^{A'} \tilde{E}^{A'},
		\qquad%
	\tilde{B}^{(2)} = \tilde{b}^{(2)}\,,
		\qquad%
	e^{\tilde{\Phi}} = \omega^{-\frac{3}{2}} \, e^{\tilde{\phi}}\,,
\end{align}
with
\begin{subequations} \label{eq:mztpfo}
\begin{align} 
	\tilde{\tau}^{\,0} &= \dd t\,,
		&%
	\tilde{E}^{A'} &= \biggl( \frac{\dd \tilde{x}^{\,1}}{\sqrt{\CB^2 + \omega^{-2}}}\,, \, \frac{\dd \tilde{x}^{\,2}}{\sqrt{\CB^2 + \omega^{-2}}}\,, \, \dd x^m \biggr), \\[4pt]
	e^{\tilde{\varphi}} &= \frac{g^{}_\text{s}}{\sqrt{\CB^{2} + \omega^{-2}}}\,,
		&%
	\tilde{b}^{(2)} &= - \frac{\CB^{-1}}{1 + \omega^{-2} \, \CB^{-2}} \, \dd \tilde{x}^{\,1} \! \wedge \dd \tilde{x}^{\,2}\,.
\end{align}
\end{subequations}
In the $\omega \rightarrow \infty$ limit, we are led to the following M0T geometry in the T-dual frame:
\begin{equation} \label{eq:nzbdmzt}
	\begin{array}{l}
		\textbf{\,\,M0T geometry} \\[4pt]
		\emph{T-dual with $\CB\neq0$} 
	\end{array}
		\qquad%
	\begin{array}{ll}
		\,\,\,\tilde{\tau}^{\,0} = \dd t\,, 
			&%
		\,\,\,\,\,\,\,\tilde{E}^{A'} = \bigl( \dd \tilde{x}^{\,1} \!/ \CB, \, \dd \tilde{x}^{\,2} \!/ \CB, \, \dd x^{m} \big)\,, \\[4pt]
		\,\tilde{b}^{(2)} = - \dd \tilde{x}^{\,1} \!\! \wedge \! \dd \tilde{x}^{\,2} / \CB,
			&%
		\quad\,\,\,\,\, e^{\tilde{\varphi}} = g^{}_\text{s} / \CB\,.
	\end{array}
\end{equation}
We thus observed that the specific M$p$T, for $p=0$ or $p=2$, that the T-dual frame describes crucially depends on the value of the internal $B$-field component $\CB$\,. 
While here we have started with M0T, it is clear that the details of the calculation are unchanged if we instead started with flat M$p$T geometry with a transverse $B$-field.

\subsubsection{Matrix 2-Brane Theory on a Longitudinal 2-Torus} 
\label{subsecM2T}

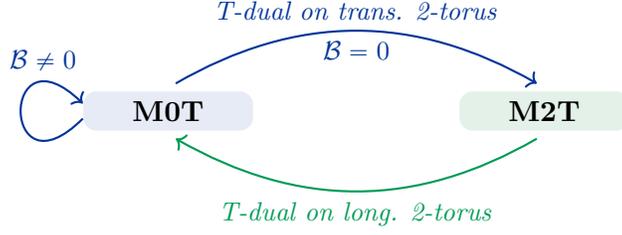
\begin{figure} [t!]
\centering
\begin{tikzpicture}

\draw (-5,0) node (A) [text width=2cm, rounded corners=5pt, fill=vub!10,align=center] {\textbf{M0T}};

\draw (0,0) node (B)[text width=2cm, rounded corners=5pt, fill=ForestGreen!10,align=center] {\textbf{M2T}};

\draw [vub,->,thick] ([xshift=0.1cm,yshift=0.1cm]A.north) to [out=30,in=150] node [pos=0.5,above=0cm] {\small \emph{T-dual on trans. 2-torus}} node [pos=0.5,below=0cm] {\small $\CB=0$} ([xshift=-0.1cm,yshift=0.1cm]B.north);

\draw [ForestGreen,<-,thick] ([xshift=0.1cm,yshift=-0.1cm]A.south)  to [out=-30,in=-150] node [pos=0.5,below=0cm] {\small \emph{T-dual on long. 2-torus}} ([xshift=-0.1cm,yshift=-0.1cm]B.south) ;

\draw [vub,<-,thick] ([xshift=0cm,yshift=0.1cm]A.west) to [out=135,in=225,loop,looseness=20] node [pos=0.2,above=0cm] {\small $\CB \neq 0$} ([xshift=0cm,yshift=-0.1cm]A.west);

\end{tikzpicture}
\caption{T-duality asymmetry between M0T on a transverse 2-torus and M2T on a longitudinal 2-torus. Under T-duality, the latter is always mapped to M0T in zero internal $B$-field. Here, $\CB$ is the component of the internal $B$-field on the 2-torus.}
\label{fig:b2da} 
\end{figure}

Conversely, we now T-dualise M2T in its longitudinal $x^1$ and $x^2$ directions compactified over a 2-torus. The M2T limiting prescription is given by
\begin{subequations}
\begin{align}
	\dd s^2 &= \omega \, \Bigl( - \dd t^2 + \dd x_1^2 + \dd x_2^2 \Bigr) + \frac{1}{\omega} \, \dd x^m \, \dd x^m\,, \\[4pt]
	B^{(2)} &= \CB \, \dd x^1 \! \wedge \dd x^2\,,
		\qquad%
	e^\Phi = \omega^{-\frac{1}{2}} \, g^{}_\text{s}\,.
\end{align}
\end{subequations}
T-dualising $x^1$ and $x^2$ leads to the following dual prescription:
\begin{subequations} \label{eq:locb}
\begin{align}
	\dd \tilde{s}^2 &= - \omega \, \dd t^2 + \frac{1}{\omega} \biggl( \frac{\dd \tilde{x}_1^{\,2} + \dd \tilde{x}_2^{\,2}}{1 + \omega^{-2} \, \CB^2} + \dd x^m \, \dd x^m \biggr), \\[4pt]
	\tilde{B}^{(2)} &= - \frac{\CB}{\omega^2 + \CB^2} \, \dd \tilde{x}^{\,1} \! \wedge \dd \tilde{x}^{\,2}\,,
		\qquad%
	e^{\tilde{\Phi}} = \frac{\omega^{-\frac{3}{2}} \, g^{}_\text{s}}{\sqrt{1 + \omega^{-2} \, \CB^2}}\,.
\end{align}
\end{subequations}
In the limit $\omega \rightarrow \infty$\,, regardless of the value of $\CB$, we are always led to the following M0T geometry in the T-dual frame:
\begin{equation} \label{eq:nzbdmzt1}
	\begin{array}{l}
		\textbf{M0T geometry} \\[4pt]
		\emph{\,in T-dual frame} 
	\end{array}
		\qquad%
	\begin{array}{ll}
		\,\,\,\tilde{\tau}^{\,0} = \dd t\,, 
			&%
		\,\,\,\,\tilde{E}^{A'} = \bigl( \dd \tilde{x}^{\,1}, \, \dd \tilde{x}^{\,2}, \, \dd x^{m} \big)\,, \\[4pt]
		\,\tilde{b}^{(2)} = 0\,,
			&%
		\quad\,\, e^{\tilde{\varphi}} = g^{}_\text{s}\,.
	\end{array}
\end{equation}
In particular, this is M0T with vanishing $B$-field, which we saw above is T-dual to M2T with vanishing $B$-field.
While M2T with a $B$-field is T-dual to M0T without a $B$-field, M0T with a $B$-field is T-dual to M0T. 
We therefore have a pair of theories, M0T and M2T, which are related by a \emph{T-duality asymmetry}, as illustrated by Figure~\ref{fig:b2da}.

\subsubsection{T-duality in Ramond-Ramond Sector} 
\label{subsecRR} 

For consistency, it must be the case that the above statements also apply when the RR sector is included. Let us verify this. We return to our example \eqref{eq:omzt}, which following the general M$p$T prescription~\eqref{eq:mptp} must be supplemented with the following RR one-form potential:
\be
	C^{(1)} = \frac{\omega^2}{g^{}_\text{s}} \, \dd t\,.
\ee
For simplicity, we take the finite part of $C^{(1)}$ and all the other RR potentials to be zero. After T-dualising on the transverse 2-torus (see Appendix~\ref{sec:ReviewOdd} for the T-duality rules for the RR fields in our conventions), we are led to the dual prescriptions,
\be \label{eq:drrf}
	\tilde{C}^{(1)} = - \omega^2 \, \frac{\CB}{g^{}_\text{s}} \, \dd t\,,
		\qquad%
	\tilde{C}^{(3)} = \frac{\omega^2}{1 + \omega^2 \, \CB^2} \, g^{-1}_\text{s} \, \dd t \wedge \dd \tilde{x}^{\,1} \wedge \dd \tilde{x}^{\,2}\,,
\ee
while all the other RR potentials remain zero. When $\CB = 0$\,, the only nonzero RR potential is $\tilde{C}^{(3)} = \omega^2 \, g^{-1}_\text{s} \, \dd t \wedge \dd \tilde{x}^{\,1} \! \wedge \dd \tilde{x}^{\,2}$\,. Together with the NSNS prescription~\eqref{eq:omzt}, we are led to the complete M2T limiting prescription as in \eqref{eq:mptp}. On the other hand, when $\CB \neq 0$ and the dual theory is M0T again, we use \eqref{eq:mztpfo} to rewrite the dual prescription~\eqref{eq:drrf} as
\be \label{eq:rrprs}
	\tilde{C}^{(1)} = - \omega^2 \, e^{-\tilde{\varphi}} \, \tilde{\tau}^{\,0} + \tilde{c}^{\,(1)}\,,
		\qquad%
	\tilde{C}^{(3)} = \tilde{c}^{(3)}\,,
\ee
where, in addition to the data already given in \eqref{eq:drrf}, we also have
\be
	\tilde{c}^{\,(1)} = \omega^2 \, \biggl( \sqrt{\CB^2 + \omega^{-2}} - \CB \biggr) \, g^{-1}_\text{s} \, \dd t\,,
		\qquad%
	\tilde{c}^{\,(3)} = \frac{g^{-1}_\text{s} \, \dd t \! \wedge \dd\tilde{x}^{\,1}\!  \wedge \dd\tilde{x}^{\,2}}{\CB^2 + \omega^{-2}}\,.
\ee
In the $\omega \rightarrow \infty$ limit, we find the following finite $\tilde{c}^{\,(1)}$ and $\tilde{c}^{\,(3)}$:
\be
	\begin{array}{l}
		\textbf{M0T geometry} \\[4pt]
		\emph{\,\,\,\,dual RR fields} 
	\end{array}
		\qquad%
	\begin{array}{l}
		\tilde{c}^{\,(1)} = \frac{1}{2} \, \CB^{-1} \, g^{-1}_\text{s} \, \dd t\,, \\[4pt]
		\tilde{c}^{\,(3)} = \CB^{-2} \, g^{-1}_\text{s} \, \dd t \wedge \dd \tilde{x}^{\,1} \wedge \dd \tilde{x}^{\,2}\,,
	\end{array}
\ee 
giving the associated Buscher rules for the dual RR potentials in M0T. This transformation has induced a shift of the potentials with coefficients which are polynomial in $\CB^{-1}$. This is a hint of a more general phenomena, which we will return to later. Meanwhile, the T-duality transformations of the RR potentials on a longitudinal 2-torus in M2T follows analogously, and are consistent with what we have seen in the NSNS sector.

Note that the prescription~\eqref{eq:rrprs} is almost identical to the M0T limit defined via \eqref{eq:mptp}, except that the sign in front of the $\omega^2$ divergence in $\tilde{C}^{(1)}$ in \eqref{eq:rrprs} is the opposite. This is because the choice of a positive $\CB > 0$ selects a different branch of the Lorentz group, while decoupling the rest. This is made manifest by looking at the probe D0-brane action in type IIA superstring theory, which takes the following form under the prescriptions~\eqref{eq:ocbexp} and \eqref{eq:rrprs}:
\begin{align}
\begin{split}
	\tilde{S}_\text{D0} = & - T_0 \int \dd \tau \, \omega^{\frac{3}{2}} \, e^{-\tilde{\phi}}\sqrt{\omega \, \p_\tau X^0 \, \p_\tau X^0 - \omega^{-1} \, \p_\tau X^{A'} \p_\tau X^{A'}} \\[4pt]
	& + T_0 \int \Bigl( - \omega^2 \, e^{-\tilde{\varphi}} \, \tilde{\tau}^{\,0} + \tilde{c}^{\,(1)} \Bigr).
\end{split}
\end{align}
The divergent term is
\be
	- \omega^2 \, T_0 \int \dd \tau \, \Bigl( \bigl| \p_\tau X^0 \bigr| + \p_\tau X^0 \Bigr)\,,
\ee
which only vanishes when $\p_\tau X^0 < 0$\,. In contrast, the derivation of the D0-brane action in M0T (see \eqref{eq:mptdp} with $p = 0$) requires that $\p_\tau X^0 > 0$\,. Therefore, the T-dual of M0T exhibits a branched structure: it maps between the branches of $\p_\tau X^0 < 0$ and $\p_\tau X^0 > 0$\,. Note that a similar branched structure has previously been seen to appear in the SL($2,\mathbb{Z}$) duals of non-relativistic string theory~\cite{Bergshoeff:2022iss, Bergshoeff:2023ogz, Ebert:2023hba}, which we will explore further shortly. Such a branched structure arises because the BPS decoupling limit could decouple either the branes or anti-branes, and these choices are related to each other via duality transformations. This is akin to the usual non-relativistic limit in the particle case, where one decouples particles with negative energies and is therefore led to the conservation of particle number. 

Another subtlety that can arise concerns the possibility of having a $B$-field which is not independent of $\omega$, but vanishes as $\CB = \omega^{-1} \, b$ with constant $b$\,, which in the $\omega \rightarrow 0$ limit is equivalent to the M0T geometry~\eqref{eq:omztg} with zero $B$-field. 
Therefore, the T-dual theory must still be M2T in the non-Lorentzian background~\eqref{eq:dmtt}. However, at finite $\omega$\,, the dual geometry \eqref{eq:ocb} now becomes
\begin{subequations} \label{eq:ocb1}
\begin{align}
	\dd \tilde{s}^{\,2} &= \omega \, \biggl( - \dd t^2 + \frac{\dd \tilde{x}_1^{\,2} + \dd \tilde{x}_2^{\,2}}{1 + b^2} \biggr) + \frac{1}{\omega} \, \dd x^m \, \dd x^m\,, \\[4pt]
	\tilde{B}^{(2)} &= - \omega \, \frac{b}{1 + b^2} \, \dd \tilde{x}^1 \! \wedge \dd \tilde{x}^2\,,
		\qquad%
	e^{\tilde{\Phi}} = \omega^{-\frac{1}{2}} \, \frac{g^{}_\text{s}}{\sqrt{1 + b^2}}\,,
\end{align}
\end{subequations}
where the $B$-field contains a linear divergence in $\omega$ and does \emph{not} match the M2T prescription. Somehow this seemingly more complicated prescription~\eqref{eq:ocb1} must lead to the same M2T geometry~\eqref{eq:dmtt} after taking the $\omega \rightarrow \infty$ limit. 
In the RR sector, \eqref{eq:drrf} becomes
\be \label{eq:drrfbdo}
	\tilde{C}^{(1)} = - \omega \, \frac{b}{g^{}_\text{s}} \, \dd t\,,
		\qquad%
	\tilde{C}^{(3)} = \omega^2 \, \frac{\dd t \! \wedge \dd \tilde{x}^{\,1} \! \wedge \dd \tilde{x}^{\,2}}{g^{}_\text{s} \, \bigl(1 + b^2\bigr)}\,.
\ee
Combined with \eqref{eq:ocb1}, we find that the T-dual of the original M0T prescription~\eqref{eq:omzt} on a transverse 2-torus can be identified with the M2T limiting procedure, except that $\tilde{B}^{(2)}$ and $\tilde{C}^{(1)}$ now contain extra linear divergences in $\omega$\,. However, the physical content of the resulting theory after sending $\omega$ to infinity has to be identical to the that of M2T on the background geometry~\eqref{eq:dmtt}. We test this by using the dual probe D2-brane in type IIA superstring theory, whose Chern-Simons term now takes the form
\be \label{eq:dtba}
	\int \Bigl( \tilde{C}^{(3)} + \tilde{C}^{(1)} \wedge \CF^{(2)} \Bigr) = \frac{1}{g^{}_\text{s}} \int \biggl( \frac{\omega^2 \, \dd t \! \wedge \dd x^1 \! \wedge \dd x^2}{1 + b^2} - \omega \, \dd t \wedge F \biggr) + O\bigl(\omega^0\bigr)\,.
\ee
Intriguingly, the $\omega$-divergences in \eqref{eq:dtba} precisely cancel the ones from the square-root kinetic term in the D2-brane action~\eqref{eq:dpa} (with $p=2$). In the $\omega \rightarrow \infty$ limit, we are led to the same M2T D2-brane action~\eqref{eq:mptdpc} (with $p=2$), in the background fields given in \eqref{eq:dmtt}. In this sense, the dual geometry in the $\omega \rightarrow \infty$ limit with the choice $\CB = b / \omega$ is equivalent to the M2T geometry~\eqref{eq:dmtt}, even though the dual limiting prescription looks rather different. 

\subsubsection{General Toroidal Compactifications} \label{sec:gtc}

Next we discuss T-duality transformations of M$p$Ts on more general toroidal compactifications, first with a higher-dimensional internal torus and then on curved backgrounds.
We continue to emphasise the conceptual aspects here -- Appendix~\ref{sec:appTMpT} can be consulted for more complete technical formulae. 

\vspace{3mm}

\noindent $\bullet$~\emph{Higher-dimensional toroidal compactification.}
We will only consider a $d$-dimensional internal torus either fully in the longitudinal or transverse sector in M$p$T.
The generalisation to the case where the internal torus includes both longitudinal and transverse directions merely introduces additional technical complexity.\,\footnote{We will see a hint of the structure later in Section~\ref{sec:sltztda} in the $O(2,2;\mathbb{Z})$ case.} 
Here we focus simply on the T-duality transformation of the dilaton, which is sufficient for us to discern which M$p$T appears as the dual theory. 
The standard Buscher transformation~\cite{Buscher:1987qj, Buscher:1987sk} of the dilaton is:
\be \label{eq:tpodd}
	e^{\tilde{\Phi}} = \Bigl[ \det \bigl( G^{}_{ij} \bigr) \, \det \bigl( G^{}_{k\ell} - B^{}_{km} \, G^{mn} \, B^{}_{n\ell} \bigr) \Bigr]^{-\frac{1}{4}} \, e^\Phi\,,
\ee
where $G_{ij}$ and $B_{ij}$ are the components of the metric and $B$-field in the toroidal directions, and $G^{ij}$ is the inverse of $G_{ij}$\,. We further assume that the internal $B$-field $B_{ij}$ has rank $r$\,, which is a non-negative even integer. Then we bring $B_{ij}$ into block-diagonal form using a special orthogonal transformation. Take the splitting $i=(a,\mathbf{m})$ with $a=1,\dots,r$ and $\mathbf{m} = r+1,\dots,d$\,, in the block-diagonal form we have $B_{\mathbf{m}\mathbf{n}} = B_{\mathbf{m} a} = 0$ and
\be
\label{bblockdiag}
B_{ab} =
    \begin{pmatrix}
        \,\,0 &\,\, \lambda_1  \\[2pt]
        \,\,- \lambda_1 &\,\, 0 \\
        & & \,\,\ddots \\
        & & & 0 &\,\, \lambda_{r/2} \\[2pt]
        & & & - \lambda_{r/2} &\,\, 0 \\[2pt]
    \end{pmatrix}.   
\ee
First, we consider M$p$T on a $d$-dimensional spatial torus in the ($p$+1)-dimensional longitudinal sector, with $d \leq p$\,. We read off from the M$p$T prescription~\eqref{eq:mptp} that 
\be \label{eq:gbvpl}
	G^{}_{ij} = \omega \, \delta_{ij}\,,
		\qquad%
	e^\Phi = \omega^{\frac{p-3}{2}} \, e^\varphi\,.
\ee 
From \eqref{eq:tpodd} we find that, at large $\omega$\,,
\be \label{eq:ddgc}
	e^{\tilde{\Phi}} = \omega^{\frac{\tilde{p} - 3}{2}} \, e^\varphi + \text{higher-order terms in $\omega^{-1}$}\,, 
		\qquad%
	\tilde{p} = p - d\,.
\ee
This is the prescription for M($p-d$)T. Therefore, as expected, T-dualising on a $d$-dimensional longitudinal torus in M$p$T leads us to M($p-d$)T. 

Next, we consider M$p$T on a $d$-dimensional torus in the ($9-p$)-dimensional transverse sector, with $d \leq 9-p$\,. The same calculation on the longitudinal torus applies, except that the reparametrisation of $g^{}_{ij}$ in \eqref{eq:gbvpl} is now replaced with 
$G^{}_{ij} = \omega^{-1} \, \delta^{}_{ij}$\,.
At large $\omega$\,, the dual dilaton $\tilde{\Phi}$ also satisfies \eqref{eq:ddgc} but now with
\be \label{eq:td}
	\tilde{\varphi} = \varphi - \ln \bigl| \lambda_1 \cdots \lambda_{r/2} \bigr|\,,
		\qquad%
	\tilde{p} = p + d - r\,.
\ee
Here, $\tilde{\varphi}$ is the dual M$\tilde{p}\,$T dilaton after the $\omega \rightarrow \infty$ limit is taken. Therefore, T-dualising on a $d$-dimensional transverse torus in M$p$T leads us to M($p$+$d$$-$$r$)T, which explicitly depends on the rank of the internal $B$-field. 

\vspace{3mm}

\noindent $\bullet$~\emph{Curved backgrounds.}
The above conclusion continues to hold when one considers more general curved M$p$T backgrounds, assuming that the isometry directions are purely transverse or purely longitudinal in a particular sense. 
The NSNS background fields in M$p$T are given by the longitudinal metric $\tau^{}_\text{MN} = \tau^{}_\text{M}{}^A \, \tau^{}_\text{N}{}^B \, \eta_{AB}$ with $A = 0\,, \, \cdots, \, p$\,, the transverse metric $E^{}_\text{MN} = E^{}_\text{M}{}^{A'} \, E^{}_\text{N}{}^{A'}$ with $A' = p+1\,, \, \cdots, \, 9$\,, the dilaton $\phi$\,, and the Kalb-Ramond potential $b^{(2)}$\,.
We split the curved index as $\text{M} = (\mu,i)$, with $i=1,\dots,d$ denoting isometry directions.
Consider T-duality between backgrounds with longitudinal isometries such that $E_i{}^{A'} = 0$ and backgrounds with transverse isometries such that $\tau_i{}^A = 0$. This means we require that the transverse (longitudinal) vielbein have \emph{no} legs in longitudinal (transverse) isometry directions.
Working in the NSNS sector alone, one show that longitudinal Buscher T-duality of M$p$T leads always to M$(p-d)$T, while transverse T-duality of M$p$T again leads to M$(p+d-r)$ depending on the rank of the $B$-field in the isometry directions. See Appendix~\ref{sec:BuscherdMpT} for the details. 

Here, we illustrate the expressions that arise by discussing briefly the cases $d=2,3$.

We start with the T-duality of M$p$T compactified over a 2-torus in the transverse sector.
As we have learned, there are two different T-dual cases to consider, depending on whether the internal $B$-field $b^{(2)} = \tfrac{1}{2} \, \CB \, \epsilon_{ij} \, \dd x^i \! \wedge \dd x^j$ vanishes:
\begin{enumerate}[(1)]

\item

\emph{From M$p$T to M$(p+2)$T when $\CB = 0$\,.} We denote the dual fields with an extra tilde. The Buscher rules associated with T-dualising the internal 2-torus are given by
\begin{subequations}
\begin{align}
    \tilde{\tau}^{}_{ij} &= E^{ij}\,, 
        &%
    \tilde{\tau}^{}_{\mu\nu} &= \tau^{}_{\mu\nu} - b^{}_{\mu i} \, E^{ij}b_{\nu j}\,,
        &%
    \tilde{b}^{}_{ij} &= 0\,, 
        \quad%
    \tilde{b}^{}_{\mu i} = E^{ij} \, E^{}_{\mu j}\,, \\[4pt]
    \tilde{\tau}^{}_{\mu i} &= E^{ij} \, b^{}_{\mu \, j}\,, 
        &%
    \tilde{E}^{}_{\mu\nu} &= E^{}_{\mu\nu} - E_{\mu i} \, E^{ij} \, E_{j \nu}\,,
        &%
    \tilde{b}^{}_{\mu\nu} &= b^{}_{\mu\nu} - 2 \, E^{ij} \, E^{}_{i[\mu} \, b^{}_{\nu]j}\,,
\end{align}
\end{subequations}
and $e^{\tilde{\varphi}} = e^\varphi / \sqrt{\det E}$\,. 
Here, $E^{ij}$ is the inverse of $E_{ij}$\,. When it comes to the RR potentials $c^{(q)}$\,, it is convenient to denote them in a collective way using the polyform, 
\be \label{eq:plf}
    \sum_q c^{(q)} \! \wedge e^{B^{(2)}} = \mathbf{C}^{}_0 + \mathbf{C}^{}_i \, \dd x^i + \tfrac{1}{2} \, \mathbf{C}^{}_{12} \, \epsilon^{}_{ij} \, \dd x^i \wedge \dd x^j\,. 
\ee
The dual polyform fields are
\be \label{eq:polyftd}
    \tilde{\mathbf{C}}^{}_0 = - \mathbf{C}^{}_{12}\,,
        \qquad%
    \tilde{\mathbf{C}}^{}_{12} = \mathbf{C}^{}_0 \,,
        \qquad%
    \tilde{\mathbf{C}}^{}_{i} = - \epsilon^{}_i{}^j \, \mathbf{C}^{}_j\,.
\ee

\item

\emph{Self-duality of M$p$T when $\CB \neq 0$\,.} The Buscher rules are
\begin{subequations}
\begin{align}
    \tilde{\tau}^{}_{\mu\nu} &= \tau^{}_{\mu\nu}\,, 
        &%
    \tilde{E}^{}_{\mu i} &= \frac{\det E}{\CB^2} \, E^{ij} \, \Bigl( b^{}_{\mu j} + b^{}_{jk} \, E^{k\ell} \, E^{}_{\ell\mu} \Bigr)\,, 
        &%
    \tilde{E}^{}_{ij} &= \frac{\det E}{\CB^2} \, E^{ij}\,, \\[4pt]
    \tilde{b}^{}_{ij} &= - \frac{\epsilon^{}_{ij}}{\CB}\,,
        &%
    \tilde{b}^{}_{\mu\nu} &= b^{}_{\mu\nu} + \frac{b^{}_{\mu i} \, \epsilon^{ij} \, b^{}_{j \nu}}{\CB}\,, 
        &%
    \tilde{b}^{}_{\mu i} &= \frac{\epsilon^{}_i{}^j}{\CB} b_{\mu j}\,, 
\end{align}
\vspace{-6mm}
\begin{align}
    \tilde{E}^{}_{\mu\nu} &= E^{}_{\mu\nu} - E^{}_{\mu i} \, E^{ij} \, E^{}_{j\nu} - \frac{\det E}{\CB^2} \, \Bigl( b^{}_{\mu i} - E^{}_{\mu k} \, E^{k\ell} \, b^{}_{\ell i}   \Bigr) \, E^{ij} \, \Bigl( b^{}_{j\nu} - b^{}_{j m} \, E^{mn} \, E^{}_{n\nu} \Bigr)\,,     
\end{align}
\end{subequations}
together with $e^{\tilde{\varphi}} = e^\varphi / \CB$ and
\begin{align}
    \tilde{\mathbf{C}}^{}_0 &= - \mathbf{C}^{}_{12} \! + \tfrac{1}{2} \, \ell^{(p+1)} \! \wedge \! e^{b^{(2)}_0} \!\!\! \wedge \! \biggl[ \epsilon^{}_{ij} A^i \! \wedge \! A^j \! + \tfrac{1}{2} \, \epsilon^{ij} \,  b^{(2)}_i \!\! \wedge \! b^{(2)}_j + \CB \, \Bigl( 1 + A^i \! \wedge \! b^{(2)}_i \Bigr)^{\!\!2} \biggr]\,, \notag \\[4pt]
    \tilde{\mathbf{C}}^{}_i &= - \epsilon_i{}^j \, \mathbf{C}^{}_j - \ell^{(p+1)} \! \wedge \! \biggl[ \tfrac{1}{2} \, \epsilon_i{}^j \, b^{(2)}_j \, \bigl( 1 - \epsilon^{}_{k\ell} \, A^k \! \wedge \! A^\ell \bigr) - \CB \, A^i \biggr]\,, \\[4pt]
    \tilde{\mathbf{C}}^{}_{12} &=  \mathbf{C}^{}_0 + \tfrac{1}{2} \, \ell^{(p+1)} \! \wedge \! e^{b^{(2)}_0} \!\!\! \wedge \! \Bigl( 1 - \CB \, \epsilon^{}_{ij} \, A^i \! \wedge \! A^j \Bigr)\,,
        \qquad%
    \ell^{(p+1)}\equiv \frac{\det E}{\CB^2 \, e^\varphi} \, \tau^0 \! \wedge \cdots \! \wedge \tau^{p}\,, \notag
\end{align}
where $A^i$ denotes the one-form with components $A_\mu{}^i \equiv E^{ij} \, E^{}_{j\mu}$\,. We have also used
\be
    b^{(2)} = b^{(2)}_0 + b^{(2)}_i \, \dd x^i + \tfrac{1}{2} \, \CB \, \epsilon^{}_{ij} \, \dd x^i \! \wedge \! \dd x^j\,. 
\ee

\end{enumerate}

\noindent Next, we consider M$p$T compactified on a 3-torus in the transverse sector. In this case, $\det (b_{ij}) = 0$ and $b_{ij}$ is at most of rank two. We focus on the rank-two case and so write $b^{}_{ij} = \epsilon^{}_{ijk} \, b^k$ for $b_k \neq 0$. We have argued that T-dualising over such a 3-torus maps M$p$T to M$(p+1)$T by examining the transformation of the dilaton in \eqref{eq:td}. Before the M$p$T limit is performed, the Buscher transformation of the internal metric $G_{ij}$ and Kalb-Ramond field $B_{ij}$ is given by
$\tilde{G} + \tilde{B} = \bigl( G + B \bigr)^{-1}$\,, which implies that the dual metric on the 3-torus is
\be
    \tilde{G}_{ij} = \omega \, \frac{b^i \, b^j}{\| b \|} + \omega^{-1} \, \frac{\det E}{\| b \|^2} \left( E^{ij} - \frac{b^i \, b^j}{\| b \|^2} \right) + O \bigl(\omega^{-3}\bigr)\,,
        \qquad%
    \| b \|^2 \equiv b^i \, E^{}_{ij} \, b^j\,.
\ee
Since $\det (E^{ij} - b^i \, b^j  \| b \|^2) = 0$\,, there exists $\tilde{E}_i{}^{2,\,3}$ such that
\be
    \tilde{G}_{ij} = \omega \, \tilde{\tau}^{}_i{}^1 \, \tilde{\tau}^{}_i{}^1 + \frac{1}{\omega} \Bigl( \tilde{E}^{}_i{}^2 \, \tilde{E}^{}_i{}^2 + \tilde{E}^{}_i{}^3 \, \tilde{E}^{}_i{}^3 \Bigr) + O \bigl( \omega^{-3} \bigr)\,,
\ee
with $\tilde{\tau}^{}_i{}^1 = b^i/\| b \|$\,. Therefore, after the $\omega \rightarrow \infty$ limit is applied, the T-dual theory has a longitudinal sector with one more dimension compared to M$p$T. This indeed implies that the T-dual theory is M($p$+1)T.

The above observations on T-duality transformations can be generalised in multiple ways. 
In Appendix~\ref{sec:BuscherdMpT}, we present the details for $d$-dimensional Buscher duality on both longitudinal and transverse tori in curved M$p$T backgrounds, restricting to the NSNS sector.
In Appendix~\ref{sec:GeneralTMpT}, we outline how the situation changes when one considers more general $O(d,d;\mathbb{Z})$ transformations, in which case the dual M$p$T depends on the rank of a matrix involving both the parameters of the transformation and the $B$-field. 
In the case $d=2$, we provide below in Section~\ref{sec:sltztda} complete details regarding the $\mathrm{SL}(2,\mathbb{Z})$ subgroup of non-trivial T-dualities on a 2-torus, where we treat general $\mathrm{SL}(2,\mathbb{Z})$ transformations and the RR sector (building on the analogous case of $\mathrm{SL}(2,\mathbb{Z})$ S-duality, which we first discuss in Section~\ref{sec:sda}).

Note that the behaviour we have identified as T-duality asymmetry is consistent with the consecutive application of the single T-duality relationships of~\cite{Blair:2023noj, Gomis:2023eav, Blair:2024aqz} when one takes into account the assumptions $\tau^{}_i{}^A = 0$ and $E^{}_i{}^{A'} = 0$ we impose on the background geometry in order to consider our isometries as wholly longitudinal/transverse. We discuss this further at the end of Appendix~\ref{sec:BuscherdMpT}.
It would also be possible to consider situations where we relax these conditions (and likewise to consider mixed longitudinal/transverse transformations).
Here one will find generically that the resulting dual theory will depend on the rank of some more convoluted combination of background fields.

\subsection{S-Duality Asymmetry} \label{sec:sda}

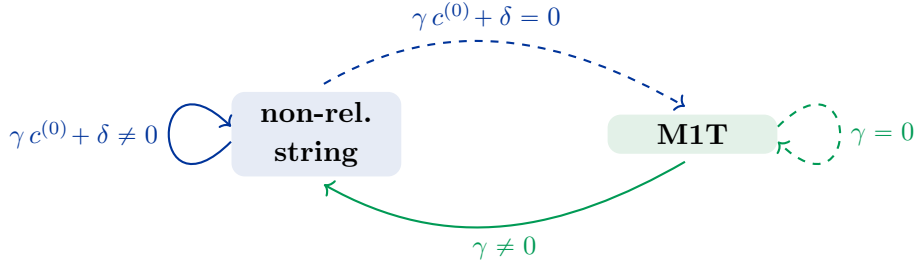
\begin{figure}[ht]
\centering
\begin{tikzpicture}

\draw (-5,0) node (A) [text width=2cm, rounded corners=5pt, fill=vub!10,align=center] {\textbf{non-rel.} \textbf{string}};

\draw (0,0) node (B)[text width=2cm, rounded corners=5pt, fill=ForestGreen!10,align=center] {\textbf{M1T}};

\draw [vub,->,dashed,thick] ([xshift=0.1cm,yshift=0.1cm]A.north) to [out=30,in=150] node [pos=0.45,above=0cm] {\small$\gamma \, c^{(0)} \! + \delta = 0$} ([xshift=-0.1cm,yshift=0.1cm]B.north);

\draw [ForestGreen,<-,thick] ([xshift=0.1cm,yshift=-0.1cm]A.south)  to [out=-30,in=-150] node [pos=0.5,below=0cm] {\small $\gamma \neq 0$}([xshift=-0.1cm,yshift=-0.1cm]B.south) ;

\draw [vub,<-,thick] ([xshift=0cm,yshift=0.1cm]A.west) to [out=135,in=225,loop,looseness=20] node [pos=0.5,left=0cm] {\small $\gamma \, c^{(0)} \! + \delta \neq 0$} ([xshift=0cm,yshift=-0.1cm]A.west);

\draw [ForestGreen,->,dashed,thick] ([xshift=0cm,yshift=0.1cm]B.east) to [out=45,in=315,loop,looseness=20] node [pos=0.5,right=0cm] {\small $\gamma=0$} ([xshift=0cm,yshift=-0.1cm]B.east);

\end{tikzpicture}
\caption{\textbf{SL($2,\mathbb{Z}$) S-duality asymmetry} between M1T and non-relativistic string theory, in terms of SL($2,\mathbb{Z}$) group parameters $\gamma,\delta$ and the RR zero-form $c^{(0)}$.
Dashed lines indicate a discrete realisation and continuous lines a polynomial realisation of the transformation. 
}
\label{fig:SL2Asymmetry} 
\end{figure}

Now, we divert our attention to S-duality. 
A similar duality asymmetry has also been observed for the S-duality transformations acting on non-relativistic string theory~\cite{Bergshoeff:2022iss, Bergshoeff:2023ogz, Ebert:2023hba}.
Here, we will extend the previous discussion to include how the SL($2,\mathbb{Z}$) group acts on M1T, which will lead us to the complete duality asymmetry illustrated in Figure~\ref{fig:SL2Asymmetry}.   

\subsubsection{\texorpdfstring{SL($2,\mathbb{Z}$)}{SL(2,Z)} S-Duality in Type IIB Superstring Theory}

We start by briefly reviewing the SL($2,\mathbb{Z}$) duality of type IIB superstring theory. From the M-theory perspective, this global symmetry group arises from the isometry group of a shrunk 2-torus, over which M-theory is compactified to give rise to the IIB theory in ten dimensions. Denote the toroidal modulus as
\be \label{eq:tomod}
	\tau \equiv \tau^{}_1 + i \, \tau^{}_2 = C^{(0)} + i \, e^{-\Phi}\,,
\ee 
where $C^{(0)}$ is the RR zero-form and $\Phi$ the dilaton in the IIB theory.  Then, the inverse toroidal metric takes the form,
\be \label{eq:tm}
	\CM^{ij} = \frac{1}{\tau^{}_2}
	\begin{pmatrix}
		|\tau|^2 &\,\, \tau^{}_1 \\[4pt]
		\tau^{}_1 &\,\, 1
	\end{pmatrix},
\ee
with `$i$' the index on the internal torus. 
The SL($2,\mathbb{Z}$) transformation matrix
\be \label{eq:sltztm}
	\Lambda^i{}_j = 
	\begin{pmatrix}
		\alpha &\quad \beta \\[0pt]
		\gamma &\quad \delta
	\end{pmatrix},
		\qquad%
	\alpha \, \delta - \beta \, \gamma = 1\,,
		\qquad%
	\alpha\,, \, \beta\,, \, \gamma\,, \, \delta \in \mathbb{Z} \,,
\ee
acts on the axion-dilaton coupling $\tau$ (and thus the toroidal metric $\CM$) as
\be \label{eq:tmsltzt}
	\tilde{\tau} = \frac{\alpha \, \tau + \beta}{\gamma \, \tau + \delta}\,,
		\qquad%
	\tilde{\CM} = \Lambda \, \CM \, \Lambda^\intercal\,.
\ee
Moreover, we have
\be \label{eq:ctm}
	\tilde{C}^{(2)}_i
		=	
	\bigl( \Lambda^{-1} \bigr){}^{\phantom{\dagger}}_i{}^j \, C^{(2)}_j\,,
		\qquad%
	C^{(2)}_i
	\equiv
	\begin{pmatrix}
		B^{(2)} \\[4pt]
		C^{(2)}
	\end{pmatrix},
\ee
\emph{i.e.}~$C^{(2)}_i$ transforms as an SL($2,\mathbb{Z}$) doublet. Finally, 
\be
	e^{-\Phi/2} \, G_{\text{M} \text{N}} \,, 
		\qquad%
	C^{(4)} + \tfrac{1}{2} \, B^{(2)} \wedge C^{(2)}\,,
\ee
are singlets invariant under SL($2,\mathbb{Z}$), with $G_{\text{M} \text{N}}$ the string-frame metric.

It is useful to repackage the above SL($2,\mathbb{Z}$) transformations in terms of vielbein fields on the internal torus, with
\be \label{eq:mve}
	\CM^{ij} = V^{i} \, V^{j} + E^i \, E^j\,,
		\qquad%
	 V^i = e^{-\frac{\Phi}{2}} 
	\begin{pmatrix}
		1 &\,\,
		0
	\end{pmatrix},
		\qquad%
	E^i = e^{\frac{\Phi}{2}}
	\begin{pmatrix}
		C^{(0)} &\,\,
		1
	\end{pmatrix}.
\ee
Then, under the SL($2,\mathbb{Z}$) transformation, the quantity 
$(V^i \, C_i^{(2)}\,,\, E^i \, C_i^{(2)})^\intercal$ 
only picks up (a discrete set of) the SO(2) rotation on the internal torus, with
\begin{subequations} \label{eq:vetrnsf}
\begin{align} 
	&
	\begin{pmatrix}
		\tilde{V}^i \\[4pt]
		\tilde{E}^i
	\end{pmatrix}
	\tilde C_i^{(2)}
        		=%
		s
	\begin{pmatrix}
		\cos\theta &\,\, - \sin\theta \\[4pt]
		\sin \theta &\,\, \cos\theta
	\end{pmatrix}
	\begin{pmatrix}
		V^i \\[4pt]
		E^i
	\end{pmatrix}
    C_i^{(2)}\,, 
\end{align}
\be
	s = \sgn \bigl( \gamma \, C^{(0)} + \delta \bigr)\,,
		\qquad%
	 \tan\theta =  \frac{\gamma \, e^{-\Phi}}{\gamma \, C^{(0)} + \delta}\,.
\ee
\end{subequations}
We take the principal value of the $\arctan$ when $\theta$ is evaluated. 
Note that the angle $\theta$ depends on both the group parameters and background fields. 
It will turn out to be convenient to recast the transformation of the doublet $C^{(2)}_i$ in terms of this $\mathrm{SO}(2)$ rotation.
Note that we have explicitly:
\be
\label{VEC}
V^i \, C^{(2)}_i = e^{-\tfrac{\Phi}{2}} \, B^{(2)} \,,
    \qquad%
E^i \, C^{(2)}_i = e^{\tfrac{\Phi}{2}} \, \Bigl( C^{(2)} + C^{(0)} B^{(2)} \Bigr) \,.
\ee
When we apply a BPS decoupling limit, both the background fields and the parameter $\tan \theta$ admit an expansion in $\omega$. 
This fact underlies the transformations rules we will present below.

\subsubsection{\texorpdfstring{SL($2,\mathbb{Z}$)}{SL(2,Z)} S-Duals of Matrix \texorpdfstring{$p$}{p}-Brane Theories} \label{sec:sdmpt}

Now, we consider how the SL($2,\mathbb{Z}$) S-duality acts on the M$p$T limits of type IIB with $p=1\,,\,3$\,.
M1T and M3T can be viewed as arising from different toroidal compactifications of non-relativistic M-theory, which is U-dual to M-theory in the DLCQ~\cite{Ebert:2023hba, Blair:2023noj}. Non-relativistic M-theory arises from a BPS decoupling limit zooming in on a background M2-brane~\cite{Gopakumar:2000ep, Bergshoeff:2000ai, Harmark:2000ff, Gomis:2000bd, Danielsson:2000gi}. As a result, the eleven-dimensional target space geometry develops a codimension-three foliation structure, with the transverse sector related to the three-dimensional longitudinal sector via a membrane Galilean boost~\cite{Blair:2021waq, Ebert:2021mfu, Ebert:2023hba}. 

Compactifying non-relativistic M-theory on a transverse 2-torus leads to M3T in ten dimensions, which then admits a standard SL($2,\mathbb{Z}$) symmetry that maps M3T to itself, inherited from the internal toroidal topology. This can be confirmed by plugging the M3T limiting prescription, with $C^{(0)} = c^{(0)}$ and $\Phi = \varphi$ from \eqref{eq:mptp}, into \eqref{eq:tomod}, under which the torus modulus becomes $\tau \rightarrow c^{(0)} + i \, e^{-\varphi}$\,. The standard SL($2,\mathbb{Z}$) transformations of the background fields thus remain unchanged. 

In contrast, M1T arises from compactifying non-relativistic M-theory on an `anisotropic' torus, with one cycle in the longitudinal sector and the other in the transverse sector~\cite{Ebert:2023hba}. Here, the transverse cycle is taken to be the M-theory circle, and compactifying on this circle leads to M2T. Then, T-dualising the longitudinal cycle maps M2T to M1T, and shrinking the original longitudinal cycle to zero decompactifies the T-dual circle in M1T, leading to M1T in ten dimensions. Due to the anisotropy of the internal torus, the SL($2,\mathbb{Z}$) transformations exhibit duality asymmetry. 

We first show how SL($2,\mathbb{Z}$) acts on M1T.  Recall the M1T limiting prescription~\eqref{eq:cmpta},
\begin{align} \label{eq:motrp}
 	\begin{array}{l}
 		\quad\,\,\,\,\textbf{M1T} \\
		\textbf{prescription} 
	\end{array}
		\qquad%
 	\begin{array}{ll}
 		G_\text{MN} = \omega \, \tau^{}_\text{MN} + \frac{1}{\omega} \, E^{}_\text{MN}\,, 
			&\qquad%
		B^{(2)} = b^{(2)}, \\[4pt]
		e^\Phi = \omega^{-1} \, e^\varphi,
			&\qquad%
		C^{(2)} = \omega^2 \, e^{-\varphi} \, \dd t \wedge \dd x^1 + c^{(2)},
	\end{array}
\end{align}
with $\tau^{}_\text{MN} = \tau^{}_\text{M}{}^A \, \tau^{}_\text{M}{}^B \, \eta^{}_{AB}$, $A = 0\,, \, 1$ and $E^{}_\text{MN} = E^{}_\text{M}{}^{A'} E^{}_\text{M}{}^{A'}$\!, $A' = 2\,,\,\cdots,\, 9$\,. Moreover, $C^{(q)} = c^{(q)}$ for $q \neq 2$\,. 
Under the above reparametrisation \eqref{eq:tomod} gives $\tau \rightarrow c^{(0)} + i \, \omega \, e^{-\varphi}$\,. It then follows that the $\omega \rightarrow \infty$ limit invalidates the metric description~\eqref{eq:tm} (of the 11-dimensional torus),\,\footnote{In the $\omega\rightarrow\infty$ limit, the torus becomes a singular ring-like object that is infinitely thin and with an infinitely large radius. This singular torus is topologically equivalent to a pinched torus, which is `almost' Riemannian except at the nodal point~\cite{Gomis:2023eav}.} but the vielbein formalism introduced in \eqref{eq:mve} still applies. Define the inverse vielbein fields
\be 
	v^i = e^{-\frac{\varphi}{2}} 
	\begin{pmatrix}
		1 \\[4pt]
		0
	\end{pmatrix},
		\qquad%
	e^i = e^{\frac{\varphi}{2}}
	\begin{pmatrix}
		c^{(0)} \\[4pt]
		1
	\end{pmatrix},
\ee 
it then follows that
\be \label{eq:vekk}
	V^i = \omega^{\frac{1}{2}} \, v^i\,, 
		\qquad%
	E^i = \omega^{-\frac{1}{2}} \, e^i\,, 
		\qquad%
	\tan \theta = \omega \, \kappa\,,
		\qquad%
	\kappa \equiv \frac{\gamma \, e^{-\varphi}}{\gamma \, c^{(0)} + \delta}\,.
\ee 
The opposite scalings of the vielbeins could be viewed as introducing a Euclidean Carroll or Galilean structure on the two-dimensional M-theory torus. Intuitively $\kappa$ could then be thought of as a Euclidean `boost' on the anisotropic torus on which non-relativistic M-theory is compactified~\cite{Ebert:2023hba}.
However, we will not pursue this point of view further.

As we now discuss, the nature of the SL($2,\mathbb{Z}$) transformations applied to M1T then differ depending on $\gamma$ is zero or not.

\vspace{3mm}

\noindent $\bullet$~\emph{From M1T to non-relativistic string theory} ($\gamma \neq 0$). In this case, the $\omega \rightarrow \infty$ limit of the  transformation~\eqref{eq:vetrnsf} gives
\be
	\begin{pmatrix}
		\tilde{v}^{\,i} \\[4pt]
		\tilde{e}^{\,i}
	\end{pmatrix}
        \tilde C_i^{(2)}
		=%
		\sgn (\gamma)
	\begin{pmatrix}
		0 &\,\, - 1 \\[4pt]
		1 &\,\, 0
	\end{pmatrix}
	\begin{pmatrix}
		v^i \\[4pt]
		e^i
	\end{pmatrix}
        C_i^{(2)} \,,
\ee
such that the two cycles of the internal anisotropic torus are switched. The S-dual frame then describes \emph{non-relativistic string theory}, defined by the following limiting prescription:
\begin{align} \label{eq:gbpcnrs}
	\begin{array}{l}
 		\textbf{non-rel. string} \\
		\textbf{\,\,\,\,prescription} 
	\end{array}
		\qquad%
	\begin{array}{ll}
		\tilde{G}_\text{MN} = \omega^2 \, \tilde{\tau}^{}_\text{MN} + \tilde{E}^{}_\text{MN}\,, 
			&\quad%
		\tilde{B}^{(2)} = - \omega^2 \, \tilde{\tau}^{\,0} \! \wedge \tilde{\tau}^{\,1} + \tilde{b}^{(2)}, \\[4pt]
		e^{\tilde{\Phi}} = \omega \, e^{\tilde{\varphi}},
			&\quad%
		\tilde{C}^{(q)} = \omega^2 \, \tilde{\tau}^{\,0}\! \wedge \tilde{\tau}^{\,1} \! \wedge \tilde{c}^{\,(q-2)} + \tilde{c}^{\,(q)},
	\end{array}
\end{align}  
with $\tau^{}_\text{MN} = \tau^{}_\text{M}{}^A \, \tau^{}_\text{N}{}^B \, \eta^{}_{AB}$, $A = 0\,, \, 1$ and $E^{}_\text{MN} = E^{}_\text{M}{}^{A'} E^{}_\text{N}{}^{A'}$\!, $A' = 2\,,\,\cdots,\, 9$\,. 
Note that $e^{-\frac{\tilde{\Phi}}{2}} \, \tilde{G}_{\text{MN}} = e^{-\frac{\Phi}{2}} \, G_{\text{MN}}$ determines how the S-dual quantity $\tilde{\tau}^{}_\text{M}{}^A$ and $\tilde{E}^{}_\text{M}{}^{A'}$ are related to $\tau^{}_\text{M}{}^A$ and $E^{}_\text{M}{}^{A'}$. 
Depending on whether $\gamma$ is positive or negative, we are mapped to sectors associated with different branches of the Lorentz group before the $\omega \rightarrow \infty$ limit is performed~\cite{Bergshoeff:2022iss}, with distinct signs in front of the $\omega^2$-divergence in the $B$-field and RR potential ansatz in \eqref{eq:gbpcnrs}. The choice that we made in \eqref{eq:gbpcnrs} corresponds to $\gamma > 0$\,. When $\gamma < 0$\,, the $\omega$-divergences in $\tilde{B}^{(2)}$ and $\tilde{C}^{(q)}$ in the non-relativistic string prescription~\eqref{eq:gbpcnrs} gain an extra minus sign. More explicitly, in the $\omega \rightarrow \infty$ limit, 
$\bigl( \tilde{\tau}^{}_\text{MN}\,, \tilde{E}{}_\text{MN} \bigr) = |\gamma| \, e^{-\varphi} \, \bigl( {\tau}{}_\text{MN}\,, {E}{}_\text{MN} \bigr)$\,,
and the dual dilaton and RR zero-form (axion) are given by
$e^{\tilde{\varphi}} = \gamma^2 \, e^{-\varphi}$ and $\tilde{c}^{\,(0)} = \alpha/\gamma$\,.

The dual $B$-field and RR two-form potential do \emph{not} form an SL($2,\mathbb{Z}$) doublet now. This asymmetry is introduced as only the RR two-form potential in the original IIB theory contains a divergence. The SL($2,\mathbb{Z}$) transformation takes the simplest form when we take our variables to be the finite parts of the expansion of the vielbein-contractions of the doublet, defined in \eqref{VEC}. 
These can be defined directly in M1T as:
\be
\label{VEC_projected}
	\begin{array}{l}
		\mathbf{B}^{(2)} \equiv  v^i \, c^{(2)}_i = e^{-\frac{\varphi}{2}} \, b^{(2)}\,, \\[4pt]
		\mathbf{C}^{(2)} \equiv  e^i \, c^{(2)}_i = e^{\frac{\varphi}{2}} \, \Bigl( c^{(2)} + c^{(0)} \, b^{(2)} \Bigr)\,,
	\end{array}
		\qquad%
	c^{(2)}_i \equiv 
	\begin{pmatrix}
		b^{(2)} \\[4pt]
		c^{(2)}
	\end{pmatrix}.
\ee
Similarly, we define in the S-dual non-relativistic string theory the quantities
\be
	\tilde{\mathbf{B}}^{\,(2)} \equiv e^{\frac{\tilde{\varphi}}{2}} \, \tilde{b}^{(2)}\,,
		\qquad%
	\tilde{\mathbf{C}}^{\,(2)} \equiv e^{\frac{\tilde{\varphi}}{2}} \, \Bigl( \tilde{c}^{\,(2)} + \tilde{c}^{\,(0)} \, \tilde{b}^{(2)} \Bigr)\,.
\ee
We further define the Einstein-frame two-form $\ell^{(2)} \equiv e^{-\frac{\varphi}{2}} \, \tau^0 \! \wedge \tau^1$.
In terms of these quantities, we then find the following polynomial realisation of SL($2,\mathbb{Z}$):
\begin{subequations} \label{eq:pinvt}
\begin{align} 
	\begin{array}{l}
		\textbf{M1T} \rightarrow \textbf{non-rel.~string} \\[4pt]
		\qquad\qquad\quad\!\! \kappa \neq 0
	\end{array}
		\qquad
	\begin{pmatrix}
		\tilde{\ell}^{(2)} \\[4pt]
		\scalebox{0.9}{$\sgn(\gamma)$} \, \tilde{\mathbf{C}}^{(2)} \\[4pt]
		\scalebox{0.9}{$\sgn(\gamma)$} \, \tilde{\mathbf{B}}^{(2)}
	\end{pmatrix}
		=
	\begin{pmatrix}
		1 &\,\, 0 &\,\,\,\, 0\, \\[4pt]
		\frac{1}{\kappa} &\,\, 1 &\,\,\,\, 0\,\\[4pt]
		\frac{1}{2 \, \kappa^2} &\,\, \frac{1}{\kappa} &\,\,\,\, 1\,
	\end{pmatrix}
	\begin{pmatrix}
		\ell^{(2)} \\[4pt]
		\mathbf{B}^{(2)} \\[4pt]
		- \mathbf{C}^{(2)}
	\end{pmatrix}\,, \label{eq:motnrspr}
\end{align}
where a branching factor $\sgn(\gamma)$ arises as explained before. 
Finally, we define the four-form $\mathbf{C}^{(4)} \equiv c^{(4)} + \tfrac{1}{2} \, b^{(2)} \!\wedge c^{(2)}$\,, which we note in passing arises from projecting the M-theory six-form using the vielbein fields on the internal metric. This four-form is no longer a singlet but receives the polynomial transformation
\begin{align}
	\tilde{\mathbf{C}}^{(4)} = \mathbf{C}^{(4)} - \tfrac{1}{2} \, \kappa^{-1} \, \mathbf{C}^{(2)} \! \wedge \ell^{(2)} + \tfrac{1}{4} \, \kappa^{-2} \, \mathbf{B}^{(2)} \! \wedge \ell^{(2)}. 
\end{align}
\end{subequations}
The transformation~\eqref{eq:pinvt} makes use of the universal form for any higher-dimensional polynomial realisations of the SL($2,\mathbb{Z}$) group, which is rooted in invariant theory~\cite{Bergshoeff:2023ogz}.
The above polynomial realisation also persists when field strengths are considered, and the related SL($2,\mathbb{Z}$) invariants that appear in supergravity can be conveniently classified using invariant theory in abstract algebra, using the same framework developed in~\cite{Bergshoeff:2023ogz}.\,\footnote{Invariant theory concerns the classification of the basis of invariants for systems of binary or multi-variable forms. This was an active field in the 19th century, which reached its peak when Hilbert published his seminal paper on an elegant proof of the finiteness theorem, alongside with the Nullstellensatz and the syzygy theorem, which form the foundation of the commutative algebraic interpretation of algebraic geometry~\cite{hilbert1890ueber}.}

\vspace{3mm}

\noindent $\bullet$~\emph{Self-duality of M1T} ($\gamma = 0$). In this case, $\kappa=0$\,, and we find that the dual prescription is again of the M1T type~\eqref{eq:motrp}. The divergent term in $C^{(2)}$ gain a sign $\sgn(\delta)$ due to the branching phenomenon, \emph{i.e.}~SL($2,\mathbb{Z}$) relates D-strings and anti-D-strings to each other, while both the original and dual frames are M1T. In the $\omega \rightarrow \infty$ limit, we find 
\begin{align}
	\begin{array}{l}
		\textbf{M1T \scalebox{1.5}{$\Circlearrowleft$}} \\[4pt]
		\,\,\,\,\kappa = 0
	\end{array}
		\qquad 
	\begin{array}{lll}
		\,\,\tilde{\tau}^{}_\text{MN} = |\delta| \, \tau^{}_\text{MN}\,,
			&\quad\,\,\,%
		e^{\tilde{\varphi}} = \delta^2 \, e^\varphi\,,
			&\quad%
		\tilde{\mathbf{B}}^{(2)} = \sgn(\delta) \, \mathbf{B}^{(2)}\,, \\[4pt]
		\tilde{E}^{}_\text{MN} = |\delta| \, E^{}_\text{MN}\,,
			&\quad%
		\tilde{c}^{\,(0)} = \delta^{-1} \, \bigl( \alpha \, c^{(0)} + \beta \bigr)\,, 
			&\quad%
		\tilde{\mathbf{C}}^{(2)} = \sgn(\delta) \, \mathbf{C}^{(2)}\,,
	\end{array}
\end{align}
while $\mathbf{C}^{(4)}$ is invariant. In this case, the polynomial realisation trivialises to a discrete transformation.\,\footnote{Note that for integer-valued transformation parameters, when $\gamma = 0$ we necessarily have $\alpha = \delta = \pm 1$. However, we do not use this to simplify the transformations here in order that they can also be read as being valid as classical $\mathrm{SL}(2,\mathbb{R})$ transformations in a supergravity description of the M1T geometry.}

\vspace{1em}
\noindent Next, we consider the action of SL($2,\mathbb{Z}$) S-duality on non-relativistic string theory, which has been studied in detail in~\cite{Bergshoeff:2022iss, Bergshoeff:2023ogz, Ebert:2023hba}. 
In this case, the transformations differ according to whether $\gamma \, c^{(0)} \! + \delta$ is zero or non-zero. 

\vspace{3mm}

\noindent $\bullet$~\emph{Self-duality of non-relativistic string theory} ($\gamma \, c^{(0)} \! + \delta \neq 0$). In this case, the SL($2,\mathbb{Z}$) transformation maps non-relativistic string theory to itself, with
\begin{subequations} \label{eq:nrstself0}
\begin{align}
	\tilde{\tau}{}^{}_\text{MN} &= \bigl| \gamma \, c^{(0)} \! + \delta \bigr| \, \tau{}^{}_\text{MN}\,, 
		&%
	e^{\tilde{\varphi}} &= \bigl( \gamma \, c^{(0)} \! + \delta \bigr)^2 \, e^\varphi\,, \\[4pt]
	\tilde{E}{}^{}_\text{MN} &= \bigl| \gamma \, c^{(0)} + \delta \bigr| \, E{}^{}_\text{MN}\,, 
		&%
	\tilde{c}{}^{\,(0)} &= \frac{\alpha \, c^{(0)} + \beta}{\gamma \, c^{(0)} + \delta}\,,
\end{align}
\end{subequations}
and the below polynomial transformation of the gauge potentials, with $s = \sgn\bigl(\gamma \, c^{(0)} \! + \delta \bigr)$\,:
\begin{subequations} \label{eq:nrstself}
\begin{align}
	\begin{array}{l}
		\textbf{non-rel.~string \scalebox{1.5}{$\Circlearrowleft$}} \\[4pt]
		\qquad\quad \frac{1}{\kappa} \neq 0
	\end{array}
		\qquad &
	\begin{pmatrix}
		\tilde{\ell}^{(2)} \\[4pt]
		- s \, \tilde{\mathbf{C}}^{(2)} \\[4pt]
		s \, \tilde{\mathbf{B}}^{(2)}
	\end{pmatrix}
		=
	\begin{pmatrix}
		1 &\,\, 0 &\,\, 0 \\[4pt]
		- \kappa &\,\, 1 &\,\, 0\\[4pt]
		\frac{\kappa^2}{2} &\,\, - \kappa &\,\, 1
	\end{pmatrix}
	\begin{pmatrix}
		\ell^{(2)} \\[4pt]
		- \mathbf{C}^{(2)} \\[4pt]
		\mathbf{B}^{(2)}
	\end{pmatrix}, 
\end{align}
\begin{align}		
	\tilde{\mathbf{C}}^{(4)} = \tilde{\mathbf{C}}^{(4)} - \tfrac{1}{2} \, \kappa \, \mathbf{B}^{(2)} \! \wedge \ell^{(2)} + \tfrac{1}{4} \, \kappa^2 \, \mathbf{C}^{(2)} \! \wedge \ell^{(2)},
\end{align}
\end{subequations}
which are analogous to \eqref{eq:pinvt}. However, importantly, $\kappa$ is now replaced with $\kappa^{-1}$ in the transformation matrix. 

\vspace{3mm}

\noindent $\bullet$~\emph{From non-relativistic string theory to M1T} ($\gamma \, c^{(0)} \! + \delta = 0$). In this case, $c^{(0)} = - \delta / \gamma$ is rational. The SL($2,\mathbb{Z}$) transformation maps non-relativistic string theory to M1T, with
\begin{subequations} 
\begin{align}
	\textbf{non-rel. string theory $\rightarrow$ M1T} \qquad \gamma \, c^{(0)} \! + \delta = 0 \notag
\end{align}
\vspace{-8mm}
\begin{align}
	\tilde{\tau}^{}_\text{MN} &= \frac{|\gamma|}{g^{}_\text{s}} \, \tau^{}_\text{MN}\,,
		&%
	e^{\tilde{\varphi}} &= \gamma^2 \, e^{-\varphi}\,,
		&%
	\tilde{\mathbf{B}}^{(2)} &= - \sgn(\gamma) \, \mathbf{C}^{(2)}\,, \\[4pt]
	\tilde{E}^{}_\text{MN} &= \frac{|\gamma|}{g^{}_\text{s}} \, E^{}_\text{MN}\,,
		&%
	\tilde{c}^{\,(0)} &= \frac{\alpha}{\gamma}\,, 
		&%
	\tilde{\mathbf{C}}^{(2)} &= \sgn(\gamma) \, \mathbf{B}^{(2)}\,, \label{eq:snrmot}
\end{align}
\end{subequations}
while $\mathbf{C}^{(4)}$ remains invariant. 

When consecutive transformations are considered, one has to make sure to start with the correct branch. For example, after we apply the rule~\eqref{eq:pinvt} mapping from M1T to non-relativistic string theory with $\kappa = \kappa_1$\,, the rule~\eqref{eq:nrstself} mapping non-relativistic string theory to itself with $\kappa = \kappa_2$ is only applicable if the branching factor in \eqref{eq:pinvt} is positive. The composed transformations are self-consistently the ones in \eqref{eq:pinvt} with $\kappa = \kappa^{-1}_1 + \kappa_2$\,. 

\subsection{\texorpdfstring{SL($2,\mathbb{Z}$)}{SL(2,Z)} T-Duality Asymmetry} \label{sec:sltztda}

\begin{figure} [ht!]
\centering
\begin{tikzpicture}

\draw (-5,0) node (A) [text width=2cm, rounded corners=5pt, fill=vub!10,align=center] {\textbf{MpT}};

\draw (0,0) node (B)[text width=2cm, rounded corners=5pt, fill=ForestGreen!10,align=center] {\textbf{M(p+2)T}};

\draw [vub,->,dashed,thick] ([xshift=0.1cm,yshift=0.1cm]A.north) to [out=30,in=150] node [pos=0.45,above=0cm] {\small$\gamma \, \CB \! + \delta = 0$} ([xshift=-0.1cm,yshift=0.1cm]B.north);

\draw [ForestGreen,<-,thick] ([xshift=0.1cm,yshift=-0.1cm]A.south)  to [out=-30,in=-150] node [pos=0.5,below=0cm] {\small $\gamma \neq 0$}([xshift=-0.1cm,yshift=-0.1cm]B.south) ;

\draw [vub,<-,thick] ([xshift=0cm,yshift=0.1cm]A.west) to [out=135,in=225,loop,looseness=20] node [pos=0.5,left=0cm] {\small $\gamma \, \CB \! + \delta \neq 0$} ([xshift=0cm,yshift=-0.1cm]A.west);

\draw [ForestGreen,->,dashed,thick] ([xshift=0cm,yshift=0.1cm]B.east) to [out=45,in=315,loop,looseness=20] node [pos=0.5,right=0cm] {\small $\gamma=0$} ([xshift=0cm,yshift=-0.1cm]B.east);

\end{tikzpicture}
\caption{\textbf{SL($2,\mathbb{Z}$) T-duality asymmetry} between M$p$T on a transverse 2-torus and M($p$+2)T on a longitudinal 2-torus, in terms of the SL($2,\mathbb{Z}$) group parameters and the internal $B$-field $\CB$\,. 
This figure is analogous to the SL($2,\mathbb{Z}$) S-duality asymmetry in Figure~\ref{fig:SL2Asymmetry}.}
\label{fig:SL2Asymmetryt} 
\end{figure}
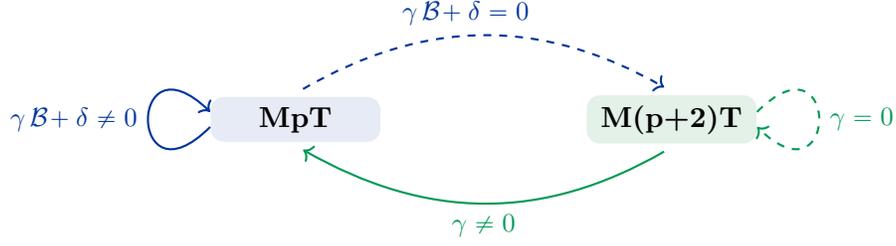

We now show that an analogous polynomial realisation of SL($2,\mathbb{Z}$) also arises from the T-duality asymmetry between M$p$T and M($p$+2)T as shown in Figure \ref{fig:SL2Asymmetryt}.

\subsubsection{\texorpdfstring{SL($2,\mathbb{Z}$)}{SL(2,Z)} from \texorpdfstring{O($2,2; \mathbb{Z}$)}{O(2,2)} Duality} \label{sec:sltzt}

We start with type II string theory, before any decoupling limit is taken. T-duality on a $d$-dimensional internal torus (with coordinates $x^i)$ consists of O$(d,d;\mathbb{Z})$ transformations, which act tensorially as linear transformations on the following combination (`generalised metric') of the internal metric $G_{ij}$ and Kalb-Ramond field $B_{ij}$\,:
\be
	\CH = 
	\begin{pmatrix}
		G^{}_{ij} - B^{}_{ik} \, G^{k\ell} \, B^{}_{\ell j} &\quad B^{}_{ik} \, G^{kj} \\[4pt]
		- G^{ik} \, B^{}_{k j} &\quad G^{}_{ij}
	\end{pmatrix}.
\ee
When $d = 2$\,, and after a proper change of basis (see Appendix~\ref{sec:ReviewOdd}), we decompose the generalised metric as
\be \label{eq:tdgm}
	\CH \rightarrow 
	\frac{G^{}_{ij}}{\sqrt{G}} \otimes
	\frac{1}{\tau^{}_2}
	\begin{pmatrix}
		\bigl|\tau\bigr|^2 &\,\, \tau^{}_1 \\[4pt]
		\tau^{}_1 &\,\, 1
	\end{pmatrix},
		\qquad%
	G \equiv \det \bigl( G_{ij} \bigr)\,,
		\qquad%
	\tau \equiv \tau^{}_1 + i \, \tau^{}_2 = \CB + i \, G^{\frac{1}{2}}\,,
\ee
where $B_{ij} = \CB \, \epsilon^{}_{ij}$\,. Here, the second factor in $\CH$ can be thought of as the modulus of a metric on a `hidden' torus, with $\tau$ the modulus akin to \eqref{eq:tomod} and the following identifications between SL($2,\mathbb{Z}$) S- and T-duality:
\be
\begin{tabular}{c|c}
	\textbf{S-duality} & \textbf{T-duality} \\[2pt]
	\hline
	$C^\text{\scalebox{0.8}{(0)}}$ & $\CB$ \\
	$e^\Phi$ & $G^{-\frac{1}{2}}$
\end{tabular}
\ee
The full T-duality group is O($2,2;\mathbb{Z}$). We are interested in the part of O($2,2;\mathbb{Z}$) that is beyond the $\mathbb{Z}_2$ transformation associated with the Buscher duality on a single circle, and therefore consider the SO($2,2;\mathbb{Z}$) subgroup that factorises as
\be
	\text{SO}(2,2;\mathbb{Z}) = \Bigl( \text{SL}(2,\mathbb{Z}) \times \text{SL}(2,\mathbb{Z}) \Bigr) / \mathbb{Z}_2\,.
\ee
One of these SL($2,\mathbb{Z}$) factors generates geometric volume-preserving transformations of the two isometry directions on the internal torus. The other SL($2,\mathbb{Z}$) subgroup acts on the hidden torus with modulus $\tau$ as prescribed in \eqref{eq:tdgm}. The Buscher transformation acting on a single circle exchanges the two SL($2,\mathbb{Z}$)'s and thus extends SO($2,2;\mathbb{Z}$) to the full T-duality group O($2,2;\mathbb{Z}$). 
This exchange reveals that the `hidden' torus with modulus $\tau$ becomes the physical torus of the theory T-dual under a single Buscher transformation. Note this torus will again be anisotropic with respect to scalings in $\omega$.

\subsubsection{Matrix \texorpdfstring{$p$}{p}-Brane Theory on a 2-Torus} \label{sec:tdampt}

We now study how the SL($2,\mathbb{Z}$) T-duality associated with the modulus $\tau$ in \eqref{eq:tdgm} acts on M$p$T.
We have learned in Section~\ref{sec:tda} that M$p$T on a transverse 2-torus is T-dualised to itself if $\CB \neq 0$ but to M($p$+2)T on a longitudinal 2-torus if $\CB = 0$\,, while M$(p+2)$T always T-dualises to M$p$T with vanishing $B$-field on the torus.
The SL($2,\mathbb{Z}$) generalisation of this Buscher T-duality asymmetry will introduce a more involved structure as in Section~\ref{sec:sdmpt}. 
To spell out the similarities, first we consider M$p$T compactified on a transverse 2-torus. 
From the BPS decoupling limit~\eqref{eq:cmpta}, we read off that 
\be \label{eq:ggg}
	G_{ij} = \frac{E_{ij}}{\omega} 
		\quad\implies\quad%
	G^{-\frac{1}{2}} = \omega \, g^{-\frac{1}{2}}\,,
        \qquad%
    g = \det \bigl( E_{ij} \bigr)\,.
\ee
This is analogous to the dilaton prescription $e^\Phi = \omega \, e^{\varphi}$ in \eqref{eq:gbpcnrs} for non-relativistic string theory.
Conversely, if we consider M($p$+2)T on a longitudinal torus, then \eqref{eq:ggg} is replaced with
\be
	G^{}_{ij} = \omega \, \tau^{}_{ij}
		\quad\implies\quad%
	G^{-\frac{1}{2}} = \frac{g^{-\frac{1}{2}}}{\omega}\,,
        \qquad%
    g = \det \bigl( \tau_{ij} \bigr)\,.
\ee
This is analogous to the dilaton prescription $e^\Phi = \omega^{-1} \, e^\varphi$ in M1T (see \eqref{eq:motrp}). We then form the following dictionary mapping between the SL($2,\mathbb{Z}$) S- and T-duality asymmetry:
\be
\begin{tabular}{c|c}
	\textbf{S-duality asymmetry} & \textbf{T-duality asymmetry}\\[2pt]
	\hline
	non-rel. string/M1T & M$p$T/M($p$+2)T \\[2pt]
	$c^\text{\scalebox{0.8}{(0)}}$ & $\CB$ \\[2pt]
	$e^\varphi$ & $g^{-1/2}$ \\[2pt]
	$\tau^{}_\text{MN} \quad E^{}_\text{MN}$ & $e^{-\varphi}$
\end{tabular}
\label{STDICT}
\ee
and, in contrast to the expression in \eqref{eq:vekk}, the asymmetry factor $\kappa$ now becomes
\be \label{eq:tdafk}
	\kappa = \frac{\gamma \, \sqrt{g}}{\gamma \, \CB + \delta}\,.
\ee
The same SL($2,\mathbb{Z}$) S-duality structure in Section~\ref{sec:sdmpt} then translates directly to T-duality asymmetry, as depicted in Figure~\ref{fig:SL2Asymmetryt}. 

To write down the explicit transformation rules for the gauge potentials requires some additional work.
The original T-duality $\mathrm{SL}(2,\mathbb{Z})$ doublets are the one-forms arising from the metric and $B$-field as well as RR polyforms with zero or two toroidal indices (see Appendix~\ref{sec:ReviewOdd}), whereas in the S-duality case we only needed to consider the two-form doublet.
The transformations of the doublet can again be conveniently expressed by considering combinations along the lines of \eqref{VEC_projected}, where we contract with vielbeins for the $\mathrm{SL}(2)/\mathrm{SO}(2)$ coset to obtain a doublet transforming under the $\mathrm{SO}(2)$ rotation of the form \eqref{eq:vetrnsf}.
Meanwhile, the T-duality singlets include a two-form arising from the $B$-field and an RR polyform with one toroidal index, whereas in the S-duality case we had just the invariant four-form. 
Starting from first principles, one needs to combine the $\mathrm{SL}(2,\mathbb{Z})$ T-duality transformations of these multiplets (following from Appendix~\ref{sec:ReviewOdd}) with the appropriate decomposition of the M$p$T background fields with longitudinal or transverse isometries (following from Appendix~\ref{sec:BuscherdMpT} specialised to $d=2$). After expanding in $\omega$ and after some algebra, we can read off the dual background. 

It can then be shown that the resulting SL($2,\mathbb{Z}$) transformations also take an elegant polynomial form using the powerful approach of invariant theory, as we have briefly commented on earlier below \eqref{eq:pinvt}. According to~\cite{Bergshoeff:2023ogz}, an $N$-dimensional polynomial realisation of SL($2,\mathbb{Z}$) can be written as a lower triangular matrix $U_N(x)$ acting on an $N$-dimensional vector field $\mathbf{V}$, such that the SL($2,\mathbb{Z}$) transformation takes the form,
\be \label{eq:polr}
    \mathbf{V} \rightarrow \mathbf{U}_{\!N}(x) \, \mathbf{V}\,,
        \qquad%
    \mathbf{U}_{\!N} (x) =
    \begin{pmatrix}
        1 & 0 & 0 &\,\, \cdots &\,\,\, 0\,\, \\[4pt]
        -x & 1 & 0 &\,\, \cdots &\,\,\, 0\,\, \\[4pt]
        \frac{x^2}{2} & -x & 1 & \cdots & 0 \\[4pt]
        \vdots &\,\, 
        \vdots & \vdots &\,\, \ddots &\,\,\, \vdots\,\, \\[4pt]
        \frac{(-x)^{N-1}}{(N-1)!} &\,\, \frac{(-x)^{N-2}}{(N-2)!} &\,\, \frac{(-x)^{N-3}}{(N-3)!} &\,\, \cdots &\,\,\, 1\,\,
    \end{pmatrix}\,.
\ee
Below, we spell out the final results of this procedure, which makes use of polynomial realisations with $N=2$ and $N=4$. 
For ease of presentation, we omit all the branching factors by committing ourselves to a particular pair of branches connected by the SL($2,\mathbb{Z}$) transformation. 
The transformations involving other branches are identical except for various changes of signs.  
It is convenient to deal with the SL($2,\mathbb{Z}$) transformations of the M$p$T (or M($p$+2)T) RR potentials $c^{(q)}$ in a collective way by introducing the polyform 
\be
    \mathbf{C}_\text{poly} = \sum_q c^{(q)} \! \wedge e^{B^{(2)}}\,. 
\ee
Decomposing with respect to the splitting of the spacetime index $\text{M} = (\mu\,,\,i)$, with $x^i$ the coordinates on the internal 2-torus, we write as in~\eqref{eq:plf},
\be
    \mathbf{C}^{}_\text{poly} = \mathbf{C}^{}_0 + \mathbf{C}^{}_i \, \dd x^i + \tfrac{1}{2} \, \mathbf{C}^{}_{12} \, \epsilon^{}_{ij} \, \dd x^i \wedge \dd x^j\,.
\ee
We further define
$$
    \mathbf{A}^i \equiv g^{\frac{1}{2}} A^i\,,
        \qquad%
    \mathbf{B}^i \equiv g^{-\frac{1}{2}} \, \epsilon^{ij} \, b^{}_j\,,
        \qquad%
    \mathbf{L} \equiv \tau^0 \! \wedge \cdots \wedge \tau^{p}\,,
        \qquad%
    (\mathbf{O}^{}_1 \, \mathbf{O}^{}_2) \equiv \tfrac{1}{2} \, \epsilon_{ij} \, \mathbf{O}_1^i \! \wedge \! \mathbf{O}_2^j\,.
$$
On the M$p$T side, where the internal torus is transverse, we define the following quantity carrying a four-dimensional polynomial realisation:
\begin{align}
	\mathbf{C}_\text{\scalebox{0.8}{\text{M$p$T}}} 
		\equiv%
	 \begin{pmatrix}
    		3 \, \bigl(\mathbf{B}\mathbf{B}\bigr) \! \wedge \! \mathbf{L} \\[4pt]
		\bigl[ - \tfrac{1}{4} \, \bigl(\mathbf{A}\mathbf{B}\bigr)^2 + 3 \, \bigl(\mathbf{A}\mathbf{B}\bigr) - 1 \bigr] \! \wedge \! \mathbf{L} \\[4pt]
       		g^{\frac{1}{2}} \, \mathbf{C}_0 \\[4pt]
    		\,- g^{-\frac{1}{2}} \bigl( \mathbf{C}_{12} - \mathbf{C}_0 \, \CB \bigr) + \bigl(\mathbf{A}\mathbf{A}\bigr) \! \wedge \! \mathbf{L}\,
	\end{pmatrix}.
\end{align}
On the M($p$+2)T side, where the internal torus is longitudinal, we instead define 
\begin{align}
	\!\!\mathbf{C}_\text{\scalebox{0.8}{\text{M($p$+2)T}}} 
		\equiv%
	 \begin{pmatrix}
    		3 \, \bigl(\mathbf{A}\mathbf{A}\bigr) \! \wedge \! \mathbf{L} \\[4pt]
		\bigl[ - \tfrac{1}{4} \, \bigl(\mathbf{A}\mathbf{B}\bigr)^2 - 3 \, \bigl(\mathbf{A}\mathbf{B}\bigr) - 1 \bigr] \! \wedge \! \mathbf{L} \\[4pt]
       		g^{-\frac{1}{2}} \bigl( \mathbf{C}_{12} - \mathbf{C}_0 \, \CB \bigr) \\[4pt]
    		g^{\frac{1}{2}} \, \mathbf{C}_0 + \bigl(\mathbf{B}\mathbf{B}\bigr) \! \wedge \! \mathbf{L}
	\end{pmatrix}.
\end{align}
We find the following T-duality asymmetry between M$p$T and M($p$+2)T:
\vspace{2mm}
\begin{gather}
\begin{align*}
	\hline \\[-30pt]
		&%
	\begin{array}{l}
        \\[20pt]
		\textbf{M($p$+2)T} \rightarrow \textbf{M$p$T} \\[4pt]
		\quad\qquad\,\,\gamma > 0
	\end{array}
		&\qquad%
	\begin{pmatrix}
		\tilde{\textbf{B}} \\[4pt]
		\tilde{\textbf{A}}
	\end{pmatrix}
		&=
	\mathbf{U}_2 \bigl(-\tfrac{1}{\kappa}\bigr)
	\begin{pmatrix}
		\textbf{A} \\[4pt]
		-\textbf{B}
	\end{pmatrix}
		\qquad%
	\tilde{\mathbf{C}}_\text{\scalebox{0.8}{\text{M($p$+2)T}}} = \mathbf{U}_4 \bigl(-\tfrac{1}{\kappa}\bigr) \, \mathbf{C}_\text{\scalebox{0.8}{\text{M$p$T}}}  \\[-10pt]
		&&%
	\tilde{\mathbf{C}}_i &= \mathbf{C}_i + \epsilon^{}_{ij} \, \mathbf{L} \! \wedge \! \Bigl[ \tfrac{1}{\kappa} \, \mathbf{B}^j \! \wedge \! e^{(\mathbf{A}\mathbf{B})} - \tfrac{1}{2 \, \kappa^2} \, \mathbf{A}^j \! \wedge \! e^{-(\mathbf{A}\mathbf{B})} \Bigr] \\[6pt]
	\hline \\[-10pt]
		&
	\begin{array}{l}
		\,\,\,\,\,\,\,\textbf{M($p$+2)T \scalebox{1.5}{$\Circlearrowleft$}} \\[4pt]
		\qquad\!\!\!\!\gamma=0 \,\, (\delta > 0)
	\end{array}
		&&\qquad%
	\begin{array}{l}
		\tilde{\textbf{A}} = \textbf{A} \\[4pt]
		\tilde{\textbf{B}} = \textbf{B}
	\end{array} 
		\qquad\qquad\,\,\,\,%
	\tilde{\mathbf{C}}_\text{poly} = \mathbf{C}_\text{poly} \\[6pt]
	\hline \\[-30pt]
		&
	\begin{array}{l}
        \\[20pt]
		\quad\quad\,\, \textbf{M$p$T \scalebox{1.5}{$\Circlearrowleft$}} \\[4pt]
		\qquad\!\!\gamma \, \CB + \delta > 0
	\end{array}
		&%
	\begin{pmatrix}
		\tilde{\textbf{B}} \\[4pt]
		\tilde{\textbf{A}}
	\end{pmatrix}
		&=
	\mathbf{U}_2 \bigl(\kappa\bigr)
	\begin{pmatrix}
		\textbf{B} \\[4pt]
		\textbf{A}
	\end{pmatrix}
		\qquad\quad\,\,%
	\tilde{\mathbf{C}}_\text{\scalebox{0.8}{\text{M$p$T}}} = \mathbf{U}_4 \bigl(\kappa\bigr) \, \mathbf{C}_\text{\scalebox{0.8}{\text{M$p$T}}} \\[-10pt]
		&&%
	\tilde{\mathbf{C}}_i &= \mathbf{C}_i + \epsilon^{}_{ij} \, \mathbf{L} \! \wedge \! \Bigl[ \kappa \, \mathbf{A}^j \! \wedge \! e^{- (\mathbf{A}\mathbf{B})} - \tfrac{\kappa^2}{2} \, \mathbf{B}^j \! \wedge \! e^{(\mathbf{A}\mathbf{B})} \Bigr] \\[6pt]
	\hline \\[-10pt]
		&
	\begin{array}{l}
		\textbf{M$p$T} \rightarrow \textbf{M($p$+2)T} \\[4pt]
		\,\gamma \, \CB + \delta = 0 \,\, (\gamma > 0)
	\end{array}
		&&%
	\begin{array}{l}
		\tilde{\mathbf{A}} = - \mathbf{B}
			\qquad%
		\tilde{\mathbf{C}}_0 = \mathbf{C}_{12}
		 \\[4pt]
		\tilde{\mathbf{B}} = \mathbf{A}
			\qquad\,\,\,\,\,%
		\tilde{\mathbf{C}}_{12} = - \mathbf{C}_0
	\end{array} 
		\qquad%
	\tilde{\mathbf{C}}_i = \mathbf{C}_i\\[6pt]
	\hline \\[-30pt]
\end{align*}
\end{gather}
In the first and third cases we indeed arrive at the desired polynomial realisations of SL($2,\mathbb{Z}$). It is also possible to combine the $\mathbf{C}_i$ with other fields to form a vector that transforms as in \eqref{eq:polr}, but this would introduce further redundancies.
Note that in order to match with the RR Buscher transformations of \eqref{eq:polyftd} one needs to take into account not only the choice of branch, but the fact that the $\mathrm{SL}(2,\mathbb{Z})$ transformation used here differs from the conventional Buscher transformation by an additional coordinate transformation (see Appendix \ref{sec:ReviewOdd}).

Finally, we briefly comment on the case where M$p$T is compactified over an anisotropic torus (or, equivalently, pinched tori), with one cycle in the longitudinal sector and the other in the transverse sector. We now revisit the BPS decoupling limit leading to M$p$T, where the toroidal metric $G_{ij}$ in \eqref{eq:tdgm} before the limit is applied becomes
\be
	G_{ij} = \omega \, \tau^{}_i \, \tau^{}_j + \frac{1}{\omega} \, E^{}_i \, E^{}_j\,,
		\qquad%
	\tau^{}_i = \sqrt{\frac{\Gamma}{\rho^{}_2}} 
	\begin{pmatrix}
		1 \\[4pt]
		-\rho^{}_1
	\end{pmatrix},
		\qquad%
	E^{}_i = \sqrt{\Gamma \rho^{}_2} 
	\begin{pmatrix}
		0 \\[4pt]
		1
	\end{pmatrix},
\ee
with the torus modulus $\tilde{\tau} = \rho^{}_1 + i \rho^{}_2 / \omega$ and $\Gamma$ the area of the torus. It then follows that $G^{\frac{1}{2}} = \Gamma$\,, and the modulus $\tau$ in \eqref{eq:tdgm} associated with the hidden torus is independent of $\omega$\,, \emph{i.e.}~$\tau = \CB + i \, \Gamma$\,. Therefore, in the $\omega \rightarrow \infty$ limit, in contrast to the T-duality asymmetry discussed above, the SL($2,\mathbb{Z}$) transformation associated with the hidden torus with the modulus $\tau$ is regular; in contrast, the other SL($2,\mathbb{Z}$) group associated with the real torus with the modulus $\rho$ is develops a T-duality `asymmetry'. However, there is \emph{no} real asymmetry in this case, because M$p$T is mapped to itself, compactified on different anisotropic tori. 
Note that this T-duality on an anisotropic torus acts on the other SL($2,\mathbb{Z}$) subgroup to the T-duality which maps between M$p$T and M$(p+2)$T discussed above, and they are exchangeable with each other via a Buscher transformation acting on a single circle. 

\subsection{IKKT Matrix Theory and Tensionless String} \label{sec:ikktm}

Another generalisation is to the case of M$p$T with $p < 0$\,. 
It has been shown in~\cite{Blair:2023noj, Gomis:2023eav} that M$p$T with $p < -1$ arises from a BPS decoupling limit that zooms in on a background Euclidean (spacelike) brane in type II${}^*$ superstring theory~\cite{Hull:1998vg}. 
In particular, M$(-1)$T arises from a BPS decoupling limit that zooms in on a background D-instanton, and the dynamics therein is described by the Ishibashi-Kawai-Kitazawa-Tsuchiya (IKKT) matrix theory~\cite{Ishibashi:1996xs}. The M$(-1)$T limit can be inferred by performing a timelike T-duality transformation of the M0T limit, and the associated BPS limit is defined by the following reparametrisation of the background fields in type IIB superstring theory:
\begin{subequations} \label{eq:mmotlp}
\begin{align}
	\dd s^2 &= \frac{1}{\omega} \,\eta^{}_{\mu\nu} \, \dd x^\mu \, \dd x^\nu  \,, 
		\qquad%
	e^{\Phi} = i \, \omega^{-2} \, e^\varphi\,, 
		\qquad%
	C^{(0)} = \omega^2 \, e^{-\varphi}\,.
\end{align}
\end{subequations}
The extra $i$ in $e^\Phi$ implies that we analytically continue to type IIB${}^*$ superstring theory. In M(-1)T, the whole target space is `transverse' as the sector longitudinal to a background instanton is pointlike in both space and time. 

It is shown in~\cite{Blair:2023noj, Gomis:2023eav} that the fundamental string in M$(-1)$T is described by tensionless string theory~\cite{Lindstrom:1990qb, Isberg:1993av}. This can be made manifest by applying the limiting prescription~\eqref{eq:mmotlp} to the Nambu-Goto action of the fundamental string, which gives rise to
\be
    S_\text{F1} = - \frac{\omega^{-1} \, T}{2} \int \dd^2 \sigma \, \sqrt{-\det \Bigl( \p_\alpha X^\mu_{\phantom{\dagger}} \, \p_\beta X^{}_\mu \Bigr)}\,.
\ee
It is manifest that for $\omega \rightarrow \infty$\, the tension $\omega^{-1} \, T$ goes to zero. Such a tensionless limit of the string worldsheet theory gives rise to a finite worldsheet action in the Polyakov formulation, where the worldsheet topology develops nodal points. This is reminiscent of the amplitude calculations~\cite{Geyer:2015bja, Geyer:2018xwu} in ambitwistor string theory~\cite{Mason:2013sva}, which is described by a chiral worldsheet action that formally arises from a singular gauge choice in tensionless string theory~\cite{Siegel:2015axg, Casali:2016atr}. 

T-dualising M$(-1)$T over a spatial toroidal compactification proceeds analogously as for M$p$T with $p \geq 0$\,. If the $B$-field on the $d$-dimensional internal torus is of rank $d$\,, then M$(-1)$T is T-dualised to itself~\cite{Banerjee:2024fbi}.\,\footnote{We would like to thank Ritankar Chatterjee for discussions on this topic.} However, if the rank $r$ of the internal $B$-field smaller than $d$\,, with $r < d$\,, then the T-dual theory is in general M$\tilde{p}\,$T with $\tilde{p} = -1-(d-r) < -1$\,, whose target space now becomes Carroll-like~\cite{Blair:2023noj, Gomis:2023eav}. In such a generalised Carrollian spacetime, a spacelike (instead of Minkowskian) sector of the target space is absolute, and the remaining Minkowskian sector is mapped to this absolute spatial sector via Carroll-like boosts.

The Carroll-like dynamics of M$p$T with $p < -1$ is expected to be captured by a new matrix theory from T-dualising IKKT matrix theory on a toroidal compactification. Moreover, it is known that M$p$T is related to M($-p-1$)T via a timelike T-duality transformation, which should be viewed as an analytic continuation instead of a genuine T-dual equivalence~\cite{Gomis:2023eav}. 
We defer the intricacies of T-duality in these novel corners of string theory for future study.\,\footnote{\emph{E.g.}~see~\cite{Chen:2025gaz} for recent work on the Carrollian superstring in M$p$T with $p<0$, which also extends the study of T-duality in one direction in~\cite{Gomis:2023eav}.} 

\section{Intermezzo: Duality Asymmetry from Matrix Theory} \label{sec:damt}

Now let us come full circle and discuss the relationship between the duality asymmetry of non-Lorentzian backgrounds and known features of matrix theory.

We began in Section~\ref{sec:tda} with the simple example of M0T with a transverse 2-torus carrying a non-trivial $B$-field.
The usual M0T limit, without a $B$-field, gives BFSS matrix theory, and can also be phrased in terms of the DLCQ of M-theory.
Adding the $B$-field on a 2-torus is equivalent to turning on the M-theory three-form with a component $C_{12-}$ in the null direction of the DLCQ. 
This configuration was famously argued in~\cite{Connes:1997cr} to correspond to matrix theory on a non-commutative torus.

\vspace{3mm}

\noindent $\bullet$~\emph{T-duality as Morita equivalence.} Let us briefly review how the above described non-commutativity is realised.
Following~\cite{Connes:1997cr}, we consider the IKKT matrix theory on a 2-torus.
This is related to the BFSS matrix theory on a 2-torus by compactifying a further direction, which, as noted above, would correspond to a timelike direction in M0T. This implies that the IKKT matrix theory lives in M$(-1)$T.
The 0-dimensional IKKT matrix theory is
\be
    S \sim \tr \Bigl( \bigl[ X_\mu\,, \, X_\nu \bigr] \, \bigl[ X_\mu\,, \, X_\nu \bigr] + \text{fermions} \Bigr)\,,
\ee
where $X_\mu$\,, $\mu = 0\,, \, \cdots, \, 9$ are $N \times N$ matrices.  Note that $\mu$ is now an Euclidean index \emph{i.e.}~we include a Wick rotation when relating M$(-1)$T and M$0$T. We compactify $X_0$ and $X_1$ on a 2-torus, whose periodicity implies that there exists invertible matrices $U_0$ and $U_1$\,, such that
\begin{subequations} \label{eq:pertt}
\begin{align}
    X^{}_0 + R^{}_0 &= U^{}_0 \, X^{}_0 \, U_0^{-1},
        &%
    X^{}_i &= U^{}_0 \, X^{}_i \, U_0^{-1}\,, \\[4pt]
    X^{}_1 + R^{}_1 &= U^{}_1 \, X^{}_1 \, U_1^{-1},
        &%
    X^{}_i &= U^{}_1 \, X^{}_i \, U_1^{-1}\,,
\end{align}
\end{subequations}
with $i = 2\,, \, \cdots, \, 9$\,. It follows that
$U_0 \, U_1 = \lambda \, U_1 \, U_0$\,,
where $\lambda = e^{2\pi i \, \theta}$ is a complex constant. 
If $\lambda = 1$, then the 2-torus is an ordinary commutative torus, while if $\lambda \neq 1$ then we have a non-commutative torus. 
Viewing the real variable $\theta$, instead of the complex variable $\lambda$, as a parameter of the non-commutative torus, it turns out that there is an equivalence between different such tori under an $\mathrm{SL}(2,\mathbb{Z})$ action:
\be 
    \tilde{\theta} = \frac{\alpha \, \theta + \beta}{\gamma \, \theta + \delta}\,,
\label{ThetaMorita}
\ee
where $\alpha \delta - \beta \gamma =1$.
When the map~\eqref{ThetaMorita} between $\theta$ and $\tilde{\theta}$ is invertible, the two algebras defined by $U_0 \, U_1 = e^{2\pi i \, \theta} \, U_1 \, U_0$ and $U_0 \, U_1 = e^{2\pi i \, \tilde{\theta}} \, U_1 \, U_0$ are equivalent to each other up to a phase factor, which is an example of \emph{Morita equivalence}~\cite{Schwarz:1998qj}.
In~\cite{Connes:1997cr} it was argued that the transformation~\eqref{ThetaMorita} is induced by $\mathrm{SL}(2,\mathbb{Z})$ T-duality given the identification $\theta \sim \CB$, where $\CB$ is the non-zero component of the $B$-field on the original geometry to which the M$(-1)$T (or M0T) limit is taken. Indeed, for $\gamma \, \CB + \delta \neq 0$, \eqref{ThetaMorita} is the fractional linear transformation of the $B$-field on a transverse 2-torus in M$p$T that follows from the results in Section~\ref{sec:da}, where the explicit form comes from applying the dictionary~\eqref{STDICT} to \eqref{eq:nrstself0}.
According to Fig.~\ref{fig:SL2Asymmetryt}, these transformations map M$p$T to itself, but in different background fields. 

When $\CB$, or $\theta$, is rational, there are two special cases to consider.
Firstly, suppose we choose the transformation parameters such that $\alpha \, \theta+\beta =0$.
In this case, the associated transformation~\eqref{ThetaMorita} maps $\theta \neq 0$ to $\tilde \theta = 0$, leading to an equivalent description in terms of matrix theory on a commutative torus. 
This is a well-known statement, and still a transformation from M$p$T to itself. 
Secondly, suppose we choose the transformation parameters such that $\gamma \, \theta + \delta = 0$.
The transformation \eqref{ThetaMorita} is singular. 
However, according to Section~\ref{sec:sltztda}, what we are really doing is mapping from M$p$T to a dual M$(p+2)$T limit, on a longitudinal 2-torus and with a $B$-field $\tilde{\CB} = \alpha/\gamma$\,.\,\footnote{This is obtained by applying the dictionary~\eqref{STDICT} to the transformation of $c^{(0)}$ in \eqref{eq:snrmot}.}
This takes us to usual matrix theory (SYM) on D$(p+2)$-branes, without non-commutativity but with a constant magnetic field on the brane. As we will see explicitly below, the M$(p+2)$T limit with longitudinal $B$-field leads to vanishing non-commutativity.

Note that more generally, we could consider M0T on a $d$-dimensional torus with a $B$-field on the torus.
This leads to matrix theory on a non-commutative $d$-torus, with non-commutativity parameters $\theta^{ij}$ transforming now under $\mathrm{SO}(d,d;\mathbb{Z})$, inherited from the T-duality of string theory~\cite{Connes:1997cr, Schwarz:1998qj, Brace:1998ku, Pioline:1999xg, Seiberg:1999vs}.

\vspace{3mm}

\noindent $\bullet$~\emph{T-duality and Seiberg-Witten map.} Now, suppose the $B$-field on the $d$-dimensional torus is of rank $r$. 
We have seen that T-dualising on the torus directions will lead to a dual M$(d-r)$T limit. 
The D0-branes on which the matrix theory live in the M0T limit transform into D$d$-branes in the M$(d-r)$T limit.
The gauge theory on these D$d$-branes is \emph{Non-Commutative Yang-Mills} (NCYM).
This T-dual picture was described by Seiberg and Witten~\cite{Seiberg:1999vs}.
It is illuminating to dive briefly into some details.
The Seiberg-Witten map defines an effective open string metric $\CG_{ij}$ and a (T-dual) non-commutativity parameter $\theta^{ij}$ via
\be
\Bigl( \CG^{-1}+\theta \Bigr)^{\!ij}
= \Bigl[ ( G+ B)^{-1} \Bigr]^{ij} \,.
\ee
Let us focus once more on the case of T-duality acting on a two-dimensional torus, taking $d=2$ and $r=0$ or $r=2$.
In the two-dimensional case, the $\mathrm{SL}(2,\mathbb{Z})$ modulus $\tau = \CB + i \, \sqrt{G}$ of \eqref{eq:tdgm} becomes in open string variables
\be
\tau = \frac{ - \theta + i \, \sqrt{\CG}}{\CG + \theta^2} \,,
\ee
where $\CG \equiv \det \CG^{ij}$ and $\theta^{ij} = \theta \, \epsilon^{ij}$.

Next, we consider the fate of these open string variables under the two T-dual decoupling limits studied in Section~\ref{sec:sltztda},
associated with M$p$T on a transverse 2-torus and M$(p+2)$T on a longitudinal 2-torus.
Here we can allow for general values of $p$.
In both cases, the open string (inverse) volume modulus scales the same way, $\sqrt{\CG} = \omega^{-1} \sqrt{\CG}$\,, where in the M$p$T case we have $\sqrt{\CG} = \sqrt{\det E_{ij}} / \CB^2$ and in the M$(p+2)$T case we have $\sqrt{\CG} = 1/\sqrt{\det \tau_{ij}}$\,.
We then find that $\theta = - 1 / \CB$ in M$p$T, indicating (for $\CB \neq 0$) a non-commutative theory from the open string perspective, while always $\theta = 0$ in M$(p+2)$T (regardless of the $B$-field on the longitudinal torus).
In both cases, we have the following induced $\mathrm{SL}(2,\mathbb{Z})$ T-duality transformations:
\begin{align} 
\sqrt{\tilde{\CG}}  = \frac{\sqrt{\CG}}{(\alpha-\beta \, \theta)^2}\,,
    \qquad%
\tilde \theta= -\frac{\gamma-\delta \,  \theta}{\alpha-\beta \, \theta} \,.
\label{openstringtransfs2d} 
\end{align}
We can then consider the effect of these transformations for the different cases displayed in Figure~\ref{fig:SL2Asymmetryt}.
For instance, if we start with M$p$T with non-vanishing $\theta = - 1 / \CB$, then transformations with $\gamma \, \CB + \delta \neq 0$, or equivalently $\gamma - \delta \, \theta \neq 0$, take us to M$p$T with non-vanishing $\tilde \theta$\,. 
If $\theta$ is non-zero and rational,
we can find a transformation such that $\gamma - \delta \, \theta = 0$\,, which maps us to an open string picture with $\tilde \theta = 0$. 
This corresponds to a transformation with $\gamma \, \CB + \delta = 0$\,, which in fact maps M$p$T to M$(p+2)$T.\,\footnote{We can also comment on the singular transformations with $\alpha - \beta \, \theta = 0$. 
From the closed string geometrical perspective, these are just the transformations that set $\tilde{\CB} = 0$ in M$p$T.
From the open string perspective, the Seiberg-Witten map is itself singular in the decoupling limit for this value.}

One can similarly check the transformations of M$(p+2)$T with $\theta = 0$\,, which give M$(p+2)$T with $\tilde \theta = 0$ for $\gamma = 0$ and M$p$T with $\tilde \theta \neq 0$ for $\gamma \neq 0$\,.

Interpreted in terms of gauge theories on D($p$+2)-branes, this tells us that (in this setup with a single $B$-field component) an M$p$T limit realises the Seiberg-Witten limit~\cite{Seiberg:1999vs} leading to non-commutative Yang-Mills, while an M$(p+2)$T limit leads to standard Yang-Mills. 
In the next section, we will see an explicit realisation of this in a holographic setting.

\section{\texorpdfstring{$B$}{B}-Field, Non-Commutativity, and Holography} \label{sec:mrh}

We have seen above that there is an interesting interplay between the presence of a $B$-field, string duality of non-Lorentzian corners of string theory, and non-commutative matrix (gauge) theories.
In this Section, we turn our attention to analysing some explicit examples of geometries with $B$-fields which have an important role in holographic duality involving non-commutative gauge theories. 

In~\cite{Blair:2024aqz}, we showed how applying M$p$T decoupling limits to curved D-brane geometries allowed one to derive the usual AdS/CFT correspondence as well as novel non-Lorentzian generalisations.
Asymptotically, applying the M$p$T limit in the form of \eqref{eq:cmpta} generates a flat non-Lorentzian background, and induces the limit leading to super-Yang-Mills (SYM) on D$p$-branes.
In the bulk curved geometry, the same limit produces the near-horizon geometry.
Further M$q$T limits can be applied leading to limits of SYM dual to non-Lorentzian bulk geometries.
In this Section, we show how this same logic carries over to D-brane solutions with non-trivial $B$-fields dual to non-commutative versions of SYM.

In particular, we will study (generalisations of) the Hashimoto-Itzhaki-Maldacena-Russo background~\cite{Hashimoto:1999ut, Maldacena:1999mh} holographically dual to four-dimensional non-commutative Yang-Mills theory (NCYM), which involves a constant $B$-field at infinity. We will discuss how duality asymmetry manifests, and further exhibit novel non-Lorentzian geometries that are conjecturally holographically dual to further BPS decoupling limits of NCYM.

\subsection{Holographic Duals for Non-Commutative Yang-Mills} \label{sec:gmrhd}

\noindent $\bullet$~\emph{D$p$-D($p$+2)-brane geometry in $B$-field.} 
The relevant geometries arise from solutions describing a D$p$-brane dissolved in a D($p$+2)-brane.
Taking a specific near-horizon limit of these solutions leads to the holographic duals of NCYM~\cite{Hashimoto:1999ut, Maldacena:1999mh, Alishahiha:1999ci}.
The $p=1$ case corresponds to the geometry analysed by Maldacena and Russo in~\cite{Maldacena:1999mh} dual to four-dimensional NCYM.
In order to generate the initial D$p$-D($p$+2) solutions, we follow~\cite{Breckenridge:1996tt} (see also~\cite{Russo:1996if}) and start with the D($p$+1)-brane solution for $p < 5$, given by:
\begin{subequations}
\begin{align}
	\dd s^2 &= \!\frac{1}{\sqrt{H}} \Bigl( - \dd X_0^2 + \cdots + \dd X_p^2 + \dd X_{p+1}^2 \Bigr) + \sqrt{H} \, \Bigl( \dd X_{p+2}^2 + \dd X_{p+3}^2 + \cdots + \dd X_9^2 \Bigr)\,, \\[4pt]
	C^{(p+1)} &= \bigl( G_\text{s} \, H \bigr)^{-1} \dd X^0 \!\!\wedge\! \cdots \!\wedge\! \dd X^p\,, 
		\qquad%
	e^\Phi = G^{}_\text{s} \, H^{\frac{2-p}{4}}\,. 
\end{align}
\end{subequations}
We smear this geometry in $X^{p+2}$ such that the harmonic function $H = H(R)$ is independent of $X^{p+2}$, with $R^2 \equiv X_{p+3}^2 + \cdots + X_9^2$\,. Rotating in the $X^{p+1}$--$X^{p+2}$ plane by a constant angle $\Theta$ and then T-dualising along one of the rotated directions, we generate the following solution with D$p$-branes delocalised in a D($p$+2)-brane worldvolume:
\begin{subequations} \label{eq:dppt}
\begin{align}
	\dd s^2 &= \!\frac{1}{\sqrt{H}} \! \left[ \Bigl(- \dd X_0^2 + \cdots + \dd X_p^2 \Bigr) + \frac{\dd X_{p+1}^2 + \dd X_{p+2}^2}{\cos^2 \Theta + H^{-1} \sin^2 \Theta} \! \right] 
	+ \sqrt{H} \, \Bigl( \dd X_{p+3}^2 + \cdots + \dd X_9^2 \Bigr)\,, \notag \\[4pt]
	B^{(2)} &= \frac{\tan \Theta \, \dd X^{p+1} \!\! \wedge \! \dd X^{p+2}}{H \cos^2 \Theta + \sin^2 \Theta}\,,
		\hspace{2.08cm}%
	e^\Phi = \frac{G_\text{s} \, H^{\frac{1-p}{4}}}{\sqrt{\cos^2 \Theta + H^{-1} \sin^2 \Theta}}\,, \\[4pt]
	C^{(p+1)} &= \frac{\sin \Theta}{G_\text{s} \, H} \, \dd X^0 \!\! \wedge \! \cdots \!\wedge\! \dd X^p\,,
		\qquad\qquad\!\!%
	C^{(p+3)} = \frac{G_\text{s}^{-1} \cos\Theta \, \dd X^0 \!\!\wedge\! \cdots \!\wedge\! \dd X^{p+2}}{H \cos^2 \Theta + \sin^2 \Theta}\,, 
\end{align}
\end{subequations}
where to minimise notation we have relabelled the new coordinates after rotating and T-dualising again as $X^{p+1}$ and $X^{p+2}$. 
Moreover, 
\be \label{eq:hlfmp}
	H = 1 + \left( \frac{L}{R} \right)^{\!\!5-p}\!, 
		\qquad%
	L^{5-p} = (4\pi)^{\!\frac{3-p}{2}} \, \frac{\Gamma(\frac{5-p}{2})}{5-p} \, \frac{N \, G_\text{s} \, \ell^{5-p}_\text{s}}{\cos\Theta}\,,
\ee
with $N$ the number of D($p$+2)-branes, $\ell_\text{s}$ the string length, and $G_\text{s}$ the string coupling at the asymptotic infinity where $R \rightarrow \infty$\,. Note that, as in~\cite{Maldacena:1999mh}, we have shifted the Kalb-Ramond two-form by a constant $\tan \Theta \, \dd X^{p+1} \!\!\wedge\! \dd X^{p+2}$ to get the solution~\eqref{eq:dppt}. 

At the asymptotic infinity where $R \rightarrow \infty$\,, we find $H \rightarrow 1$ and the solution~\eqref{eq:dppt} gives
 \begin{subequations} \label{eq:dpptf}
\begin{align}
	\dd s^2 &= - \dd X_0^2 + \dd X_1^2 + \cdots + \dd X_9^2\,, 
		&%
	C^{(p+1)} &= \frac{\sin \Theta}{G_\text{s}} \, \dd X^0 \!\! \wedge \! \cdots \!\wedge\! \dd X^p\,,	
		\qquad%
	e^\Phi = G_\text{s}\,, \\[4pt]
	B^{(2)} &= \tan \Theta \, \dd X^{p+1} \!\! \wedge \! \dd X^{p+2}\,,
		&%
	C^{(p+3)} &= \frac{\cos\Theta}{G_\text{s}} \, \dd X^0 \!\!\wedge\! \cdots \!\wedge\! \dd X^{p+2}\,. 
\end{align}
\end{subequations}
This shows that there is indeed a constant $B$-field whose value is given by $\tan \Theta$\,, originating from the shift we have performed.  

\vspace{3mm}

\noindent $\bullet$~\emph{Asymptotic BPS decoupling limit.}
We now follow the logic of our earlier paper~\cite{Blair:2024aqz}, and seek asymptotic BPS decoupling limits, following the prescription~\eqref{eq:mptp}, such that the ten-dimensional geometry asymptotes to the flat non-Lorentzian M$q$T geometry.
This means grouping the coordinates into longitudinal and transverse sets, $X^A$ and $X^{A'}$, and rescaling them as $X^A = \omega^{1/2} \, x^A$ and $X^{A'} = \omega^{-1/2} \, x^{A'}$\,. Simultaneously we are required to rescale the string coupling $G^{}_\text{s} = \omega^{(q-3)/2} g^{}_\text{s}$\,. The asymptotic BPS decoupling
limit is implemented by sending $\omega$ to infinity. This procedure will in turn induce a limit of the full bulk solution.
We are interested in the cases where this limit is well-defined leading to either a near-horizon limit (which remains a Lorentzian geometry) or a new non-Lorentzian M$q$T geometry.

In the present case, a natural choice would be to apply an asymptotic M$(p+2)$T limit aligned with the D$(p+2)$-brane configuration of the solution \eqref{eq:dppt}, \emph{i.e.}~choosing $X^A = ( X^0,\dots, X^{p+2})$ to be the longitudinal sector.
This simply generates the standard near-horizon limit of this brane geometry, with a vanishing $B$-field in the associated $\omega \rightarrow \infty$ limit. 
This is as expected from~\cite{Maldacena:1999mh} and the discussion in Section~\ref{sec:damt}.

Instead, we consider the asymptotic M$p$T limit aligned with the D$p$-brane configuration of the solution~\eqref{eq:dppt}, \emph{i.e.}~choosing $X^A = ( X^0,\dots, X^{p})$ to be the longitudinal sector.
In this case, we have to ensure that the $B$-field is finite. 
In particular, we require that, after taking the $\omega \rightarrow \infty$ limit, the resulting M$p$T geometry at asymptotic infinity is described by  
\begin{align} \label{eq:dpptfpre}
    \hspace{-9mm}
	\begin{array}{c}
        \qquad
        \bigl(\tau^0\,,\,\cdots,\,\tau^p\bigr) = \bigl( \dd t\,, \, \dd x^1, \, \cdots, \, \dd x^p \bigr)\,, \\[6pt]
	   \bigl( E^{p+1}\,,\,\cdots,\,E^9\bigr) = \bigl( \dd x^{p+1}, \, \cdots, \, \dd x^9 \bigr)\,, 
    \end{array}
        \qquad\quad%
    b^{(2)} = \CB \, \dd x^{p+1} \!\!\wedge\! \dd x^{p+2}\,, 
\end{align}
together with a constant string coupling $g^{}_\text{s}$ and some RR potentials, whose values we do not specify \emph{a priori}. Comparing with the flat-spacetime M$p$T prescription~\eqref{eq:mptp}, we find the following prescription:
\begin{subequations} \label{eq:amptl}
\begin{align}
	\bigl( X^0,\,\cdots,\, X^p \bigr) &= \sqrt{\omega} \, \bigl( t\,,\,x^1,\,\cdots,\, x^p \bigr)\,, 
		&%
	G^{}_\text{s} &= g^{}_\text{s} \, \omega^{\frac{p-3}{2}}\,, \\[4pt]
	\bigl( X^{p+1},\,\cdots,\, X^9 \bigr) &= \frac{1}{\sqrt{\omega}} \bigl( x^{p+1},\,\cdots,\, x^9 \bigr)\,,
		&%
	\Theta &= \arctan\bigl(\omega \, \CB \bigr)\,.
\end{align}
\end{subequations}
Without loss of generality, we assume that $\CB > 0$\,. 
We emphasise that our convention, as in~\cite{Blair:2024aqz}, is that $\alpha'$ is kept fixed. 
One may translate conventions, using for instance that $\dd s^2/\alpha'$ is fixed, to confirm that the decoupling limit we use matches the $\alpha' \rightarrow 0$ limit taken in~\cite{Hashimoto:1999ut, Maldacena:1999mh, Alishahiha:1999ci}.

Then, in the $\omega \rightarrow \infty$ limit of the asymptotic geometry~\eqref{eq:dpptf}, we are led to the M$p$T geometry~\eqref{eq:dpptfpre}, with the RR potentials
\be \label{eq:cptb}
	c^{(p+1)} = - \frac{1}{2 \, g_\text{s} \, \CB^2} \, \dd x^0 \!\wedge\! \cdots \!\wedge\! \dd x^{p}\,,\quad
    c^{(p+3)} = \frac{1}{g^{}_\text{s} \, \CB} \, \dd x^0 \!\wedge\! \cdots \!\wedge\! \dd x^{p+2}\,.
\ee
The complete asymptotic M$p$T geometry is then described by \eqref{eq:dpptfpre} and \eqref{eq:cptb}.

\vspace{3mm}

\noindent $\bullet$~\emph{Generating the bulk near-horizon limit.} Next, we apply the asymptotic BPS decoupling limit prescribed by \eqref{eq:amptl} to the bulk supergravity solution~\eqref{eq:dppt}. It can be checked that the $\omega \rightarrow \infty$ limit generates the near-horizon limit in the bulk, leading to the following geometry:
\begin{subequations} \label{eq:mr}
\begin{align}
	\dd s^2 &= \mathbb{\Omega} \, \Bigl( - \dd t^2 + \cdots + \dd x_p^2 \Bigr) + \frac{1}{\mathbb{\Omega}} \left[ \frac{\dd x_{p+1}^2 \! + \dd x_{p+2}^2}{1 + ( \mathbb{\Omega} \, \CB )^{-2}} + \dd x^2_{p+3} + \cdots + \dd x^2_{9} \right]\!, \\[4pt]
	B^{(2)} &= \frac{\CB}{1 + ( \mathbb{\Omega} \, \CB )^{-2}} \, \dd x^{p+1} \!\!\wedge\! \dd x^{p+2}\,,
		\qquad%
	e^\Phi = \frac{\mathbb{\Omega}^{\frac{p-3}{2}} \, g^{}_\text{s}}{\sqrt{1 + ( \mathbb{\Omega} \, \CB )^{-2}}}\,, \\[4pt]
	C^{(p+1)} &= \frac{\mathbb{\Omega}^2}{g^{}_\text{s}} \, \dd t \! \wedge \! \dd x^1 \! \wedge \cdots \!\wedge\! \dd x^p\,,
		\quad\,\,\,\,\,\,%
	C^{(p+3)} = \frac{\bigl(g^{}_\text{s} \, \CB \bigr)^{-1}}{1 + ( \mathbb{\Omega} \, \CB )^{-2}} \, \dd t \! \wedge \! \dd x^1 \!\wedge\! \cdots \!\wedge\! \dd x^{p+2}\,, 
\end{align}
\end{subequations}
where
\be \label{eq:ortlfmp}
	\mathbb{\Omega} = \left( \frac{r}{\ell} \right)^{\!\!\frac{5-p}{2}}\!\!,
		\qquad%
	r^2 = x^2_{p+3} + \cdots + x^2_9\,,
		\qquad%
	\ell^{5-p} = (4\pi)^{\!\frac{3-p}{2}} \, \frac{\Gamma\bigl(\frac{5-p}{2}\bigr)}{5-p} \, N \, G_\text{s} \, \ell^{5-p}_\text{s} \, \CB\,.
\ee
In the special case of $p = 1$\,, we recover the Maldacena-Russo geometry~\cite{Maldacena:1999mh}, whose NSNS sector was obtained in~\cite{Hashimoto:1999ut}.
When $p \neq 1$, we recover the backgrounds studied in~\cite{Alishahiha:1999ci}.
These are the holographic duals of NCYM on the D$(p+2)$-brane.

In the $\mathbb{\Omega} \rightarrow \infty$ limit of the holographic geometry \eqref{eq:mr}, \emph{i.e.} when $(r/\ell) \rightarrow \infty$, we recover the asymptotic M$p$T geometry~\eqref{eq:dpptfpre} up to a shift of the RR potential $c^{(p+1)}$.
This shift can be understood by noting that the limit of the asymptotic RR potentials of \eqref{eq:dpptf} is sensitive to subleading $O(\omega^{-2})$ terms in $\sin\Theta$\,. 
As in the simpler near-horizon geometries discussed in~\cite{Blair:2024aqz}, it should be possible to view the full bulk Lorentzian near-horizon geometry as a deformation of the asymptotic flat non-Lorentzian geometry, with the geometric deformation determined by the replacement $\omega \rightarrow \mathbb{\Omega}$\,.
To ensure this is the case, we are led to the following modification of the asymptotic limiting prescription~\eqref{eq:amptl}: 
\begin{align} \label{eq:amptlmod}
	\bigl( X^0,\,\cdots,\, X^p \bigr) &= \sqrt{\omega} \, \bigl( t\,,\,x^1,\,\cdots,\, x^p \bigr)\,, 
		&%
	& G^{}_\text{s} = \frac{g^{}_\text{s} \, \omega^{\frac{p-3}{2}}}{\sqrt{1 + \omega^{-2} \, \CB^{-2}}} \,, 
        \quad%
    \Theta = \arctan\bigl(\omega \, \CB \bigr)\,, \notag \\[4pt]
	\bigl( X^{p+3},\,\cdots,\, X^9 \bigr) &= \frac{1}{\sqrt{\omega}} \bigl( x^{p+3},\,\cdots,\, x^9 \bigr)\,,
		&%
	&\bigl( X^{p+1},\, X^{p+2} \bigr) = \frac{\bigl( x^{p+1},\, x^{p+2} \bigr)}{\sqrt{\omega + \omega^{-1} \, \CB^{-2}}}\,.
\end{align}
Now, in the $\omega \rightarrow \infty$ limit, the asymptotic M$p$T geometry is described by
\begin{subequations} \label{eq:dpptffinal}
\begin{align}
	\bigl(\tau^0\,,\,\cdots,\,\tau^p\bigr) &= \bigl( \dd t\,, \, \dd x^1, \, \cdots, \, \dd x^p \bigr)\,, 
		&%
	b^{(2)} &= \CB \, \dd x^{p+1} \!\!\wedge\! \dd x^{p+2}\,, \\[4pt]
	\bigl( E^{p+1}\,,\,\cdots,\,E^9\bigr) &= \bigl( \dd x^{p+1}, \, \cdots, \, \dd x^9 \bigr)\,, 
		&%
	c^{(p+3)} &= \frac{1}{g^{}_\text{s} \, \CB} \, \dd x^0 \!\wedge\! \cdots \!\wedge\! \dd x^{p+2}\,.
\end{align}
\end{subequations}
This is almost identical to \eqref{eq:dpptfpre} and \eqref{eq:cptb} except that now $c^{(p+1)} = 0$ in the asymptotic M$p$T geometry. 
After the above replacement, the $\omega$-deformed prescription of the asymptotic M$p$T geometry (which is flat before the deformation) precisely gives rise to the bulk curved geometry~\eqref{eq:mr} upon replacing $\omega$ with $\mathbb{\Omega}$\,. We will show later in Section~\ref{sec:asdbgttbar} that, after an S-duality transformation, the deformation parametrised by the background-dependent quantity $\mathbb{\Omega}$ receives an interpretation as the $T\bar{T}$-deformation of non-relativistic string theory, conforming to the expectation from \cite{Blair:2024aqz}.

\vspace{3mm}

\noindent $\bullet$~\emph{Non-commutative Yang-Mills theory.} The field theory dual is described by the gauge theory on a stack of D($p$+2)-branes that extend in $(t,\,x^1,\,\cdots,\,x^{p+2})$ at the asymptotic infinity. These branes are orthogonal to $(x^{p+3},\,\cdots,\,x^9)$ that enter in the definition of $r$ in \eqref{eq:ortlfmp}. Due to the presence of the $B$-field in the $(x^{p+1},\,x^{p+2})$-plane, the gauge theory on such D($p$+2)-branes is NCYM. The non-commutativity between $x^{p+1}$ and $x^{p+2}$ is made manifest after applying the Seiberg-Witten map~\cite{Seiberg:1999vs}, discussed briefly in Section~\ref{sec:damt}, to pass to the effective description in terms of the open string. Before sending $\omega$ to infinity, the limiting prescription leading to the asymptotic geometry is given in \eqref{eq:amptlmod}. The reparametrised background fields in the asymptotic infinity are essentially given by \eqref{eq:mr} except that $\mathbb{\Omega}$ is replaced with $\omega$\,, so that in particular the asymptotic NSNS sector is:
\begin{subequations} \label{eq:bpsd}
\begin{align}
	\dd s^2 &= {\omega} \, \Bigl( - \dd t^2 + \cdots + \dd x_p^2 \Bigr) + \frac{1}{{\omega}} \left[ \frac{\dd x_{p+1}^2 \! + \dd x_{p+2}^2}{1 + ( {\omega} \, \CB )^{-2}} + \dd x^2_{p+3} + \cdots + \dd x^2_{9} \right]\!, \\[4pt]
	B^{(2)} &= \frac{\CB}{1 + ( {\omega} \, \CB )^{-2}} \, \dd x^{p+1} \!\!\wedge\! \dd x^{p+2}\,,
		\qquad%
	e^\Phi = \frac{g^{}_\text{s} \, {\omega}^{\frac{p-3}{2}}}{\sqrt{1 + ( {\omega} \, \CB )^{-2}}}\,. 
\end{align}
\end{subequations}
The Seiberg-Witten map says that, in the effective propagator of open strings, the non-commutativity between the worldvolume coordinates is determined by
\be
    \bigl[ x^\alpha, x^\beta \bigr] \, \propto \left( \frac{1}{G + B} \, B \, \frac{1}{G - B} \right)^{\!\!\alpha\beta}\!,
        \quad%
    \alpha = 0\,,\, \cdots, \, p+2
        \quad\xRightarrow{\omega \rightarrow \infty}\quad
    \bigl[ x^{p+1}, x^{p+2} \bigr] \, \propto \, \frac{1}{\CB}\,.
\ee
Therefore, the worldvolume coordinates $x^{p+1}$ and $x^{p+2}$ do \emph{not} commute with each other on the brane. Moreover, the open string coupling from the Seiberg-Witten map is given by
\be \label{eq:osc}
    g^2_\text{o} = e^\Phi \sqrt{\frac{\det(G+B)}{\det G}} = \omega^{\frac{p-1}{2}} \, g^{}_\text{s} \, \CB\,,
\ee
which implies that the effective NCYM coupling is proportional to $\sqrt{\CB}$\,. 
When $p=1$ as in~\cite{Maldacena:1999mh} we automatically have a finite NCYM coupling (before taking the large $N$ limit). 
For general $p$, \eqref{eq:osc} can be viewed as defining the $\omega$ expansion of the open string coupling in this M$p$T limit prescribed by \eqref{eq:bpsd}.
To see that this is sensible, we check explicitly that the D-brane action is finite and extract the effective NCYM coupling, which is also given by $g^{}_\text{YM} \sim \sqrt{g^{}_\text{s} \, \CB}$.
Namely, we apply the limiting prescription directly to a single probe D($p$+2)-brane 
extending in $(t,\,x^1,\,\cdots,\,x^{p+2})$ at the asymptotic infinity, where there is a U(1) gauge field strength $F_{\alpha\beta}$ residing on the brane. 
As a particularly simple example, we consider the case where $p = 0$\,. A short calculation shows that the $\omega\rightarrow\infty$ limit of the D2-brane action leads to a finite action, 
\be
    S^{}_{\text{D2}} \sim \frac{1}{2 \, \CB^2 \, g^{}_\text{s}} \int \dd^3 \sigma \, \frac{\CB^2 \, \bigl( F_{01}^2 + F_{02}^2 \bigr) - F_{12}^2}{\CB + F_{12}}\,.
\ee
Recall that we have assumed $\CB > 0$\,, which can be thought of as a vacuum expectation value of the field strength $F_{12}$\,. 
After rescaling $F_{12} \rightarrow \CB F_{12}$, we expand the action for small $F_{12}$ as
\be
    S^{}_{\text{D2}} \sim -  \frac{1}{4 \, g_\text{YM}^2} \int \dd^3 \sigma\,\sum_{n=0}^\infty \bigl( - F_{12} \bigr)^n \, F_{\alpha\beta} \, F^{\alpha\beta}\,,
\ee
where the effective gauge coupling is indeed $g^{}_\text{YM} \sim \sqrt{g^{}_\text{s} \, \CB}$\,. 
The $p > 1$ cases work similarly. 

In the non-abelian case and in terms of the open string data, NCYM involves the field strength $F_{\mu\nu} = \p_\mu A_\nu - \p_\nu A_\mu - i \, ( A_\mu \star A_\nu - A_\nu \star A_\mu)$\,, where `$\star$' denotes the Moyal product and involves the non-commutative parameter $\theta$ as introduced in Section~\ref{sec:damt}~\cite{Seiberg:1999vs}.

\subsection{Asymptotic Duality Asymmetry}

Having described the holographic duals for NCYM in Section~\ref{sec:gmrhd}, with the new emphasis on its non-Lorentzian geometric aspect, we now apply the insights of duality asymmetry developed in Section~\ref{sec:da} to study how duality transformations act on these backgrounds. 
In this scenario, it is always the asymptotic geometry that becomes non-Lorentzian, while the bulk geometry remains Lorentzian, which constitutes an example of the DLCQ${}^0$/DLCQ${}^1$ correspondence as proposed in~\cite{Blair:2024aqz}. Therefore, it is the `asymptotic' duality asymmetry that we will focus on in this case. Later in Section~\ref{sec:nlbbf}, we will study examples with a non-Lorentzian bulk geometry, also in the presence of a $B$-field. 

\subsubsection{Asymptotic T-duality} 

\noindent $\bullet$~\emph{Self-duality of asymptotic M$p$T.} Now, we take a closer look at the near-horizon geometry~\eqref{eq:mr}, which asymptotes to the non-Lorentzian M$p$T geometry \eqref{eq:dpptffinal} with a constant $B$-field in the $x^{p+1}$ and $x^{p+2}$ directions. 
Our general results tell us that T-dualising these directions will lead to a dual M$p$T description asymptotically, as long as the $B$-field is not zero. Intriguingly, T-dualising $x^{p+1}$ and $x^{p+2}$ generates a particularly simple bulk geometry,
\begin{subequations} \label{eq:tdbg}
\begin{align}
    \dd s^2 &= \mathbb{\Omega} \, \Bigl( - \dd t^2 + \dd x_1^2 + \cdots + \dd x_p^2 \Bigr) + \frac{1}{\mathbb{\Omega}} \left( \frac{\dd x_{p+1}^2 + \dd x_{p+2}^2}{\CB^2} + \dd x_{p+3}^2 + \cdots + \dd x_{9}^2 \right)\!, \\[4pt]
    B^{(2)} &= - {\CB}^{-1} \, \dd x^{p+1} \!\! \wedge \! \dd x^{p+2},
        \quad%
    C^{(p+1)} = - \mathbb{\Omega}^2 \, \frac{\CB}{g^{}_\text{s}} \, \dd t \! \wedge \! \dd x^1 \!\! \wedge \! \cdots \! \wedge \! \dd x^p,
        \quad%
    e^\Phi = \mathbb{\Omega}^{\frac{p-3}{2}} \, \frac{g^{}_\text{s}}{\CB}\,, 
\end{align}
\end{subequations}
with the same data as given in \eqref{eq:ortlfmp}. In particular, recall that $\mathbb{\Omega} = (r/\ell)^{\frac{5-p}{2}}$\,. Note that $C^{(p+3)} = 0$ in this dual frame. 
In the asymptotic limit where $\mathbb{\Omega}$ tends to infinity, we are led to the non-Lorentzian M$p$T geometry described by
\begin{subequations} \label{eq:tebpmpt}
\begin{align}
    \tau^{0\,,\,\cdots,\,p} &= \bigl( \dd t\,, \, \dd x^1\,, \, \cdots, \, \dd x^p \bigr)\,, 
        &%
    e^\varphi &= \frac{g^{}_\text{s}}{\CB}\,, \\[4pt]
    E^{p+1\,, \, \cdots, \, 9} &= \left( \frac{\dd x^{p+1}}{\CB}\,, \, \frac{\dd x^{p+2}}{\CB}\,, \, \dd x^{p+3}, \, \cdots, \, \dd x^9 \right),   
        &%
    B^{(2)} &= - \CB^{-1} \, \dd x^{p+1} \! \wedge \! \dd x^{p+2}\,.
\end{align}
\end{subequations}
As expected, this asymptotic M$p$T geometry~\eqref{eq:tebpmpt} arises from T-dualising the $x^{p+1}$ and $x^{p+2}$ directions in the other asymptotic M$p$T geometry that we found earlier in \eqref{eq:dpptffinal}, which can be checked by using the T-duality rules derived in Section~\ref{sec:tda}. This is an example where an M$p$T geometry compactified on a transverse 2-torus is T-dualised to another M$p$T geometry, when the internal $B$-field on the 2-torus is nonzero. 
In this T-dual frame, the corresponding field theory is NCYM with the non-commutativity $[x^{p+1}, x^{p+2}] \sim \CB$\,. 

\vspace{3mm}

\noindent $\bullet$~\emph{From asymptotic M$p$T to asymptotic M($p$+2)T.} 
We have also learned in Section~\ref{sec:sltztda} that there exist T-duality transformations that map M$p$T to M$(p+2)$T in the asymptotic infinity. These are the $\mathrm{SL}(2,\mathbb{Z})$ transformations such that $\gamma \, \CB + \delta = 0$\,,
which can only happen when $\CB$ is rational. Taking this to be so, we now apply the T-duality transformation to the M$p$T geometry~\eqref{eq:dpptffinal} that arises from an asymptotic limit of the near-horizon geometry~\eqref{eq:mr}. We focus on the positive $\gamma$ branch here.  This dual geometry is 
\begin{align} \label{eq:dpptfdual}
	& \bigl({\tau}^0, \cdots, {\tau}^{\,p+2} \bigr) = \lr \dd t, \cdots, \dd x^p, \frac{\dd x^{p+1}}{\gamma},\frac{\dd x^{p+2}}{\gamma} \rr, 
		\qquad%
	\bigl( {E}^{p+1}\!,\cdots,{E}^9\bigr) = \bigl( \dd x^{p+1}\!,\cdots,\dd x^9 \bigr)\,, \notag \\[4pt]
    & \qquad e^{{\varphi}} = \frac{g^{}_\text{s}}{\gamma}\,,
        \qquad%
    {b}^{(2)} = \frac{\alpha}{\gamma} \, \dd x^{p+1} \!\!\wedge\! \dd x^{p+2}\,,
        \qquad%
	{c}^{\,(p+1)} = - \frac{\gamma}{g^{}_\text{s} \, \delta} \, \dd t \! \wedge \! \dd x^1 \!\wedge\! \cdots \!\wedge\! \dd x^{p}\,. 
\end{align}
Hence this is a geometry generated by an asymptotic M($p$+2)T limit. The field theory residing in the ($p$+3)-dimensional longitudinal sector is ordinary commutative Yang-Mills: 
although there is still a constant $B$-field in \eqref{eq:dpptfdual}, it is longitudinal and, according to the discussion of the Seiberg-Witten map in Section~\ref{sec:damt}, this does \emph{not} correspond to any non-commutativity.

We can connect the above T-dualities to the discussion in~\cite{Maldacena:1999mh}.
There it was noted that asymptotically the physical size of the $x^{p+1}$ and $x^{p+2}$ direction in the geometry~\eqref{eq:mr} (for $p=1$) shrinks in string units. 
If these are compact this leads to a singularity associated with winding modes becoming light.  
As noted there, T-dualising does not remove this singularity -- because the dual background \eqref{eq:tdbg} again shrinks in these directions asymptotically.
The exception is when the asymptotic $B$-field is rational, in which case the singularity can be removed by dualising to the geometry~\eqref{eq:dpptfdual} in M($p$+2)T.   
This has the side effect of going to a description with no non-commutativity on the field theory side.
All these translate elegantly into our M$p$T/M($p$+2)T framework and dovetails with the discussion in Section~\ref{sec:damt}. 

Moreover, for rational $B$-field, the singular behaviour before T-dualising now receives an interpretation that the ($p$+3)-dimensional field theory resides on a non-Lorentzian geometry described by 
\be
    \bigl(\tau^0\,, \cdots, \, \tau^{p}\bigr) = \bigl(\dd t\,, \, \dd x^1, \, \cdots, \, \dd x^p\bigr)\,,
        \qquad%
    \bigl(E^{p+1}\,,\, E^{p+2}\bigr) = \bigl(\dd x^{p+1}\,, \, \dd x^{p+2} \bigr)\,,
\ee
which is a ($p$+3)-dimensional non-Lorentzian submanifold of the ten-dimensional asymptotic M$p$T geometry~\eqref{eq:dpptffinal}. The new physical side of the $x^{p+1}$ and $x^{p+2}$ direction is kept finite. It is only when compared with the bulk relativistic geometry, where a metric is in use, that this submanifold of M$p$T appears to be singular. As expected from the T-dual relation, the original M$p$T frame is equally valid as the dual M($p$+2)T frame, and the relativistic description of the asymptotic field theory in M($p$+2)T is not necessarily privileged. 

\subsubsection{Asymptotic S-Duality: Bulk Geometry from \texorpdfstring{$T\bar{T}$}{TTbar}} \label{sec:asdbgttbar}

We now focus on the $p=1$ case and consider the S-dual of the background~\eqref{eq:mr}. Instead of S-dualising the geometry~\eqref{eq:mr} with $p=1$ as in~\cite{Maldacena:1999mh}, we focus on the S-dual of the simpler bulk geometry~\eqref{eq:tdbg} with $p=1$\,, which gives
\begin{align} \label{eq:mrsd}
	\dd s^2 &= \frac{\CB}{g_\text{s}} \biggl\{ \mathbb{\Omega}^2 \Bigl( - \dd t^2 + \dd x_1^2 \Bigr) + \Bigl[ \mathbb{\CB}^{-2} \, \bigl( \dd x_2^2 + \dd x_3^2 \bigr) + \dd x_4^2 + \cdots + \dd x^2_{9} \Bigr] \biggr\}\,,
        \qquad%
    e^\Phi = \mathbb{\Omega} \, \frac{\CB}{g^{}_\text{s}}\,, \notag \\[4pt]
	B^{(2)} &= - \mathbb{\Omega}^2 \, \frac{\CB}{g_\text{s}} \, \dd t \! \wedge \! \dd x^1\,, 
		\qquad%
    C^{(2)} = \frac{\dd x^2 \!\!\wedge\! \dd x^3}{\CB}\,,
        \qquad%
	C^{(4)} = \frac{\mathbb{\Omega}^2}{g^{}_\text{s}} \, \dd t \! \wedge \! \dd x^1 \!\wedge\! \dd x^2 \!\wedge\! \dd x^3\,. 
\end{align}
Asymptotically, the spacelike non-commutativity in NCYM is mapped to non-commutativity between space and time in the S-dual \emph{non-commutative open string theory} (NCOS)~\cite{Gopakumar:2000na}. 
In order for the associated NCOS to be well defined, the field strength $F_{01}$ has to acquire a vacuum expectation value $b$\,, which can be equivalently thought of as adding an $O(\mathbb{\Omega}^0)$ term to $B^{(2)}$ of \eqref{eq:mrsd}.
This shift follows from considering S-duality on the D3 worldvolume~\cite{Gopakumar:2000na}, where one works as usual with the gauge invariant sum of the gauge field strength and the $B$-field. Now, the Seiberg-Witten map implies that $[t\,,\,x^1] \, \propto \, b^{-1}$ in the S-dual NCOS, with the effective coupling $\sim g^{}_\text{s} \, \sqrt{b}$\,.  

Note that S-duality is the $\mathbb{Z}_2$ part of SL($2,\mathbb{Z}$) parametrised by $\alpha=\delta=0$ and $\beta=-\gamma=1$\,. According to Section~\ref{sec:sdmpt}, this falls in the $\gamma \neq 0$ case and M1T is mapped to non-relativistic string theory. The dual asymptotic non-Lorentzian geometry is encoded by
\begin{subequations} \label{eq:snc}
\begin{align} 
    \tau^{0\,,\,1} &= \sqrt{\frac{\CB}{g_\text{s}}} \, \bigl( \dd t\,, \, \dd x^1 \bigr)\,, 
        &%
    E^{2\,,\,3} &= \frac{\bigl( \dd x^2\,, \, \dd x^3 \bigr)}{\sqrt{g_\text{s} \, \CB}}\,,
        &%
    e^\varphi &= \frac{\CB}{g^{}_\text{s}}\,, \\[4pt]
        &&%
    E^{4\,,\,\cdots,\,9} &= \sqrt{\frac{\CB}{g_\text{s}}} \bigl( \dd x^2\,, \, \dd x^9 \bigr)\,, 
        &%
    C^{(2)} &= \frac{ \dd x^2 \!\wedge\dd x^3}{\CB}\,.
\end{align}
\end{subequations}
This non-Lorentzian geometry~\eqref{eq:snc} arises from the $\mathbb{\Omega} \rightarrow \infty$ limit of the bulk geometry~\eqref{eq:mrsd}, which fits the general non-relativistic string limiting prescription~\cite{Andringa:2012uz, Bergshoeff:2019pij, Ebert:2021mfu},
\begin{subequations}
\begin{align}
    \dd s^2 &= \omega^2 \bigl( - \tau^0 \, \tau^0 \! + \tau^1 \, \tau^1 \bigr) + E^2 E^2 + \cdots + E^9 E^9\,, 
        &%
    B^{(2)} &= - \omega^2 \, \tau^0 \!\! \wedge \! \tau^1 + b^{(2)}\,, \\[4pt]
    e^\Phi &= \omega \, e^{\varphi}\,, 
        &%
    C^{(q)} &= \omega^2 \, \tau^0 \!\! \wedge \! \tau^1 \!\! \wedge \! c^{(q-2)} + c^{(q)}\,,
\end{align}
\end{subequations}
after replacing $\omega$ with $\mathbb{\Omega}$\,.\footnote{Note that while the RR potentials of \eqref{eq:mrsd} are non-zero, and have the correct form for this limit, they give rise to vanishing fluxes: as follows from the fact that the RR five-form field strength in our conventions is $F^{(5)} = \dd C^{(4)} + H^{(3)} \wedge C^{(2)}$, where $H^{(3)} = \dd B^{(2)}$.} Here, the $\omega^2$ divergence in $C^{(q)}$ is induced by the criticality of the $B$-field~\cite{Ebert:2021mfu}. See \emph{e.g.}~\cite{Andringa:2012uz, Harmark:2017rpg, Kluson:2018egd, Bergshoeff:2018yvt, Harmark:2018cdl,  Bergshoeff:2019pij, Harmark:2019upf, Bidussi:2021ujm} for studies of general non-Lorentzian target space geometry in non-relativistic string theory.

Consider a probe string in this S-dual holographic setup. Asymptotically where $\mathbb{\Omega} \rightarrow \infty$\,, we are led to the Gomis-Ooguri action~\cite{Gomis:2000bd} describing the non-relativistic string,
\be \label{eq:go}
    S = - \frac{1}{4\pi\alpha'} \int \dd^2 \sigma \, \Bigl[ \p_\alpha X^2 \, \p^\alpha X^2 + \cdots + \p_\alpha X^9 \, \p^\alpha X^9 + P \, \bar{\p} \bigl( X^0 + X^1 \bigr) + \bar{P} \, \p \bigl( X^0 - X^1 \bigr) \Bigr]\,,
\ee
where $X^\text{M} (\sigma)$ are the embedding coordinates mapping the worldsheet with coordinates $\sigma^\alpha$, $\alpha = 0\,,\,1$ to the ten-dimensional target space. Moreover, $P$ and $\bar{P}$ are one-form worldsheet fields playing the role of Lagrange multipliers; they are the momenta conjugate to $X^0 + X^1$ and $X^0 - X^1$, respectively. We also defined the worldsheet light-cone derivatives $\p = \p_0 + \p_1$ and $\bar{\p} = \p_0 - \p_1$\,. 
Now, for this worldsheet action there exists a marginal deformation
\be \label{eq:ccd}
    \frac{1}{4\pi\alpha'} \int \dd^2 \sigma \, U(X) \, P \bar{P}\,, 
\ee
where $U(X)$ is a background field. 
Integrating out the non-dynamical fields $P$ and $\bar{P}$ in the deformed action leads us back to the full type IIB superstring theory. 
Undoing the conformal gauge in \eqref{eq:go} followed by integrating out the worldsheet zweibein field gives rise to the Nambu-Goto formulation of the non-relativistic string~\cite{Bergshoeff:2018yvt}, in which case the current-current deformation~\eqref{eq:ccd} maps to the $T\bar{T}$-deformation.  
This identification with a $T \bar T$-deformation was made in~\cite{Blair:2020ops} and elaborated on in~\cite{Blair:2024aqz}.

The deformation~\eqref{eq:ccd} is sometimes known as the torsional deformation~\cite{Yan:2021lbe}, as it is generically generated from quantum corrections unless a sufficiently strong symmetry principle is invoked that constrains the torsion of the longitudinal vielbein fields $\tau^{0,1}$~\cite{Gomis:2019zyu, Gallegos:2019icg, Yan:2019xsf, Yan:2021lbe}. In the context of backgrounds sourced by strings and branes, it has been argued recently in~\cite{Avila:2023aey, Guijosa:2023qym, Guijosa:2025mwh} that it is necessary to allow such deformations.

Now let us return to the study of the background~\eqref{eq:mrsd}.
Starting with non-relativistic string theory in the asymptotic non-Lorentzian geometry \eqref{eq:snc}, with action \eqref{eq:go}, choosing $U(X) = \mathbb{\Omega}^{-2}$ deforms the asymptotic flat non-Lorentzian geometry to be the bulk fundamental string geometry~\eqref{eq:mrsd}. 
In this manner, and as discussed more generally in~\cite{Blair:2024aqz}, the $T\bar{T}$-deformation generates the Lorentzian near-horizon geometry from the asymptotic non-Lorentzian geometry, which is a recurring phenomenon that we have seen through this section. 
Here we have used dualities to obtain the background \eqref{eq:mrsd}
which asymptotes to a non-relativistic string theory background, in which case the deformation can really be viewed as the original $T \bar T$-deformation. 
More generally, one can view the deformation which generates the near-horizon limit from other non-Lorentzian geometries as particular (higher-dimensional) generalisations of $T \bar T$, as explained in~\cite{Blair:2024aqz}. Such a generalised $T\bar{T}$ deformation essentially undoes the associated BPS decoupling limit.  

Note that the second quantisation of non-relativistic string theory leads to \emph{matrix string theory}~\cite{Motl:1997th, Dijkgraaf:1997vv}, which is described by a two-dimensional $\CN = 8$ SYM. See~\cite{Blair:2024aqz, Harmark:2025ikv} for more recent studies. In particular, \cite{Harmark:2025ikv} contains further studies of the holographic correspondence between the large $N$ limit of matrix string theory and gravitational string solitons in non-relativistic string theory.  

\subsection{Non-Lorentzian Holographic Duals} \label{sec:nlbbf}

So far, we have only been concerned with known holographic duals to non-commutative Yang-Mills, for which we have highlighted the appearance of asymptotic non-Lorentzian geometry. 
In the language of~\cite{Blair:2024aqz}, these duals are of the DLCQ${}^0$/DLCQ${}^1$ type, \emph{i.e.}~they pair a relativistic bulk geometry in the DLCQ${}^0$ orbit dual to a field theory in the DLCQ${}^1$ orbit. In this subsection, we show evidence for new holographic duals with a non-Lorentzian curved geometry in the bulk, which are of the DLCQ${}^1$/DLCQ${}^2$ type. 

\subsubsection{Example: A Bulk 3-Brane Decoupling Limit}

For concreteness, we start with the $p=1$ case associated with the D1-D3 brane configuration. We consider general bulk M$p$T limits of the D1-D3 background~\eqref{eq:dppt}, such that we obtain not a near-horizon limit but a genuine non-Lorentzian curved geometry.

In~\cite{Blair:2024aqz}, we discussed possible bulk M$p$T limits of individual D$q$-brane geometries. We can characterise these limits in terms of the $n$ M$p$T longitudinal directions that are transverse to the D$q$-brane, and the $m$ M$p$T transverse directions that are longitudinal to the D$q$-brane.
Requiring that the RR $(q+1)$-form appearing in the brane geometry was finite led to the condition $m+n=4$. In particular, we showed that one could take a bulk M1T limit of the D3-brane with $(m\,,n)=(3\,,1)$\,, a bulk M3T limit of the D1-brane with $(m\,,n)=(1\,,3)$ and a bulk M3T-limit of the D3-brane with $(m\,,n)=(2\,,2)$\,.

This suggests that the D1-D3 solution obtained from setting $p=1$ in \eqref{eq:dppt} may admit a further M3T limit in the bulk. 
In this limit, we want the $B$-field to be finite. Assuming that $\Theta$ is fixed, this requires one of the directions $(X^2,X^3)$ to be longitudinal and the other to be transverse.
This fixes the allowed limit to have the following form: 
\begin{center}
\begin{tabular}{c||cc|cc|cccccc}
& 0 & 1 & 2 & 3 & 4 & 5 & 6 & 7 & 8 & 9 \\
\hline\hline
D1-brane& $\times$ & $\times$ & -- & -- & -- & -- & -- & -- & -- & -- \\
D3-brane& $\times$ & $\times$ & $\times$ & $\times$ & -- & -- & -- & -- & -- & -- \\
bulk M$3$T limit& $\times$ & -- & $\times$ & -- & $\times$  & $\times$ & -- & -- & -- & -- \\
\end{tabular} 
\end{center}
This M3T limit then corresponds to letting
\begin{align} \label{M3TforD1D3}
    \begin{array}{c}
        \!\!\!\!\bigl( X^0, X^2, X^4, X^5 \bigr) = \omega^\frac{1}{2} \, \bigl( t, x^2, x^4, x^5 \bigr), \\[6pt]
        \,\,\,\bigl(X^1,X^3,X^6,\cdots,X^9\bigr) = \omega^{-\frac{1}{2}} \, \bigl( x^1,x^3,x^6,\cdots,x^9\bigr),
    \end{array}
        \qquad%
    G^{}_\text{s} = g^{}_\text{s} \,.
\end{align}
In order to obtain a well-defined M3T geometry that admits a further near-horizon limit, the harmonic function $H$ in \eqref{eq:dppt} must be finite under the bulk M3T limit. There are two different possibilities:
\begin{enumerate}[(1)]

\item

\emph{Smearing.} We smear in the $X^{4,\,5}$ directions, which are taken to be compactified over circles of radii $R_{4,\,5}$\,, respectively. After this smearing, $H$ becomes 
\be \label{eq:smhf}
    H = 1 +  \frac{L^4}{4 \pi \, R_4 \, R_5 \, \bigl( X^2_6 + \cdots + X^2_9 \bigr)}\,.
\ee
Substituting $R^{}_{4,\,5} = \sqrt{\omega} \, r^{}_{4,\,5}$ and $X^{6,\,\cdots,\,9} = x^{6,\,\cdots,\,9} / \sqrt{\omega}$\,, it follows that $H$ in \eqref{eq:smhf} is manifestly finite in $\omega$\,, with
\be \label{eq:shf}
    \textbf{smearing:} 
        \qquad%
    H = 1 + \frac{\ell^2}{x^2_6 + \cdots + x^2_9}\,, 
        \qquad%
    \ell^2 = \frac{\ell^4_\text{s} \, N \, g^{}_\text{s}}{r_4 \, r_5 \, \cos \Theta}\,.
\ee 
This gives a background localised in the M3T transverse directions $(x^6,\, \cdots,\,x^9)$\,. 

\item

\emph{Large $N\!$ rescaling.} Alternatively, similarly to examples considered in \cite{Lambert:2024uue, Fontanella:2024rvn, Fontanella:2024kyl, Blair:2024aqz}, we introduce an additional formal rescaling of the parameter $N$ as $N \rightarrow \omega^2 \, N$, such that 
\be \label{eq:fNScaled}
    \textbf{large $N$:} 
        \qquad%
    H = 1 +  \frac{\ell^4}{(x^2_4 + x^2_5)^2} \,,
        \qquad%
    \ell^4 = \frac{4 \pi \ell_\text{s}^4 \, N \, g^{}_\text{s}}{\cos \Theta} \,.
\ee
This gives a background localised in the M3T transverse directions $(x^4,\,x^5)$\,. One caveat to note here is that this mechanism of generating supergravity solutions by introducing an additional rescaling of $N$ is \emph{not} part of our intrinsic definition of an M$p$T limit.
The physical significance of this additional scaling, and in particular its BPS nature, is not entirely clear.

\end{enumerate}
In either case, we find that the bulk geometry~\eqref{eq:dppt} becomes,
\begin{align} \label{eq:mttrp}
    \dd s^2 &= \omega \, \Bigl( - \tau^0 \, \tau^0 + \tau^2 \, \tau^2 + \tau^4 \, \tau^4 + \tau^5 \, \tau^5 \Bigr) + \frac{1}{\omega} \, \Bigl( E^1 \, E^1 + E^3 \, E^3 + E^6 \, E^6 + \cdots + E^9 \, E^9 \Bigr)\,, \notag \\[4pt]
    B^{(2)} &= b^{(2)}\,, 
        \qquad%
    C^{(2)} = c^{(2)}\,,
        \qquad%
    C^{(4)} = \frac{\omega^2}{e^\varphi} \, \tau^0 \!\!\wedge\! \tau^2 \!\!\wedge\! \tau^4 \!\!\wedge\! \tau^5 + c^{(4)}\,, 
        \qquad%
    e^\Phi = e^\varphi\,,
\end{align}
which is the M3T limiting prescription with the M3T geometric data
\begin{subequations} \label{eq:D1D3_M3Tgeo}
\begin{align}
    \bigl( \tau^0,\,E^1 \bigr) = \frac{(\dd t,\,\dd x^1)}{H^{\frac{1}{4}}} \,,
        \qquad%
    \bigl( \tau^2,\,E^3 \bigr) = \frac{H^{\frac{1}{4}} \, (\dd x^2,\,\dd x^3)}{\sqrt{\sin^2 \Theta+ H^{-1} \cos^2 \Theta}}\,, \\[-30pt] 
    \notag
\end{align}
\begin{align}
    \bigl(\tau^4,\,\tau^5,\,E^6,\,\cdots,\,E^9\bigr) = H^{\frac{1}{4}} \, \bigl(\dd x^4,\,\cdots,\,\dd x^9\bigr),
\end{align}
together with
\begin{align}
    b^{(2)} & = \frac{\tan \Theta \, \dd x^2 \!\! \wedge \! \dd x^3}{\sin^2 \Theta+ H \cos^2 \Theta}\,,
        &%
    c^{(2)} &= \frac{\sin \Theta}{g^{}_\text{s} \, H} \, \dd t \!\wedge\! \dd x^1\,, \\[4pt]
    e^\varphi &=  \frac{g^{}_\text{s}}{\sqrt{\cos^2\Theta+ H^{-1} \sin^2 \Theta}}\,,
        &%
    c^{(4)} &= \frac{g_\text{s}^{-1} \cos \Theta}{\sin^2\Theta + H \cos^2 \Theta} \, \dd t \!\wedge\! \dd x^1 \!\!\wedge\! \dd x^2 \!\!\wedge\! \dd x^3\,.
\end{align}
\end{subequations} 
Here, $H$ is either given by \eqref{eq:shf} or \eqref{eq:fNScaled}, depending on whether we follow the smearing or large $N$ rescaling procedure. 
Note that we have shifted $C^{(4)}$ from the original geometry~\eqref{eq:dppt} by an $\omega^2$ term in \eqref{eq:mttrp} that takes the form 
\be \label{eq:flles}
    \frac{\omega^2}{e^\varphi} \, \tau^0 \!\!\wedge\! \tau^2 \!\!\wedge\! \tau^4 \!\!\wedge\! \tau^5 = g^{-1}_\text{s} \, \dd t \! \wedge \! \dd x^2 \!\! \wedge \! \dd x^4 \!\! \wedge \! \dd x^5\,,
\ee
which is closed, meaning that this required critical term can be added to the original background without modifying the fact that it is a solution to IIB supergravity. In the $\omega \rightarrow \infty$ limit, we are led to the curved non-Lorentzian M3T geometry encoded by \eqref{eq:D1D3_M3Tgeo}.

The geometry~\eqref{eq:D1D3_M3Tgeo} allows for a further asymptotic M1T limit of the form \eqref{eq:bpsd} to be taken.
This leads to a `near-horizon' geometry of the M3T type, which should be dual to a non-relativistic limit of NCYM (induced by the M3T limit in ten dimensions).
Using the terminology we introduced in~\cite{Blair:2024aqz}, this is an example of a DLCQ${}^1$/DLCQ{}$^2$ correspondence.

To describe this further asymptotic M1T limit that corresponds to a near-horizon limit of the bulk M3T geometry, we use the same parametrisation as \eqref{eq:bpsd} but in terms of a new parameter $\tilde \omega$ that replaces $\omega$\,, such that, asymptotically,
\begin{subequations}
\begin{align}
    \bigl(t,\,x^1\bigr) &= \sqrt{\tilde{\omega}} \, \bigl( \mathbb{t}\,,\,\mathbb{x}^1\bigr)\,, 
        &%
    & \bigl( x^2,\,x^3 \bigr) = \frac{\bigl( \mathbb{x}^2,\,\mathbb{x}^3 \bigr)}{\sqrt{\tilde{\omega} + \tilde{\omega}^{-1} \, \CB^{-2}}}\,, \\[4pt]
    \bigl(x^4,\,\cdots,\,x^9\bigr) &= \frac{1}{\sqrt{\tilde{\omega}}} \, \bigl( \mathbb{x}^4,\,\cdots,\,\mathbb{x}^9\bigr)\,,
        &%
     & g^{}_\text{s} = \frac{\mathbb{g}^{}_\text{s}}{\sqrt{\tilde{\omega}^2 + \CB^{-2}}}\,,
        \qquad%
    \Theta = \arctan \bigl(\tilde{\omega} \, \CB \bigr)\,.
\end{align}
\end{subequations}
It then follows that $H \sim \bigl( \tilde{\omega} / \tilde{\mathbb{\Omega}} \bigr)^2$, with
\begin{align}
    \tilde{\mathbb{\Omega}} =
    \begin{cases}
         \tilde{\ell}^{-1} \, \sqrt{\mathbb{x}^2_6 + \cdots + \mathbb{x}^2_9}\,, 
            \qquad%
        \tilde{\ell}^{\,2} = 4\pi \, \ell^4_\text{s} \, N \, \mathbb{g}^{}_\text{s} \, \CB\, \bigl(r^{}_4 \, r^{}_5\bigr)^{-1}
            \qquad%
        \text{(smearing)} \\[6pt]
        \tilde{\ell}^{-2} \, \bigl( x^2_4 + x^2_5 \bigr)\,,
            \qquad\qquad\,%
        \tilde{\ell}^{\,4} = 4\pi \, \ell^4_\text{s} \, N \, \mathbb{g}^{}_\text{s} \, \CB
            \qquad\qquad\qquad\,\,%
        \text{(large $N$ rescaling)}
    \end{cases}
\end{align}
Finally, the $\tilde{\omega} \rightarrow \infty$ limit of \eqref{eq:D1D3_M3Tgeo} gives the following bulk M3T geometry:
\begin{subequations} \label{eq:D1D3_M3Tgeomtt}
\begin{align}
    \bigl(\tau^0,\,E^1\bigr) &= \tilde{\mathbb{\Omega}}^{\frac{1}{2}} \bigl(\dd \mathbb{t}\,,\,\dd \mathbb{x}^1\bigr)\,, 
        &%
    \bigl(\tau^4,\,&\tau^5,\,E^6,\,\cdots,\,E^9\bigr) = \tilde{\mathbb{\Omega}}^{-\frac{1}{2}} \, \bigl(\dd \mathbb{x}^4,\,\cdots,\,\dd \mathbb{x}^9\bigr)\,, \\[4pt]
    \bigl(\tau^2,\,E^3\bigr) &= \frac{(\dd \mathbb{x}^2,\,\dd \mathbb{x}^3)}{\sqrt{\tilde{\mathbb{\Omega}} + \tilde{\mathbb{\Omega}}^{-1} \CB^{-2}}}\,, 
        &%
    c^{(2)} &= \frac{\tilde{\mathbb{\Omega}}}{\mathbb{g}^{}_\text{s}} \, \dd \mathbb{t} \!\wedge\! \dd \mathbb{x}^1\,, 
        \qquad%
    e^\varphi = \frac{\mathbb{g}^{}_\text{s}}{\sqrt{\tilde{\mathbb{\Omega}}^2 + \CB^{-2}}}\,, \\[4pt]
    b^{(2)} &= \frac{\CB \, \dd \mathbb{x}^2 \!\! \wedge \! \dd \mathbb{x}^3}{1 + (\tilde{\mathbb{\Omega}} \, \CB)^{-2}}\,,
        &%
    c^{(4)} &= \frac{\mathbb{g}_\text{s}^{-1}}{\tilde{\mathbb{\Omega}} + \tilde{\mathbb{\Omega}}^{-1} \CB^{-2}} \, \dd \mathbb{t} \!\wedge\! \dd \mathbb{x}^1 \!\!\wedge\! \dd \mathbb{x}^2 \!\!\wedge\! \dd \mathbb{x}^3\,.
\end{align}
\end{subequations}
The dual field theory conjecturally arises from an M3T limit of four-dimensional NCYM. Compactifying the $\mathbb{x}^2$ and $\mathbb{x}^3$ directions leads to an anisotropic torus, and the SL($2,\mathbb{Z}$) T-duality associated with this torus exhibits the asymmetric structure that we have commented on at the end of Section~\ref{sec:tdampt}. 

\subsubsection{General Brane Geometries}

The above construction for the D1-D3 configuration can be generalised to the D$p$-D($p$+2) case, which has to take the following form (under the assumption that $\Theta$ is fixed):
\begin{center}
\begin{tabular}{c||c|c|c|c|c|c}
$\sharp$ of dim. & $p-m+1$ (incl.~$t$) & $m$ & 1 & 1 & $n$ & $7-p-n$ \\
\hline\hline
D$p$-brane& $\times$ & $\times$ & -- & -- & -- & -- \\
D($p$+2)-brane& $\times$ & $\times$ & $\times$ & $\times$ & -- & -- \\
bulk M$q$T limit& $\times$ & -- & $\times$ & -- & $\times$  & -- 
\end{tabular} 
\end{center}
where we consider a bulk M$q$T limit with $q = p + n - m + 1$\,. Another constraint is that $m + n = 3$\,, such that the $\omega^2$-divergence in the RR potential $C^{(q+1)}$ required by the M$q$T limit is closed, which generalises \eqref{eq:flles}. Note that $n \leq 3$ and $p + n \geq 3$\,. Moreover, the harmonic function
\be
    H = 1 + \frac{\omega^{\frac{q-3}{2}} \, L^{5-p}}{\bigl( \omega \, x^u \, x^u + \omega^{-1} \, x^i \, x^i \bigr)^{\!\frac{5-p}{2}}}\,,
        \qquad%
    L^{5-p} = \frac{g^{}_\text{s} \, N}{\cos \Theta} \frac{(2\pi L^{}_\text{s})^{5-p}}{(5-p) \, \omega^{6-p}}\,.
\ee
has to be finite such that a near-horizon limit can be taken. There are three possibilities:
\begin{align}
    H = 
    \begin{cases}
        1 + \ell^{5-p}_{} \, \bigl(x^i \, x^i\bigr)^{\!\frac{p-5}{2}}\,,
            \hspace{1.5cm}%
        n = 0\,, \\[4pt]
        1 + \ell^{5-p-n}_{} \, \bigl(x^i \, x^i\bigr)^{\!\frac{p+n-5}{2}}\,,
            \qquad%
        n > 0\,, \text{\qquad smearing in } x^u\,, \\[4pt]
        1 + \ell^{5-p}_{} \, \bigl( x^u \, x^u \bigr)^{\!\frac{p-5}{2}}\,, 
            \hspace{1.45cm}%
        n > 0\,, \text{\qquad} N \rightarrow \omega^\frac{8-p-q}{2} \, N\,.
    \end{cases}
\end{align}
In the case where $p=1$ and $n = 2$\,, we recover the D1-D3 configuration considered earlier in this subsection. More generally, we obtain a series of other non-Lorentzian geometries that are dual to various field theories that arise from further BPS decoupling limits of NCYM in various dimensions. 

\section{Conclusions} \label{sec:concl}

In this paper, we have initiated the systematic study of the phenomenon of duality asymmetry between non-Lorentzian geometries coupled to matrix theories.
Working within the setting of Matrix $p$-brane Theory, we saw how in the simplest cases the presence or absence of a $B$-field dictated the nature of the T-dual theory. For more general $\mathrm{SL}(2,\mathbb{Z})$ duality transformations, we saw that both S- and T-duality asymmetry share similar structures and involve intricate algebraic features linked to invariant theory, building on \cite{Bergshoeff:2023ogz}.
We also revisited the holographic duals of non-commutative Yang-Mills, showing how to interpret these geometries in the holographic framework we introduced in~\cite{Blair:2024aqz}, and demonstrating that by iterating the decoupling limit we can obtain new non-Lorentzian holographic duals.
We hope that this work helps establish a modern geometric foundation for the study of matrix theories and non-commutativity.

The paper \cite{Bergshoeff:2023ogz} proposed a  program of `non-Lorentzian bootstrap', in which bosonic non-Lorentzian symmetries are bootstrapped to constrain Lorentzian supergravity by undoing the BPS decoupling limit.
In this program, novel duality asymmetries may help us determine higher-order terms in Lorentzian supergravity efficiently.

The new non-Lorentzian bulk geometries of Section~\ref{sec:nlbbf} should be dual to non-relativistic limits of non-commutative Yang-Mills. These limits can be engineered, in principle, using the knowledge of the underlying bulk limit and its overlap with the brane on which the field theory lives.
In previous (commutative) examples~\cite{Lambert:2024uue, Lambert:2024yjk, Lambert:2024ncn} it has been found that the limit of the field theory localises the dynamics onto a moduli space of a field theory BPS configuration. A natural conjecture would be that the field theories dual to our non-Lorentzian geometries should relate to non-commutative instantons~\cite{Nekrasov:1998ss, Seiberg:1999vs}. 

It would be interesting to apply the techniques developed in this paper to matrix theories on more general compactifications.
One could speculate about the possible appearance of (mirror) duality asymmetry for general Calabi-Yau compactifications of non-Lorentzian geometries.
It would also be interesting to explore generalisations to corners related to heterotic string theory~\cite{Bergshoeff:2023fcf, hete,Lescano:2024url}.

We finish with some remarks about U-duality asymmetry in M-theory.
The U-duality group of M-theory on a 3-torus is $\text{SL}(3,\mathbb{Z}) \times \text{SL}(2,\mathbb{Z})$.
In non-relativistic M-theory this should provide a unified framework for the discussions in Sections~\ref{sec:sda} and~\ref{sec:sltztda} on SL($2,\mathbb{Z}$) S- and T-duality asymmetries.
Note for instance that the T-duality asymmetry between M0T and M2T can be lifted to eleven dimensions, where it is interpreted as a U-duality between M-theory in the DLCQ and non-relativistic M-theory, with the T-duality $\mathrm{SL}(2,\mathbb{Z})$ being the non-geometric $\mathrm{SL}(2,\mathbb{Z})$ factor of the U-duality group, and the $B$-field on the torus lifting to the M-theory 3-form on a 3-torus.

Similarly, one can consider the U-duality between non-relativistic M-theory on a transverse 3-torus and its magnetic dual on a longitudinal 3-torus. This magnetic dual is based on an M5-brane decoupling limit (see \emph{e.g.}~\cite{Blair:2023noj}), for which an associated supergravity was recently constructed in~\cite{Bergshoeff:2025grj}. U-duality between these limits uplifts the T-duality asymmetry between M2T and M4T: non-relativistic M-theory is either self U-dual or dualises to its magnetic dual depending on the value of the three-form on the transverse torus.

\acknowledgments

We would like to thank Ritankar Chatterjee, Jorge Russo and Matthew Yu for useful discussions. 
The work of C.B. is supported through the grants CEX2020-001007-S and PID2021-123017NB-I00, funded by MCIN/AEI/10.13039/501100011033 and by ERDF A way of making Europe. The work of N.O. is supported in part by VR project Grant 2021-04013 and Villum Foundation Experiment Project No.~00050317.  
The work of Z.Y. is
supported in part by Olle Engkvists Stiftelse Project Grant 234-0342, VR Project Grant 2021-04013, and the European Union’s Horizon 2020 research and innovation programme
under the Marie Sk{\l}odowska-Curie Grant Agreement No.~31003710.  
Nordita is supported in part by NordForsk. 
The Center of Gravity is a Center of Excellence funded by the Danish National Research Foundation under grant No.~184.

\appendix

\section{T-duality of Lorentzian Background Fields}
\label{sec:ReviewOdd} 

\noindent $\bullet$ \emph{$O(d,d)$ T-duality.} Here we summarise our conventions for T-duality of the NSNS and RR background fields.
Write the ten-dimensional coordinates as $x^{\text{M}}=(x^\mu,x^i)$, where the $x^i$, $i=1,\dots,d$, are adapted coordinates for $d$ isometry directions. 
Decompose the 10-dimensional string frame metric as
\be
\label{GforT}
    G_{\text{M} \text{N}} = 
    \begin{pmatrix}
        g^{}_{\mu\nu} + G^{}_{ij} \, A^{}_\mu{}^i \, A^{}_\nu{}^j &\quad G^{}_{j\ell} \, A^{}_\mu{}^\ell \\[4pt]
        G^{}_{ik} \, A^{}_\nu{}^k &\quad G_{ij}
    \end{pmatrix} \,,
\ee
and let $B_{\mu\nu}$, $B_{\mu i}$ and $B_{ij}$ denote the components of the $B$-field.
Then in the NSNS sector, the T-duality multiplets with non-trivial transformation are 
\be
    \CH = 
    \begin{pmatrix}
        G^{}_{ij} - B^{}_{ik} \, G^{k\ell} \, B^{}_{\ell j} &\quad B^{}_{ik} \, G^{kj} \\[4pt]
        - G^{ik} \, B^{}_{kj} &\quad G^{ij}
    \end{pmatrix},
        \qquad%
    \CA_\mu = 
    \begin{pmatrix}
        A^{}_\mu{}^i \\[4pt]
        B^{}_{\mu i} - A^{}_\mu{}^j \, B^{}_{ji}
    \end{pmatrix},
\label{defHA}
\ee
while
$g^{}_{\mu\nu}$\,,
$B_{\mu\nu} + A_{[\mu}{}^i \, B_{\nu]i}$\,, 
and
$e^{-2\Phi} \sqrt{\det \bigl( G_{ij} \bigr)}$
are $O(d, d)$ singlets.
For $\boldsymbol{\Lambda}$ a generic $O(d,d)$ transformation we write,  
\be
\boldsymbol{\Lambda} = \begin{pmatrix}
\om &\,\, \on  \\[4pt]
\op &\,\, \oq
\end{pmatrix},
    \qquad%
\boldsymbol{\Lambda}\, \eta \, \boldsymbol{\Lambda}^T = \eta \,,
    \qquad%
\eta = \begin{pmatrix} 0 &\,\, \mathbb{1} \\
\mathbb{1} &\,\, 0 \end{pmatrix},
\label{defineLambdaOdd}
\ee
where the $d \times d$ matrices $\om,\on,\op,\oq$ obey 
\be
    \om \, \oq^T + \on \, \op^T = \mathbb{1}\,,
        \qquad%
    \om \, \on^T + \on \, \om^T = \op \, \oq^T + \oq \, \op^T = 0\,.
\ee
The field combinations of \eqref{defHA} transform as $\cH \rightarrow \Lambda \, \cH \, \Lambda^\intercal$ and $\mathcal{A} \rightarrow \Lambda^{-\intercal} \mathcal{A}$\,.
A Buscher T-duality transformation on all $d$ directions simultaneously corresponds to $\om=\oq=0$ and $\op=\on=\mathbb{1}$\,,
which has the effect of swapping  upper and lower $d$-dimensional indices in $\cH$ and $\mathcal{A}_\mu$\,. 
For the field components appearing in \eqref{defHA}, this results in
\be
    \tilde{G} + \tilde{B} = \bigl( G + B \bigr)^{-1}\,,
        \qquad%
    \tilde{A}_{\mu}{}^i = B_{\mu i} - A_\mu{}^j B_{ji} \,, 
        \qquad%
    \tilde{B}^{}_{\mu i} - \tilde{A}^{}_\mu{}^j \, \tilde{B}^{}_{ji} =A_\mu{}^i\,,
\label{compactBuscherd} 
\ee
From these expressions we determine the transformed fields explicitly.
Define
\be
    \mathscr{G}^{ij} \equiv \Bigl[ \bigl( G- B \, G^{-1} \, B)^{-1} \Bigr]^{ij}\,.
\label{mathscrGinv}
\ee
Note that `$G$' and `$B$' in the above expressions \eqref{compactBuscherd} and \eqref{mathscrGinv} have their indices being $i$ on the internal $d$-torus. 
It then follows that the full T-dual metric and dilaton are
\begin{subequations} \label{eq:tgrl}
\begin{align}
\begin{split}
    \tilde G_{\mu\nu} &= G_{\mu\nu} - G_{ij} \, A_\mu{}^i \, A_\nu{}^j + \bigl( B_{\mu i} - A_\mu{}^k \, B_{ki} \bigr) \, \mathscr{G}^{ij} \, \bigl( B_{\nu j} - A_\nu{}^l B_{lj} \bigr) \,, 
        \qquad%
    \tilde G_{ij} = \mathscr{G}^{ij} \,, \\[4pt]
    \tilde G_{\mu i} & = \mathscr{G}^{ij} \, \bigl( B_{\mu j} - A_\mu{}^k \, B_{kj} \bigr) \,,
        \hspace{3.1cm}%
    \tilde{\Phi} = \Phi - \frac{1}{4} \ln \Bigl[ \det \bigl( G_{ij} \bigr) \det (\mathscr{G}_{ij}) \Bigr]\,,
\end{split}
\end{align} 
and the T-dual $B$-field is
\begin{align} \label{eq:tbrl} 
\tilde B_{ij} & = -G^{ik} \, B_{kl} \, \mathscr{G}^{lj} \,,
    \hspace{2.85cm}%
\tilde B_{\mu i}  = A_\mu{}^i + G^{ik} \, B_{kl} \, \mathscr{G}^{lj} \bigl( B_{\mu j} + B_{jm} \, A_\mu{}^m \bigr) \,, \\[4pt]
\tilde B_{\mu\nu} & = B_{\mu\nu} + 2 \, A_{[\mu}{}^i B_{\nu]i} + A_{\mu}{}^i \, B_{ij} \, A_\nu{}^j - \bigl( B_{\mu i} + B_{ik} \, A_\mu{}^k\bigr) \, G^{il} \, B_{lm} \, 
\mathscr{G}^{mj} \, \bigl( B_{\nu j} + B_{jl} \, A_\nu{}^l\bigr)\,, \notag
\end{align} 
\end{subequations}
When $d=1$\,, the above transformations recover the usual Buscher rules~\cite{Buscher:1987qj, Buscher:1987sk}. \emph{E.g.} letting $y$ denote the isometry direction, we have $\tilde G_{yy} = G^{-1}_{yy}$\,.

The transformation of the RR sector is most naturally expressed in terms of the polyform 
\be
\CCpoly = e^{B^{(2)}} \!\! \wedge \sum_q C^{(q)} \,,
\label{CCpoly} 
\ee
which appears naturally in the Wess-Zumino part of the D-brane worldvolume action (our conventions are as in~\cite{Ebert:2021mfu}).
Splitting the coordinates as above, this decomposes as:
\be
\CCpoly = \sum_{k=0}^d \frac{1}{k!} \, \mathcal{C}_{i_1\dots i_k} \, \dd x^{i_1} \!\! \wedge \dots \wedge \! \dd x^{i_k} \,,
\label{CCPoly_decompose} 
\ee
where each $\mathcal{C}_{i_1 \dots i_k}$ is a polyform in $10-d$ dimensions.
These polyforms transform together as spinors of the double cover of $O(d,d)$\,: in particular the polyforms $(\mathcal{C}\,, \, \mathcal{C}_{i_1 i_2}\,, \, \dots)$ and $(\mathcal{C}_i\,, \, \mathcal{C}_{i_1i_2i_3}\,, \,\dots)$ correspond to chiral spinors.
When $d=1$, again letting $y$ denote the isometry direction, one can write simply $\CCpoly = \mathcal{C} + \mathcal{C}_y \!\wedge\! \dd y$\,, 
and the Buscher transformation for RR fields takes the form: 
\be
\tilde{\mathcal{C}} = \mathcal{C}_y \,,
    \qquad%
\tilde{\mathcal{C}}_y = \mathcal{C} \,,
\label{RRPolyNicer1} 
\ee
which in terms of the original RR polyforms $C = \sum_q C^{(q)}$ implies the expressions
\begin{align}
\tilde C & = C_y + C \wedge B_y + \frac{C_y \wedge B_y \wedge G_y}{G_{yy}} \,,\quad
\tilde C_y  = C - \frac{C_y \wedge G_y}{G_{yy}} \,,
\label{RRpolyT1}
\end{align}
where we use the shorthand notation $B_y \equiv B_{\mu y} \, \dd x^\mu$ and $G_y \equiv G_{\mu y} \, \dd x^\mu$\,.
The Buscher transformation on $d$ directions follows most simply by repeated application of \eqref{RRPolyNicer1}.

\vspace{1em}
\noindent $\bullet$ \emph{$O(2,2)$ T-duality and $\mathrm{SL}(2)$.}
Here we give further details of the special case with $d=2$\,.
Given $V= (V^i, \tilde V_i)$ and $W= (W_i \,, \tilde W^i)$ transforming in the vector and covector representations of $O(2,2)$, consider the following change of basis: $\tilde V^i = \epsilon^{ij} \, \tilde V_j$\,, $\tilde W_i = \epsilon_{ij} \, \tilde W^j$, such that the natural pairing $V \cdot W$ is preserved.
Here, $\epsilon^{}_{12} = \epsilon^{12} = 1$\,, so $\epsilon^{ik} \, \epsilon_{jk} = \delta^i_j$\,.
After making this change of basis, the non-trivial $O(2,2)$ multiplets of \eqref{defHA} can be expressed as
\begin{align}
\label{genmetSL2SL2}
    \cH & = \frac{G_{ij}}{\sqrt{|G|}} \otimes \gM 
\,,
        &%
    \gM &\equiv 
    \frac{1}{\sqrt{|G|}}
    \begin{pmatrix}
        |G| + \CB^2 &\,\, \CB \\[4pt]
        \CB &\,\, 1
    \end{pmatrix},
        &%
    \mathcal{A}_\mu{}^i &  = 
    \begin{pmatrix}
        A_\mu{}^i \\[4pt]
        B_{\mu}{}^i - \CB \, A_\mu{}^i 
    \end{pmatrix},
\end{align}
where we let $B_{ij} = \CB \, \epsilon_{ij}$ and $B_\mu{}^i \equiv \epsilon^{ij} B_{\mu j}$\,.
Similarly, a transformation $\Lambda \in O(2,2)$ in this basis has the general form following from \eqref{defineLambdaOdd},
\be
\boldsymbol{\Lambda} = \begin{pmatrix}
\om_i{}^j & \quad \on_{ik} \, \epsilon^{jk} \\[4pt]
\epsilon_{ik} \, \op^{kj} &\quad \epsilon_{ik} \, \oq^k{}_l \, \epsilon^{jl} 
\end{pmatrix}.
\ee
Consider the subclass of transformations where $\om_i{}^j = \alpha \, \delta_i^j$, $\oq^i{}_j = \delta 
\, \delta^i_j$, $\on_{ij} = \beta \, \epsilon_{ij}$ and $\op^{ij} = - \gamma \, \epsilon^{ij}$.
Then we obtain a factorisation,
\be
\boldsymbol{\Lambda}  = \delta_i{}^j \otimes \Lambda \,,\quad
\Lambda \equiv \begin{pmatrix} 
    \alpha &\,\, \beta \\
\gamma &\,\, \delta 
\end{pmatrix},
        \qquad%
    \alpha \, \delta - \beta \, \gamma = 1 \,,
\label{hSL2}
\ee
which is an $\mathrm{SL}(2)$ subgroup of $\mathrm{O}(2,2)$ that acts solely on the second matrix $\CM$ in the tensor product in \eqref{genmetSL2SL2}, or equivalently in the usual manner on the complex scalar $\tau \equiv B + i |G|^{1/2}$\,.
We can explicitly write the transformation of the NSNS fields of \eqref{genmetSL2SL2} as $\gM \rightarrow \tilde{\gM} = \Lambda \, \gM \, \Lambda^\intercal_{\phantom{\dagger}}$ and $\mathcal{A}_\mu{}^i \rightarrow \tilde{\mathcal{A}}_\mu{}^i = \Lambda^{-\intercal}_{\phantom{\dagger}} \, \mathcal{A}_\mu{}^i$.

Another $\mathrm{SL}(2)$ subgroup follows from transformations with $\on = \op = 0$\,, and with $\om= \oq^{-\intercal}$ restricted to have unit determinant. This other $\mathrm{SL}(2)$ subgroup acts on the two-dimensional indices on the internal torus as volume preserving transformations.
This reveals that 
\be
    \mathrm{SO}(2,2) = \Bigl( \mathrm{SL}(2) \times \mathrm{SL}(2) \Bigr) / \, \mathbb{Z}_2\,,
\ee
The $\mathbb{Z}_2$ quotient reflects the fact that the transformation $\boldsymbol{\Lambda} = (- \delta_i{}^j) \otimes (-\mathbb{I}_2)$ acts trivially on $\mathrm{SO}(2,2)$ vectors.
Hence, $\mathrm{SL}(2) \times \mathrm{SL}(2)$ is the double cover of $\mathrm{SO}(2,2)$\,.
The full T-duality group $\mathrm{O}(2,2)$ also includes a $\mathbb{Z}_2$ transformation which exchanges the $\mathrm{SL}(2)$ factors. This realises the Buscher transformation on a circle.
Note that the usual Buscher transformation on both directions corresponds to the $\mathrm{SL}(2)$ inversion with $\alpha=\delta=0$, along with a reordering of the coordinates, such that $x^1 \rightarrow \pm x^2$ and $x^2 \rightarrow \mp x^1$\,. The latter is a geometric $\mathrm{SL}(2)$ transformation. When describing the T-duality on both directions in $\mathrm{SL}(2)$ language we will omit this coordinate reordering, however.

The RR fields transform as spinors under T-duality.
We can construct the spinor representation of $O(2,2)$ in a standard way by introducing creation and annihilation operators
\be
\psi_M = ( \psi_i\,,  \psi^i ) \,,\quad
\{ \psi_i\,,  \psi^j \} = \delta_i^j\,,\quad
\{\psi_i\,,  \psi_j \} = 0 \,,\quad
\{  \psi^i, \psi^j \} = 0 \,,
\ee
realising the Clifford algebra, with the gamma matrices of $O(2,2)$ being $\Gamma = \sqrt{2} (\psi_i, \psi^i)$.
Introducing a state $|0\rangle$ such that $\psi_i |0\rangle =0$, then 
an $O(2,2)$ spinor can be expressed as 
\be
|\mathcal{C}\rangle = \mathcal{C}_0 \, |0\rangle + \mathcal{C}_{12} \, \psi^1 \, \psi^2 \, |0\rangle  + \mathcal{C}_i \, \psi^i \, |0\rangle \,.
\ee
A $\mathrm{Pin}(2,2)$ transformation $S: |\mathcal{C} \rangle \rightarrow S \, |\mathcal{C}\rangle$ defines an $O(2,2)$ transformation $\boldsymbol{\Lambda}$ which can be derived from the invariance of the gamma matrices as usual:
\be \label{SGammaSh}
	S \, ( \boldsymbol{\Lambda} \, \Gamma) S^{-1} =  \Gamma \,,
\ee
where here $S$ acts on the spinorial indices of $\Gamma$ and $\boldsymbol{\Lambda}$ acts on the covector index.
The Weyl spinors of definite chirality are $\mathcal{C}_{i}$, which transforms solely under the geometric $\mathrm{SL}(2)$
and $\mathcal{C} \equiv ( \mathcal{C}_{12}\,,\,\mathcal{C}_0 )^\intercal$, transforming solely under the other $\mathrm{SL}(2)$.
Taking $\mathcal{C}$ to transform under $\Lambda \in \mathrm{SL}(2)$ as $\mathcal{C} \rightarrow \Lambda \, \mathcal{C}$, the formula \eqref{SGammaSh} can be verified to produce an associated $O(2,2)$ transformation of the form \eqref{hSL2}.
Note this means that $\mathcal{C}$ transforms covariantly while $\mathcal{A}_\mu{}^i$ transforms contravariantly under $\mathrm{SL}(2)$.

Specialising the RR polyform \eqref{CCPoly_decompose} to the $d=2$ case, we obtain a set of polyforms in the eight undualised directions, which can be identified directly with $\mathcal{C}_i$ and $(\mathcal{C}_{12}, \mathcal{C}_{0})$, i.e.
\be
\CCpoly = \CC_0 + \CC_i \wedge \dd x^i + \CC_{12} \dd x^1 \wedge \dd x^2 \,.
\label{mathcalCRR_here}
\ee
In practice, it is convenient to repackage the transformations of the $\mathrm{SL}(2)$ doublets using the vielbein $\mathcal{V}$ such that $\gM = \mathcal{V} \, \mathcal{V}^\intercal$\,. 
Making the analogous vielbein choice to that appearing in \eqref{eq:mve}, this means we consider the combinations 
\be
\mathcal{V}^\intercal \mathcal{A}_\mu{}^i  = \begin{pmatrix}
    |G|^{\frac{1}{4}} \, A_\mu{}^i \\[4pt]
    |G|^{-\frac{1}{4}} \, B_\mu{}^i
\end{pmatrix},
        \qquad%
    \mathcal{V}^{-1} \, \mathcal{C} 
= 
\begin{pmatrix}
    |G|^{-\frac{1}{4}}  \bigl(\mathcal{C}_{12} - \CB \, \mathcal{C}_{0} \bigr)
\\[4pt]
    |G|^{\frac{1}{4}} \, \mathcal{C}_0 
\end{pmatrix},
\label{ACflat}
\ee
which transform in the same manner under the compensating local $\mathrm{SO}(2)$ transformation needed to preserve the vielbein parametrisation, 
\be
    \wt{\mathcal{V}}^T \wt{\mathcal{A}}_{\mu}{}^i 
        = h^\intercal \, \mathcal{V}^\intercal \, \mathcal{A}_{\mu}{}^i\,,
            \qquad%
    \wt{\mathcal{V}}^{-1} \, \wt{\mathcal{C}} 
        = h^{-1} \, \mathcal{V}^{-1} \, \mathcal{C} \,,
            \qquad%
    h \equiv 
        \begin{pmatrix} 
            \cos \theta &\,\, \sin \theta \\[4pt]
            - \sin\theta &\,\, \cos \theta 
        \end{pmatrix} \in \mathrm{SO}(2) \,,
\ee
with the angle $\theta$ satisfying $\tan \theta = \gamma \, |G|^{\frac{1}{2}} / (\gamma B + \delta)$\,.

\section{T-duality of Matrix \texorpdfstring{$p$}{p}-Brane Theory} 
\label{sec:appTMpT}

\subsection{Buscher Duality of Matrix \texorpdfstring{$p$}{p}-Brane Theory}
\label{sec:BuscherdMpT}

In this appendix, we discuss further details of the T-duality transformation rules for the NSNS background fields of Matrix $p$-Brane Theory (M$p$T). As before, we write the 10-dimensional coordinates as $x^\text{M}= ( x^\mu\,, \, x^i)$, where $i=1,\dots,d$ denote the isometry directions.

\vspace{1em}
\noindent $\bullet$~\emph{Gauge fixing for transverse isometries.} 
First, we focus on a $d$-dimensional toroidal compactification within the transverse sector, and thus split the transverse index as $A' = (a'\,, \, I)$\,, where $I$ is the $d$-dimensional toroidal index. We assume that the isometry is wholly transverse in the sense that the longitudinal vielbein contains no legs in the isometry directions.
With this assumption, we use the local longitudinal SO$(1\,, \, p)$ and transverse SO$(9-p)$ symmetries to gauge fix the boost invariant vielbein fields as\,\footnote{A brief justification: if $\tau'^{A}{}_\text{M}{}$ is some other choice of longitudinal vielbein also such that $\tau'^A{}_i =0$, then $\Lambda^A{}_B = \tau'^A{}_\text{M} \, \tau{}^\text{M}{}_B$ is by construction an $SO(1,p)$ rotation such that $\tau'{}^A{}_\text{M} = \Lambda^A{}_B \, \tau{}^B{}_\text{M}$. Similarly for $E^\text{M}{}_{A'}$. A careful counting of independent components, taking into account the surviving boost freedom (with parameters $l^{I}{}_A$), shows that the \eqref{MpTfixtrans1} and \eqref{MpTfixtrans2} contain the correct number of independent components, \emph{i.e.}~equal to that of a relativistic vielbein.
}
\be
 \tau^A{}_\text{M}  = \begin{pmatrix}  
    \tau^A{}_\mu &\,\, 0 
 \end{pmatrix}\,,
    \qquad%
 E^\text{M}{}_{A'} = 
 \begin{pmatrix}
    E^\mu{}^{}_{a'} & 0 \\[4pt]
    - A^{}_\nu{}^i \, E^\nu{}^{}_{a'} &\,\,  E^i{}^{}_I 
\end{pmatrix}.
\label{MpTfixtrans1}
\ee
Here, $A_\mu{}^i$ is the Kaluza-Klein vector. We assume $E^i{}_I$ to be invertible, and $\tau^A{}_\mu\, E^\mu{}{}_{a'} =0$\,.
The general form of the boost covariant transverse vielbein is then fixed to be
\be
    E^A{}_\text{M}{} = 
    \begin{pmatrix}
        E^{a'}{}_\mu & 0 \\[4pt]
        l^I{}_A \, \tau^A{}_\mu + A_\mu{}^j \, E^I{}_j &\,\,  E^I{}_i
    \end{pmatrix}, 
\label{MpTfixtrans2}
\ee
where $E^I{}_i$ is the inverse of $E^i{}^{}_I$, and $E^{a'}{}_\mu{} \, E^\mu{}{}^{}_{b'}=\delta^{a'}_{b'}$.
The arbitrary functions $l^I{}_A$ can be removed by a boost transformation (or by a field redefinition of $ A_\mu{}^i$) and we drop them henceforth. To carry out T-duality, we embed this configuration into a relativistic metric $G_{\text{M}\text{N}}$\,, which is decomposed as in \eqref{GforT} 
with
\be \label{eq:gaioexp}
    g^{}_{\mu\nu} = \omega \, \tau^{}_{\mu\nu} + \omega^{-1} \, ( E_{\mu \nu} - E_{ij} \, A_\mu{}^i A_\nu{}^j )\,, 
        \qquad%
    G^{}_{ij} = \omega^{-1} \, E^{}_{ij}\,.
\ee
Here, $\tau^{}_{\mu\nu} \equiv \tau^A{}_\mu \tau^B{}_\nu \, \eta^{}_{AB}$\,, $E_{\mu \nu} - E_{ij} A_\mu{}^i A_\nu{}^j = E^{a'}{}_\mu \, E^{a'}{}_\nu$, and $E_{ij} = E^I{}_i E^I{}_j$ is invertible. 

\vspace{1em}
\newcommand{\Ap}{\underline{A}}
\noindent $\bullet$~\emph{Gauge fixing for longitudinal isometries.} 
Next, we focus on a $d$-dimensional toroidal compactification within the longitudinal sector, and thus split the longitudinal index as $A = (a\,, \, I)$\,, where $I$ is the $d$-dimensional toroidal index. 
We assume that the isometry is wholly longitudinal in the sense that the transverse vielbein contains no legs in the isometry directions.
In analogy to \eqref{MpTfixtrans1}, we gauge fix the boost invariant vielbeins as follows:
\be
    \tau^A{}_{\text{M}} = 
        \begin{pmatrix} 
            \tau^a{}_\mu &\,\, 0 \\[4pt]
            A_\mu{}{}^j \, \tau{}^I{}_j &\,\, \tau{}^I{}_i 
        \end{pmatrix},
    \quad%
    E^\text{M}{}_{A'} = 
        \begin{pmatrix}
            E^\mu{}_{A'} \\[4pt] 
            - A_\nu{}^i \, E^\nu{}_{A'} 
        \end{pmatrix},
\label{MpTfixlong1}
\ee
where $\tau{}_\mu{}^{A} \, E^\mu{}_{A'} = 0$
and we take $\tau^I{}_i$ 
to be invertible.
To discuss T-duality of the metric, we then need to know the form of the boost covariant longitudinal vielbein $E^{A'}{}_{\text{M}}$\,.  
Using invertibility of $\tau^I{}_i$ and the boost symmetry, we can put this in the form
\be
    E^{A'}{}_\text{M} = 
        \begin{pmatrix}
            E^{A'}{}_\mu &\,\, 0 
        \end{pmatrix}. 
\label{MpTfixlong2}
\ee
We then embed this gauge fixing in a relativistic metric of the same form as \eqref{GforT}, with
\be
    g_{\mu\nu} = \omega \, \Bigl( \tau^{}_{\mu\nu} - A^{}_\mu{}^i \, \tau^{}_{ij} \, A^{}_\nu{}^j \Bigr) + \omega^{-1} \, E_{\mu\nu}\,,
        \qquad%
    G_{ij} = \omega \, \tau_{ij}\,.
\label{metric_long_decomp}
\ee
where we have
$\tau_{\mu\nu} - \tau_{ij} A_\mu{}^i A_\nu{}^j = \tau^a{}_\mu \, \tau^b{}_\nu \, \eta^{}_{ab}$, $E^{}_{\mu\nu} = E^{A'}_\mu{} \, E^{A'}{}_\nu$, and $\tau^{}_{ij} = \tau^{}_i{}^I \, \tau^{}_j{}^I$ is invertible.

\vspace{1em}\noindent
Assuming the M$p$T background fields take the form given above, we derive the following T-duality rules.

\vspace{1em} 
\noindent 
$\bullet$~\emph{Longitudinal T-duality.} 
Using the metric prescription of \eqref{metric_long_decomp}, we find that the transformation rules \eqref{eq:tgrl} gives rise to the T-dual metric
\begin{subequations} \label{BuscherMpTdLong}
\begin{align} \label{eq:mBuscherMpTdLong}
\begin{split}
    \tilde{G}_{\mu\nu} & = \omega \, \Bigl(\tau_{\mu\nu} - \tau_{ij} A_\mu{}^i A_\nu{}^j \Bigr)
    + \omega^{-1} \, 
    \Bigl[ 
        E_{\mu \nu} + 
        \bigl(b_{\mu i} - A_\mu{}^k \, b_{ki}\bigr)\, \mathscr{E}^{ij}  \, 
       \bigl(b_{\nu j} - A_\nu{}^l \, b_{lj}\bigr)
    \Bigr]\,, \\[4pt]
    {G}_{\mu i} & = \omega^{-1} \, \mathscr{E}^{ij} \, 
    \bigl( b_{\mu j} - A_\mu{}^k \, b_{kj} \bigr)\,,
        \qquad\qquad\qquad\qquad\!\!%
    {G}_{ij} =\omega ^{-1}\, \tilde{\mathscr{E}}^{ij} \,,
\end{split}
\end{align}
the T-dual $B$-field,
\begin{align}
    \tilde{B}_{\mu\nu} & = b_{\mu\nu} + 2 \, A_{[\mu}{}^i \,  b_{\nu]i} +  b_{ij} \, A_{\mu}{}^i \, A_\nu{}^j 
    - \omega^{-2} \, \tau^{ik} \, b_{kl} \, \mathscr{E}^{lj} \, \bigl( b_{\mu i} + b_{im} \, A_\mu{}^m \bigr) \, \bigl( b_{\nu j} + b_{jn} \, A_{\nu}{}^n \bigr) 
    \,, \notag \\[4pt]
    \tilde{B}_{\mu i}& = A_\mu{}^i + \omega^{-2} \, \tau^{ik} \, b_{kl} \, \mathscr{E}^{lj} \, \bigl( b_{\mu j} + b_{jm} \, A_\mu{}^m \bigr)\,,
        \qquad%
    \tilde{B}_{ij} = -\omega^{-2} \, E^{ik} \, b_{kl} \, {\mathscr{E}}^{l j}\,, 
\label{eq:BFromLongT}
\end{align}
and the T-dual dilaton,
\begin{align} \label{eq:DilFromTTrans}
    \tilde{\Phi} = \varphi + \frac{1}{2} \, \bigl( p -d - 3 \bigr) \, \ln \omega - \frac{1}{4} \, \ln  \Bigl[ \det \bigl(\tau_{ij}\bigr) \, \det \bigl({\mathscr{E}}_{ij}\bigr) \Bigr]\,,
\end{align} 
\end{subequations}
where
\be \label{eq:mscre_before}
    \mathscr{E}_{ij} = \bigl(\tau -\omega^{-2} \,\, b \, \tau^{-1} \, b \bigr)_{ij}\,,
\ee
and ${\mathscr{E}}^{ij}$ is the inverse of $\mathscr{E}_{ij}$.
Note that, in \eqref{eq:mscre_before}, when $\tau$\,, $E$\,, and $b$ appear without indices they implicitly carry the toroidal index $i$ and $\tau^{-1}$ refers to the inverse of $\tau_{ij}$\,. 
For $\omega \rightarrow\infty$\,, $\mathscr{E}^{}_{ij} \rightarrow \tau^{}_{ij}$\,, 
and we recover the defining prescription for an M$(p-d)$T limit, gauge fixed as in \eqref{MpTfixtrans1} and \eqref{MpTfixtrans2} with 
\begin{subequations} \label{BuscherMpTdLongzb}
\begin{align}
    \tilde{G}^{}_{\mu\nu} & =\omega \, \tilde{\tau}^{}_{\mu\nu} + \omega^{-1} \, \tilde{E}^{}_{\mu\nu}\,,
        &%
    \tilde{B}^{}_{\mu\nu} & = \tilde{b}^{}_{\mu\nu}\,,
        &%
    \tilde{\Phi} & = \tilde{\varphi} + \frac{1}{2} \, \bigl( p - d-3 \bigr) \, \ln \omega\,,
\end{align} 
\end{subequations}
where the components of the new M$(p-d)$T background fields are:
\begin{subequations} \label{eq:MpTFromLongT}
\begin{align}
    \tilde{E}^{}_{\mu\nu} & = E_{\mu\nu} +  
    \bigl(b_{\mu i} - A_\mu{}^k \, b_{ki}\bigr) \,
    \tau^{ij} \,
    \bigl(b_{\nu j} - A_\nu{}^l \, b_{lj}\bigr)\,,
        &%
    \tilde{\varphi} & = \varphi - \tfrac{1}{2} \ln  \det \bigl( \tau^{}_{ij} \bigr)\,,  \\[4pt]
    \tilde{E}^{}_{\mu i }&  = \tau^{ij} \, 
    \bigl(b_{\mu j} - A_\mu{}^k \, b_{kj}\bigr)\,,&
   \tilde{E}^{}_{ij} & = \tau^{ij}\,,\\[4pt]
    \tilde{\tau}^{}_{\mu\nu} & = \tau^{}_{\mu\nu} - \tau^{}_{ij} \, A^{}_\mu{}^i \, A^{}_\nu{}^j\,,
        &%
    \tilde{\tau}^{}_{\mu i} & = \tilde{\tau}^{}_{ij} = \tilde{b}^{}_{ij} = 0\,, \\[4pt]
    \tilde{b}^{}_{\mu\nu} & = b_{\mu\nu} + 2 \, A_{[\mu}{}^i \,b_{\nu]i} + b_{ij} \, A_\mu{}^i \, A_\nu{}^j\,,
        &%
    \tilde{b}^{}_{\mu i} & = A^{}_\mu{}^i\,.
\end{align}
\end{subequations}
where $\tau^{ij}$ denotes the inverse of $\tau^{}_{ij}$\,, $A_\mu{}^i \equiv \tau^{ij} \, \tau_{\mu j}$\,, and we note that 
$\tilde{\tau}^{}_{\mu\nu} = \tau^a{}_\mu \, \tau^b{}_\nu \, \eta^{}_{ab}$\,.
More specifically, this is M($p-d)$T with zero $B$-field on the transverse isometry directions.

\vspace{1em}
\noindent 
$\bullet$~\emph{Transverse T-duality.} 
In this case, the presence of the transverse $B$-field induces a much richer set of dual limits.
Using the metric prescription of \eqref{eq:gaioexp}, we find that the transformation rules \eqref{eq:tgrl} gives rise to the T-dual metric
\begin{subequations} \label{BuscherMpTdTrans}
\begin{align} \label{eq:mBuscherMpTdTrans}
\begin{split}
    \tilde{G}_{\mu\nu} & =\omega \, \Bigl[  \tau_{\mu\nu} + \mathscr{E}^{ij} \, \bigl( b_{\mu i} -  A_\mu{}^k \, b_{ki} \bigr) \, \bigl( b_{\nu j} -  A_\nu{}^l \, b_{lj} \bigr) \Bigr] + \omega^{-1} \,( E_{\mu \nu} - E_{ij} A_\mu{}^i A_\nu{}^j )\,,
    \\[4pt] 
    \tilde{G}_{\mu i} & = \omega \, \mathscr{E}^{ij} \, \bigl( b_{\mu j} -  A_\mu{}^k \, b_{kj} \bigr)\,,
        \qquad\qquad\qquad\qquad\,\,\,\,%
    \tilde{G}_{ij} =\omega \, \mathscr{E}^{ij} \,,
\end{split}
\end{align}
the T-dual $B$-field,
\begin{align}
    \tilde{B}_{\mu\nu} & = b_{\mu\nu} + 2 \, A_{[\mu}{}^i \, b_{\nu]i} +  b_{ij} \, A_{\mu}{}^i \, A_\nu{}^j 
   - \omega^2 \, E^{ik} \, b_{kl} \, \mathscr{E}^{lj} \, \bigl( b_{\mu i} + b_{im} \, A_\mu{}^m \bigr) \, \bigl( b_{\nu j} + b_{jn} \, A_\mu{}^n \bigr)\,, \notag \\[4pt]
    \tilde{B}_{\mu i} &  = A_\mu{}^i+\omega^2 \, E^{ik} \, b_{kl} \, \mathscr{E}^{lj} \, \bigl( b_{\mu j} + b_{jm} \, A_\mu{}^m \bigr) \,,
        \qquad%
    \tilde{B}_{ij} = -\omega^{2} \, E^{ik} \, b_{kl} \, \mathscr{E}^{\ell j}\,, 
\end{align}
and the T-dual dilaton,
\begin{align} \label{eq:dit}
    \tilde{\Phi} = \varphi + \frac{1}{2} \, \bigl( p + d - 3 \bigr) \, \ln \omega - \frac{1}{4} \, \ln  \Bigl[ \det (E_{ij}) \, \det(\mathscr{E}_{ij}) \Bigr]\,,
\end{align} 
\end{subequations}
where now we have
\be \label{eq:mscre}
    \mathscr{E}_{ij} = \bigl(E-\omega^2 \, b \, E^{-1} \, b \bigr)_{ij}\,. 
\ee
Again, in \eqref{eq:mscre}, both $E$ and $b$ carry the toroidal index $i$ and $(E^{-1})^{ij} = E^{ij}$ is the inverse of $E$\,. 
Now, the $\omega \rightarrow \infty$ limit is not immediate, as $\mathscr{E}_{ij}$ diverges.

Firstly, we can consider a general $d$-torus but with $B_{ij} = b_{ij} = 0$ on the torus.
Then we find $\mathscr{E}^{ij} = E^{ij}$ and
\begin{subequations} \label{BuscherMpTdTranszb}
\begin{align}
    \tilde{G}^{}_{\mu\nu} & =\omega \, \tilde{\tau}^{}_{\mu\nu} + \omega^{-1} \, \tilde{E}^{}_{\mu\nu}\,,
        &%
    \tilde{B}^{}_{\mu\nu} & = \tilde{b}^{}_{\mu\nu}\,,
        &%
    \tilde{\Phi} & = \tilde{\varphi} + \frac{1}{2} \, \bigl( \tilde{p} - 3 \bigr) \, \ln \omega\,,
\end{align} 
\end{subequations}
where $\tilde{p} = p + d$ and where we can express the components of the $\tilde \tau_{\mu\nu}$, $\tilde E_{\mu\nu}$ and $\tilde b_{\mu\nu}$ directly in terms of those of $\tau_{\mu\nu}$, $E_{\mu\nu}$ and $b_{\mu\nu}$ as follows:
\begin{subequations} \label{eq:mptmpntbr}
\begin{align}
    \tilde{\tau}^{}_{\mu\nu} & = \tau^{}_{\mu\nu} + E^{ij} \, b^{}_{\mu i} \, b^{}_{\nu j}\,,
        &%
    \tilde{\tau}^{}_{\mu i} & = E^{ij} \,b^{}_{\mu j}\,,
        &%
    \tilde{\tau}^{}_{ij} & = E^{ij}\,, \\[4pt]
    \tilde{E}^{}_{\mu\nu} & = E_{\mu\nu} - E_{ij} \, A_\mu{}^i \, A_\nu{}^j\,,
        &%
    \tilde{E}^{}_{\mu i } & = 0\,,
        &%
    \tilde{E}^{}_{ij} & = 0\,, \\[4pt]
    \tilde{b}^{}_{\mu\nu} & = b_{\mu\nu} + 2 \, A_{[\mu}{}^i \, b_{\nu]i}\,,
        &%
    \tilde{b}^{}_{\mu i} & = A^{}_\mu{}^i\,, 
        &%
    \tilde{b}^{}_{ij} & = 0\,,
    \\[4pt]
    \tilde{\varphi} & = \varphi - \tfrac{1}{2} \, \ln  \det \bigl( E_{ij} \bigr)\,,
\end{align}
\end{subequations}
where $E^{ij}$ denotes the inverse of $E_{ij}$ and $A_\mu{}^i = E^{ij} E_{\mu j}$.

In the $\omega \rightarrow \infty$ limit, this T-duality transformation maps M$p$T to M($p$+$d$)T with the related Buscher rules given by \eqref{eq:mptmpntbr}. 
The dual internal $B$-field $\tilde{b}_{ij}$ on the torus is consistently zero. 
If we make the restriction $\tilde b_{ij}$ in the inverse transformations \eqref{eq:MpTFromLongT}, then the multiple direction Buscher rules \eqref{eq:MpTFromLongT} and \eqref{eq:mptmpntbr} express the T-duality between M$p$T and M$(p+d)$T with the restriction that $b_{ij} = \tilde b_{ij} =0$.
This also includes the case $d=1$, for which the $B$-field is trivially zero, and we recover the Buscher transformation rules between M$p$T and M$(p\pm 1)$T given previously in~\cite{Gomis:2023eav, Blair:2024aqz} and used to generate the duality web of~\cite{Blair:2023noj}.

Now, we suppose that the $B$-field has non-zero components in the internal transverse directions, such that \emph{not} all components of $b_{ij}$ vanish in M$p$T. 
Suppose that the internal $B$-field $b_{ij}$ has rank $r$.
Then we can bring it into block-diagonal form using a special orthogonal transformation, such that splitting $i=(a,\mathbf{m})$ with $a=1,\dots,r$ and $\mathbf{m} = r+1,\dots,d$ we have $b_{\mathbf{m}\mathbf{n}} = b_{\mathbf{m} a} = 0$ and $b_{ab}$ is given by \eqref{bblockdiag}, 
so that $\det b_{ab} = \lambda_1^2 \dots \lambda_{r/2}^2 \neq 0$. 
We can figure out most easily what the T-dual theory is in the $\omega \rightarrow \infty$ limit by examining the dilaton transformation~\eqref{eq:dit}.
We need to compute
\be
\begin{split} 
    \det \mathscr{E}_{ij}& = \det 
    \begin{pmatrix}
        E_{ab} - \omega^2 \, b_{ac} \, \bigl( E^{-1} \bigr)^{cd} \, b_{db} & E_{a \mathbf{n}} \\[4pt]
        E_{\mathbf{m} a} & E_{\mathbf{m} \mathbf{n}}
    \end{pmatrix} \\[4pt]
    & = \omega^{r} \Big[  \det \bigl( E_{\mathbf{m} \mathbf{n}} \bigr)^2 \bigl(\det E_{ij}\bigr)^{-1} \bigl(\det b_{ab}\bigr)^2 + \mathcal{O}\bigl(\omega^{-2}\bigr) \Bigr] \,.
\end{split} 
\ee
Consequently we find 
\begin{align} 
    \tilde{\Phi} = \tilde{\varphi} + \frac{1}{2} \bigl( p + d - r - 3 \bigr) \ln \omega \,,
        \qquad%
    \tilde{\varphi} = \varphi - \frac{1}{2} \, \ln  \det \bigl( E_{\mathbf{m}\mathbf{n}} \bigr) - \ln \bigl| \lambda_1 \, \cdots \lambda_{r/2} \bigr|\,.
\end{align}
Furthermore, the new internal $B$-field can be shown to be finite, taking the form
\be
\tilde b_{ij} = 
    \begin{pmatrix}
        (b_{ab})^{-1} &\,\, (b_{ac})^{-1} \, \beta_c{}^{\mathbf{n}} \\[4pt]
        (b_{bc})^{-1} \, \beta_c{}^{\mathbf{m}} &\,\, (b_{cd})^{-1} \, \beta_c{}^{\mathbf{m}} \, \beta_d{}^{\mathbf{n}}
    \end{pmatrix},
        \quad%
    \beta_a{}^{\mathbf{m}} \equiv - \bigl(E^{ab}\bigr)^{-1} E^{b \mathbf{m}}\,.
\ee
It follows that the full $B$-field in \eqref{eq:mBuscherMpTdTrans} is likewise finite.
We conclude that when the rank of the internal $B$-field is $r$\,, M$p$T is mapped to M($p+d-r$)T under the T-duality transformation along all the directions on the $d$-torus. 

It may seem that this result is in tension with the fact that we have shown both above and in~\cite{Blair:2024aqz} that a single Buscher duality is sufficient to map M$p$T to M$(p+1)$T, with the $B$-field components playing no role in the validity of this statement. Applying $d$ consecutive T-duality on different isometry transformations it would then seem to imply that one should be able to map M$p$T to M$(p+d)$T directly. 
The resolution of this puzzle is to note that our transverse T-duality rules above are derived with the assumption that $\tau^A$ has no legs on the isometry directions, and that T-duality with respect to a subset of these isometries does not lead to a background obeying this restriction. 
Indeed, suppose one has two isometry directions, denoted $y$ and $z$, and one T-dualises first on the $y$ direction alone, assuming the condition $\tau^A{}_y= 0$.
From \eqref{eq:mptmpntbr} specialised to the $d=1$ case we read off the dual component $\tilde \tau_{zy} = B_{zy} / E_{yy}$\,.
Hence if there is initially a $B$-field component $B_{zy}$ then the condition $\tilde \tau^A{}_z = 0$ does not hold, and we cannot continue to apply the $d=1$ Buscher rules following from \eqref{eq:mptmpntbr}.
This can be further illustrated explicitly in the example of M0T on a transverse 2-torus studied in Section~\ref{sec:mztttt}.
T-dualising on the direction $x_1$ of the background \eqref{eq:omzt} 
leads to the intermediate geometry
\begin{subequations} \label{eq:1mzt}
\begin{align}
	\dd s^2 &= \omega \, \Bigl[ - \dd t^2 + \bigl( \dd \tilde x_1 - \CB \, \dd x_2 \bigr)^2 \Bigr]
+ \frac{1}{\omega} \Bigl(
\dd x_2^2 + \dd x^m \, \dd x^m
\Bigr) \,,
\end{align}
with $B^{(2)} = 0$ and $e^\Phi = \omega^{-1} \, g^{}_\text{s}$
\end{subequations}
describing an M1T limit. 
In the present case, wanting to T-dualise on the direction $x_2$ of \eqref{eq:1mzt}, it is clear that $\tau^A{}_{x_2} \neq 0$ unless $\CB = 0$.
So therefore the naive Buscher T-duality rules which would relate M1T to M2T do not apply.
We leave further extension of the T-duality transformations of M$p$T, allowing for arbitrary background field configurations respecting the isometry, to future work. 

\subsection{General \texorpdfstring{$O(d,d; \mathbb{Z})$}{O(d,d)} Transformations} 
\label{sec:GeneralTMpT}

Here we outline how to extend the analysis to general $O(d,d; \mathbb{Z})$ transformations of the form \eqref{defineLambdaOdd}.
We focus just on the dual dilaton and the dual metric in the isometry directions as these are sufficient to determine the nature of the dual theory.
These are given by:
\be
    \tilde G = \Bigl[  \op \, \bigl(G+B\bigr) + \oq \Bigr]^{-\intercal} G\, \Bigl[ \op \, \bigl(G+B\bigr)+ \oq \Bigr]^{-1},
        \quad%
e^{\tilde \Phi} = e^\Phi \, \det \Bigl[ \op \, \bigl(G+B\bigr)+\oq \Bigr]^{-\frac{1}{2}}.
\label{TDualGeneral}
\ee
Recall that the Buscher transformation has $\op=\mathbb{1}$ and $\oq=0$.
We then consider the results of longitudinal and transverse T-duality transformations of M$p$T.

\vspace{1em}
\noindent $\bullet$ \emph{Longitudinal $O(d,d; \mathbb{Z})$ transformation.} Given $G = \omega \, \tau$ and $e^\Phi = \omega^{\tfrac12(p-3)} \, e^\varphi$,  the transformation rules \eqref{TDualGeneral} give
\begin{align}
    \tilde G &  =  \omega^{-1} \Bigl[\op + \omega^{-1} \bigl(\op B +\oq\bigr) \, \tau^{-1} \Bigr]^{-\intercal} \,  \tau^{-1}  \, \Bigl[ \op + \omega^{-1} \, \bigl( \op \, B +\oq \bigr) \, \tau^{-1} \Bigr]^{-1}, \\[4pt]
    e^{\tilde \Phi} & = \omega^{\frac{p-3-d}{2}} \, e^\varphi \, \bigl(\det \tau\bigr)^{-\frac{1}{2}} \, \det\Bigl[\op + \omega^{-1} \, \bigl(\op B +\oq\bigr) \, \tau^{-1} \Bigr]^{-\frac{1}{2}}.
\end{align} 
The nature of the dual decoupling limit then depends on the rank of $\op$:

\begin{enumerate}[(1)]
\item {\bf$\op=0$\,: from M$p$T to M$p$T.} For $\op=0$, we obtain
\begin{align}
\tilde G &  = \omega \, \oq^{-\intercal} \, \tau \, \oq^{-1}  \,,
    \qquad%
e^{\tilde \Phi}  = \omega^{\tfrac{p-3}{2}} \, e^\varphi \, \bigl( \det \oq \bigr)^{-\frac{1}{2}}\,,
\end{align} 
and the transformation therefore takes us to a dual M$p$T prescription.

\item {\bf$\det \op \neq 0$\,: from M$p$T to M$(p-d)$T.} If $\op$ is invertible, then in the limit $\omega \rightarrow \infty$, we get
\begin{align}
\tilde G &  =  \omega^{-1} \, \op^{-\intercal} \, \tau^{-1} \, \op^{-1} \,,
    \qquad%
e^{\tilde \Phi}  = \omega^{\frac{p-3-d}{2}} \, e^\varphi \, (\det \tau)^{-\frac{1}{2}} \, (\det \op)^{-\frac{1}{2}}\,.
\end{align} 
which gives a transformation to a dual M$(p-d)$T prescription.

\item {\bf $\text{rank}(\op) = r$\,, $0<r<d$\,: from M$p$T to M$(p-r)$T.} 
This can be argued from the dilaton transformation rule.
Let $\chi = ( \op \, B + \oq) \, \tau^{-1}$.
We need to evaluate the following determinant:
\be
    \det \! \lr\op+ \frac{\chi}{\omega} \rr
    = \omega^{r-d} \, \Bigl[ 
    \mathcal{Q} + O\bigl(\omega^{-1} \bigr)
 \Bigr]\,,
\ee
where
\be
    \mathcal{Q} \equiv \frac{\epsilon_{i_1 \dots 
        i_{d}} \, 
    \epsilon_{j_1 \dots 
        j_{d}} \, 
    \op^{i_1j_1} \cdots \op^{i_r j_r} \chi^{i_{r+1} j_{r+1}} \cdots \chi^{i_d j_d}}{r! \, (d-r)!}\,.
\label{detrank1}
\ee
Subadditivity of the matrix rank implies  $
\text{rank}\bigl(\op+ \omega^{-1} \, \chi \bigr) \leq \text{rank}(\op) + \text{rank}\bigl(\omega^{-1} \, \chi \bigr)$.
We assume that $\op+\omega^{-1}\chi$ is invertible so that the metric T-duality transformation does not break down.
Hence $\text{rank}(\omega^{-1} \, \chi ) \geq d - r$, and
so the expression in \eqref{detrank1} is the leading divergence, and 
\be
e^{\tilde \Phi}  = \omega^{\frac{p-3-r}{2}} \, e^\varphi \,  (\det \tau)^{-1/2} \, \mathcal{Q}^{-1/2} \,,
\ee
signals that we are in M$(p-r)$T.
\end{enumerate}

\noindent  $\bullet$ \emph{Transverse $O(d,d; \mathbb{Z})$ transformation.}
Given $G = \omega^{-1} E$, $e^\Phi = e^\varphi \omega^{\tfrac12(p-3)}$, the transformation rules \eqref{TDualGeneral} give:
\begin{subequations}
\begin{align}
    \tilde G &  =  \omega \, \Bigl[\op \, E  + \omega \, \bigl( \op \, B +\oq \bigr) \Bigr]^{-\intercal} \, E \, \Bigl[ \op \, E + \omega^{1} \, \bigl( \op \, B +\oq \bigr) \Bigr]^{-1}, \\[4pt]
    e^{\tilde \Phi} & = \omega^{\tfrac{p-3+d}{2}} \, e^\varphi \, \det \Bigl[\op \, E + \omega \, \bigl( \op \, B + \oq \bigr) \Bigr]^{-\frac{1}{2}}.
\end{align} 
\end{subequations}
The nature of the dual decoupling limit now depends on the rank of $\op B+\oq$:

\begin{enumerate}[(1)]

\item {\bf$\op B+\oq=0$\,: from M$p$T to M$(p+d)$T.} If $\oq= -\op \, B$ then the condition $\om \, \oq^\intercal + \on \, \op^\intercal = \mathbb{1}$ implies that $(\om \, B+\on) \, \op^\intercal = \mathbb{1}$.
Therefore $\op$ is invertible in this case, and we get:
\begin{align}
    \tilde G = \omega \, \op^{-\intercal} \, E^{-1} \, \op^{-1}\,,
        \qquad%
    e^{\tilde{\Phi}} = \omega^{\tfrac{p-3+d}{2}} \, e^\varphi \, \bigl(\det E\bigr)^{-\frac{1}{2}} \, \bigl(\det \op\bigr)^{-\frac{1}{2}}\,.
\end{align} 
This is a transformation to a dual M$(p+d)$T prescription.

\item {\bf$\det (\op B+\oq)\neq0$\,: from M$p$T to M$p$T.}
In this case, we find 
\begin{align}
\tilde G   =  \omega^{-1} \, \bigl(\op B +\oq\bigr)^{-\intercal} \, E \, \bigr( \op B +\oq \bigr)^{-1}  \,,
    \qquad%
e^{\tilde \Phi}  = \omega^{\tfrac{p-3}{2}} \, e^\varphi \, \det \bigl( \op B +\oq \bigr)^{-\frac{1}{2}}\,,
\end{align} 
which implies that the dual frame is still M$p$T. 

\item {\bf$\text{rank} (\op B+\oq)=r$\,, $0<r<d$\,: from M$p$T to M$(p+d-r)$T.}
We can again see this behaviour simply by extracting the leading power of $\omega$ arising from the determinant appearing in the dilaton transformation rule.
Letting $\boldsymbol{\kappa} \equiv \op B + \oq$, we have
\begin{align} 
    \det  \bigl( \op E + \omega \, \boldsymbol{\kappa} \bigr) 
    = \omega^r \, \Bigl[ \CR + \mathcal{O}\bigl(\omega^{-1}\bigr) \Bigr]\,,
\end{align}
where
\be
    \CR = \frac{\epsilon_{i_1 \dots i_d} \, \epsilon^{j_1 \dots j_d} \, \boldsymbol{\kappa}^{i_{1}}{}_{j_{1}} \dots
\boldsymbol{\kappa}^{i_{r}}{}_{j_{r}} 
( \op E)^{i_{r+1}}{}_{j_{r+1}} \dots (\op E)^{i_{d-r}}{}_{j_{d-r}} 
}{r!(d-r)!}\,.
\ee
This then implies
\be
	e^{\tilde \Phi} = \omega^{\frac{(p+d-r)-3}{2}} \, e^\varphi \,  \mathcal{R}^{-\frac{1}{2}}\,,
\ee
signalling that we are in M$(p+d-r)$T.

\end{enumerate}

\bibliographystyle{JHEP}
\addcontentsline{toc}{section}{References}
\bibliography{dnlb}

\end{document}